\documentclass[sigconf]{acmart}

\usepackage{textcomp}
\usepackage{setspace}
\usepackage{latexsym,fancyhdr,url}
\usepackage{enumerate}
\usepackage[linesnumbered,ruled]{algorithm2e}
\usepackage{graphics}
\usepackage{xparse} %
\usepackage{xspace}
\usepackage{multirow}
\usepackage{csvsimple}
\usepackage{balance}
\usepackage{amsmath}
\usepackage{booktabs}
\usepackage{enumitem}
\usepackage{colortbl}
\usepackage{fancyhdr}
\usepackage{makecell}
\usepackage{xcolor, eucal}
\usepackage{array}
\usepackage{subfigure}
\usepackage{graphicx}
\usepackage{caption}
\usepackage{fix-cm}
\usepackage{tabularx}
\usepackage{dsfont}
\usepackage{bbding}
\usepackage{pifont}
\usepackage{url}
\definecolor{gray0}{gray}{0.9}
\definecolor{mine}{RGB}{205, 232, 248} 
\usepackage{algpseudocode}

\usepackage{graphicx}
\usepackage{textcomp}
\usepackage{booktabs}
\usepackage{hyperref}
\usepackage{multirow}
\usepackage{makecell}
\usepackage{tikz}

\usepackage{amsmath,amsthm}
\usepackage{epsfig,endnotes}
\usepackage{xspace}
\usepackage{enumitem}
\usepackage{cleveref}
\usepackage{float}
\usepackage{xspace}

\def \toolname{DPImageBench}

\newcommand\gc[1]{{\color{black}  #1}}

\newtheorem{definition}{\bf Definition}[]

\setcopyright{acmlicensed}
\copyrightyear{2025}
\acmYear{2025}
\setcopyright{acmlicensed}\acmConference[CCS '25]{Proceedings of the 2025 ACM SIGSAC Conference on Computer and Communications Security}{October 13--17, 2025}{Taipei, Taiwan, China}
\acmBooktitle{Proceedings of the 2025 ACM SIGSAC Conference on Computer and Communications Security (CCS '25), October 13--17, 2025, Taipei, Taiwan, China}

\begin{document}
\pagestyle{plain}
\def\thetitle{\toolname: A Unified Benchmark for Differentially Private Image Synthesis}

\title{\thetitle}

\author{Chen Gong}
\authornote{Co-first author. Kecen is with the Chinese Academy of Science, and this work is done in his personal capacity as a remote intern at UVA.}
\affiliation{
    \institution{University of Virginia, USA}\country{}
}

\author{Kecen Li}
\authornotemark[1]
\affiliation{
    \institution{University of Virginia, USA}\country{}
}

\author{Zinan Lin}
\affiliation{
    \institution{Microsoft Research, USA}\country{}
}

\author{Tianhao Wang}
\affiliation{
    \institution{University of Virginia, USA}\country{}
}

\date{}
\fancyhf{} 
\fancyfoot[C]{\thepage}

\begin{abstract}
Differentially private (DP) image synthesis aims to generate artificial images that retain the properties of a sensitive image dataset while protecting the privacy of individual images within the dataset. Despite recent advancements, we find that inconsistent--and sometimes flawed--evaluation protocols have been applied across studies. This not only impedes the understanding of current methods but also hinders future advancements in the field.

To address the issue, this paper introduces \toolname{} for DP image synthesis, with thoughtful design across several dimensions: (1) \textit{Methods.} We study twelve prominent methods and systematically characterize each based on model architecture, pretraining strategy, and privacy mechanism. (2) \textit{Evaluation.} We include nine datasets and seven fidelity and utility metrics to thoroughly assess these methods. Notably, we find that the common practice of selecting downstream classifiers based on the highest accuracy on the sensitive test set not only violates DP but also overestimates the utility scores. \toolname{} corrects for these mistakes. (3) \textit{Platform.} Despite the wide variety of methods and evaluation protocols, \toolname{} provides a standardized interface that accommodates current and future implementations within a unified framework. 
With \toolname{}, we have several noteworthy findings. For example, contrary to the common wisdom that pretraining on public image datasets is usually beneficial, we find that the distributional similarity between pretraining and sensitive images significantly impacts the performance of the synthetic images and does not always yield improvements. In addition, adding noise to low-dimensional features, such as the high-level characteristics of sensitive images, is less affected by the privacy budget compared to adding noise to high-dimensional features, like weight gradients. 
The source code is available.\footnote{\label{link:opensource}\url{https://github.com/2019ChenGong/DPImageBench}}

\end{abstract}
\vspace{-2mm}



\settopmatter{printacmref=false}
\maketitle

\section{Introduction}
\label{sec:intro}
\vspace{-1mm}
Privacy-preserving synthetic data aims to generate artificial data that maintains the properties of real data, enabling data sharing within and across organizations while minimizing privacy risks~\cite{hu2023sok,lin2022data}. Differentially private (DP) dataset synthesis~\cite{hu2023sok,symster} provides a theoretical guarantee to quantify privacy leakage of real data using the synthetic dataset. In recent years, there has been significant development of DP dataset synthesis methods across various domains, including images~\cite{li2023PrivImage,dpsda,lin2025differentially,dpdm}, tabular data~\cite{privmrf,aim}, text~\cite{yue-etal-2023-synthetic,xiedifferentially,yu2023selective}, networking data~\cite{lin2020using,yin2022practical,sun2024netdpsyn} and more.

Among them, DP image synthesis has witnessed impressive progress~\cite{dpsda,lin2025differentially,li2023PrivImage,dp-kernel,dplora} but lacks a comprehensive understanding of the strengths and weaknesses of existing synthesizers. 
While there are several studies summarizing advances in this field, 
they either focus on discussions~\cite{chen2023unified,hu2023sok} and do not perform much evaluation, or only cover subfields~\cite{Dpmlbench,pate-bench}.
This paper focuses on benchmarking DP image dataset synthesis, addressing several challenges:

\begin{itemize}[leftmargin=*]
    \item \emph{Platform:} The complex codebases from different methods make it challenging to evaluate different methods easily. For example, failing to standardize the pretraining dataset and model architectures may lead to unfair comparisons. 
    \item \emph{Evaluation:} The lack of unified metrics makes it difficult to evaluate the quality of synthetic images. 
    Existing methods, such as PE~\cite{dpsda} and PrivImage~\cite{li2023PrivImage}, both utilize FID~\cite{fid} for fidelity evaluation. However, FID does not assess whether the generated images align with human perception, as described in Section~\ref{subsec:fidelity}.
    Even worse, the standard practice that evaluates synthetic image quality by tuning classifiers on the {\it sensitive test set} to select optimal parameters (used in several methods~\cite{dplora,gs-wgam}) is incorrect: they both violate DP and inflate the results.
\end{itemize}

\noindent \textbf{Contributions.} Without a clear understanding of existing methods and an easy-to-use benchmarking platform, it is challenging for the community to make meaningful and reproducible advancements in this field. To bridge this gap, we establish \toolname, a benchmark focused on DP image synthesis. \toolname{} addresses the above challenges by making the following contributions:

\noindent \textbf{\emph{Taxonomy}}. We taxonomize methods from two orthogonal dimensions, covering twelve state-of-the-art DP image synthesis algorithms. (1) Private Data Perspective. Based on where noise is added to satisfy DP, we categorize methods into \textit{(a) Input-level}: adding noise to (some format of) the sensitive images; \textit{(b) Model-level}: incorporating noise into the gradient or loss during model training; and \textit{(c) Output-level}: introducing noise during the synthesis process. (2) Public Data Perspective. Existing works~\cite{li2023PrivImage,dpldm} use publicly available data from open platforms without privacy concerns to enhance DP image synthesis. We categorize them into three types based on how the public data is utilized: leveraging pretrained models, public datasets without selection, and public datasets with selection. Besides, we argue that the state-of-the-art method DP-Promise~\cite{dppromise} does not strictly satisfy DP in Appendix~\ref{app:dppromise} to prevent others from wrongly using it.

\noindent \textbf{\emph{Platform.}} We establish a platform, \toolname{}, with the following contributions: 
    (1) Fair comparison. \toolname{} standardizes the public images, hyper-parameters, and model architectures to ensure fair comparisons. \toolname{} offers an interface to pretrain the synthesizer, even for the ones that were not originally designed to use a public dataset for pretraining. (2) Straightforward implementation.
    \toolname{} features an abstract interface compatible with any synthesis algorithm. With its modular design, \toolname{} provides a flexible tool that enables users to select public and sensitive datasets, and to integrate new algorithms by adding functional code to the relevant modules.  Since \toolname{} isolates the parameter configurations from implementations, users typically only need to modify the configuration files to tune algorithms. (3) Open Source. \toolname{} is open-sourced.$^{\text{\ref{link:opensource}}}$ It serves as a starting point for the implementation and evaluation of future methods as well. 
    
\noindent \textbf{\emph{Evaluation.}} We have conducted comprehensive evaluations, and some key highlights include: (1) Datasets. \toolname{} implements synthesis methods on \textbf{seven}  sensitive datasets and \textbf{two} public dataset. (2) {Fidelity.} We establish metrics to assess synthetic images based on similarity, diversity, novelty, and alignment with human perception~\cite{xu2024imagereward}. (3) {Utility.} {We correct the common evaluation mistakes of using \emph{sensitive test} images for hyper-parameter tuning. Instead, \toolname{} adopts the standard of using distinct training, validation, and test sets for utility evaluations. Downstream classifier training is performed on the training set, hyper-parameter selection is done on the validation set, and final results are reported on the test set. \toolname{} provides two choices of validation set: (1) the synthetic images, and (2) images having the same distribution as the test set but not overlapping with it. In the latter, \toolname{} adds noise to the validation results to satisfy DP, which is detailed in Section~\ref{subsec:utility_error}. }
    
\noindent \gc{\textbf{\emph{Insights to new techniques.}} The proposed taxonomies and straightforward implementations enable us to analyze the potential of combining different methods and bring insight into developing new technology. We consider some methods mostly in isolation,  though they can be effectively combined. For example, the public dataset selection method from PrivImage~\cite{li2023PrivImage} and creating shortcuts for DP training in DP-FETA \cite{dp-feta}, as described in Section~\ref{sec:taxonomy}, can be paired with other methods. We present details in Section~\ref{subsec:combining_improvement}.}

\noindent \textbf{Main findings.} From experimental results, we observe several interesting findings, which are described as follows. 
\vspace{-1mm}
\begin{itemize}[leftmargin=*]
    \item Current methods tend to overestimate the utility of synthetic images. Compared to directly reporting the highest accuracy of classifiers on test sets of the sensitive dataset, (1) using synthetic images as validation sets or (2) adding noise to validation results of the sensitive dataset to ensure DP and avoid bias for classifier tuning, leads to an accuracy decrease of 4.48\% and 1.33\% under the privacy budget $\epsilon=1$, averaged across seven sensitive datasets.
    \item Current methods using FID~\cite{fid} and IS~\cite{is} are limited in their ability to comprehensively evaluate fidelity. Neither of the metrics considers the novelty or alignment with human perception of synthetic images.
    \item {Pretraining with public datasets does not always benefit DP image synthesis. The effectiveness of pretraining depends heavily on the choice of public dataset and the pretraining strategy (i.e., conditional or unconditional pretraining). Conditional pretraining yields better synthetic performance than unconditional pretraining. A pretraining dataset similar to the sensitive images can enhance synthetic performance. }
    \item The performance of adding noise to low-dimensional features, such as high-level characteristics of sensitive images in DP-MERF~\cite{dp-merf} and training loss in DP-Kernel~\cite{dp-kernel}, is less sensitive to privacy budgets compared to adding noise to high-dimensional features, like weight gradients using DP-SGD. The former methods perform better than the latter under a low privacy budget.
    \item  %
    \gc{The combinations of algorithmic improvements, such as integrating DP-FETA~\cite{dp-feta} into other methods, do not consistently enhance synthetic performance. This inconsistency underscores the need for further research to effectively leverage the strengths of diverse algorithms.}
\end{itemize}

\vspace{-1mm}
\section{Preliminaries}
\subsection{Privacy Concepts and Mechanisms}
\label{sub:dp}

\toolname{} concentrates on image synthesis under \textit{differential privacy} (DP) constraint. DP~\cite{dp} provides mathematical guarantees that quantify and limit the amount of information one can derive from the algorithm's output.

\begin{definition}[($\epsilon,\delta$)-DP~\cite{dp}] A randomized algorithm $\mathcal{A}$ satisfies ($\epsilon,\delta$)-DP if and only if, for any two neighboring datasets $D$ and $D'$ and all possible sets of outputs $\mathcal{O}$ of the mechanism, the following condition holds: $\Pr[\mathcal{A}(D) \in \mathcal{O}] \leq e^\varepsilon \Pr[\mathcal{A}(D') \in \mathcal{O}] + \delta$.
\end{definition}

The privacy budget $\epsilon$ quantifies how much information can be exposed by the algorithm $\mathcal{A}$. 
A smaller value of $\epsilon$ implies that the algorithm provides stronger privacy guarantees. The $\delta$ can be intuitively understood as failure probability~\cite{dp}. Given two datasets $D$ and $D'$, they are considered neighbors (denoted as $D\sim D'$) if we can obtain one from the other by adding, removing, or replacing a single image. We define the $\ell_2$ sensitivity (intuitively, a stability measure) of a function $f$ as ${\Delta_f} = \max_{D \sim D'} \left| f(D) - f(D') \right|_2$. As presented in Table~\ref{tab:notion} of Appendix~\ref{app:privacy-notations}, two widely used definitions for neighboring datasets exist in existing DP image synthesis: the `replace-one' (bounded DP) and `add-or-remove-one' (unbounded DP) definitions. Specifically, given two datasets $D$ and $D'$, in the `replace-one' definition, neighboring datasets are defined by \textit{replacing} one image sample, i.e., $D' \cup \{x'\} = D \cup \{x\}$ for image samples $x$ and $x'$. In the `add-or-remove-one' definition, neighboring datasets are constructed by \textit{adding} or \textit{removing} a single image sample, e.g., $D' = D \cup \{x\}$. We show in Appendix \ref{app:privacy-notations} that all methods that used the replace-one definition can be adapted to add-or-remove-one without affecting their privacy guarantees. As a result, all methods can be compared fairly under the same privacy notion.

\noindent \textbf{Gaussian Mechanism}. Gaussian mechanism (GM) adds noise sampled from a Gaussian distribution.
For an arbitrary function $f$ with sensitivity ${\Delta_f}$, the GM $\mathcal{A}$ is defined as, 
\begin{equation}\label{eq:gaussian}
    \mathcal{A}(D) = f(D) + \mathcal{N}(0, \Delta_f^2 \sigma^2 \mathbb{I}).
\end{equation}
$\mathcal{N}(0, \Delta_f^2 \sigma^2 \mathbb{I})$ is a multi-dimensional random variable drawn from an i.i.d. normal distribution with a mean of 0 and a variance of $\Delta_f^2 \sigma^2$, where $\sigma$ is a hyper-parameter known as the noise multiplier~\cite{dp}.

\noindent \textbf{DP-SGD}. 
To train machine learning models with DP guarantees, %
DP-SGD~\cite{dpsgd} extends standard SGD by controlling 
the gradient contributions from individual data 
and applying the Gaussian mechanism.
Specifically, in each training iteration,
DP-SGD samples a mini-batch from the whole training data $D$ and bounds the sensitivity of each gradient using a clipping operation: $\text{Clip}_{C}\left(\mathbf{g}\right) = \text{min}\left\{1,\frac{C}{||\mathbf{g}||_2}\right\}\mathbf{g},$ where $\mathbf{g}$ indicates the original gradient and $C$ is a hyper-parameter. The `$\text{Clip}_{C}$' operation scales the gradient norm to be at most $C$, and the model parameters $\phi$ are then updated with:
\begin{align}
\small
\eta \mathbb{E}_{x_i \in x} \left[ \text{Clip}_{C}\left(\nabla {\mathcal{L}}(\phi, x_i)\right) + C \mathcal{N}(0, \sigma^2 \mathbb{I}) \right],
\label{eq:DP-SGD}
\end{align}
where $\eta$ is the learning rate, $\nabla \mathcal{L}(\theta, x_i)$ denotes the gradient computed from the loss function $\mathcal{L}$ for the sample $x_i$, and $x$ refers to a mini-batch sampled from the dataset using Poisson sampling with a rate $q$.
Besides, $\mathcal{N}(0, \sigma^2 \mathbb{I})$ is i.i.d. Gaussian noise with variance $\sigma^2$.

As mentioned above, DP-SGD uses only one batch of the sensitive dataset for each iteration of the model parameter update. However, the training process involves many iterations of DP-SGD to ensure that the target model converges. Composition theorems (privacy accounting), e.g., R\'{e}nyi DP~\cite{sgm}, can be used to account for the privacy cost for each iteration and derive the final values of $\epsilon$~\cite{sgm}.

\begin{figure}[!t]
\vspace{-1mm}
    \centering
    \setlength{\abovecaptionskip}{0pt}
    \includegraphics[width=1.0\linewidth]{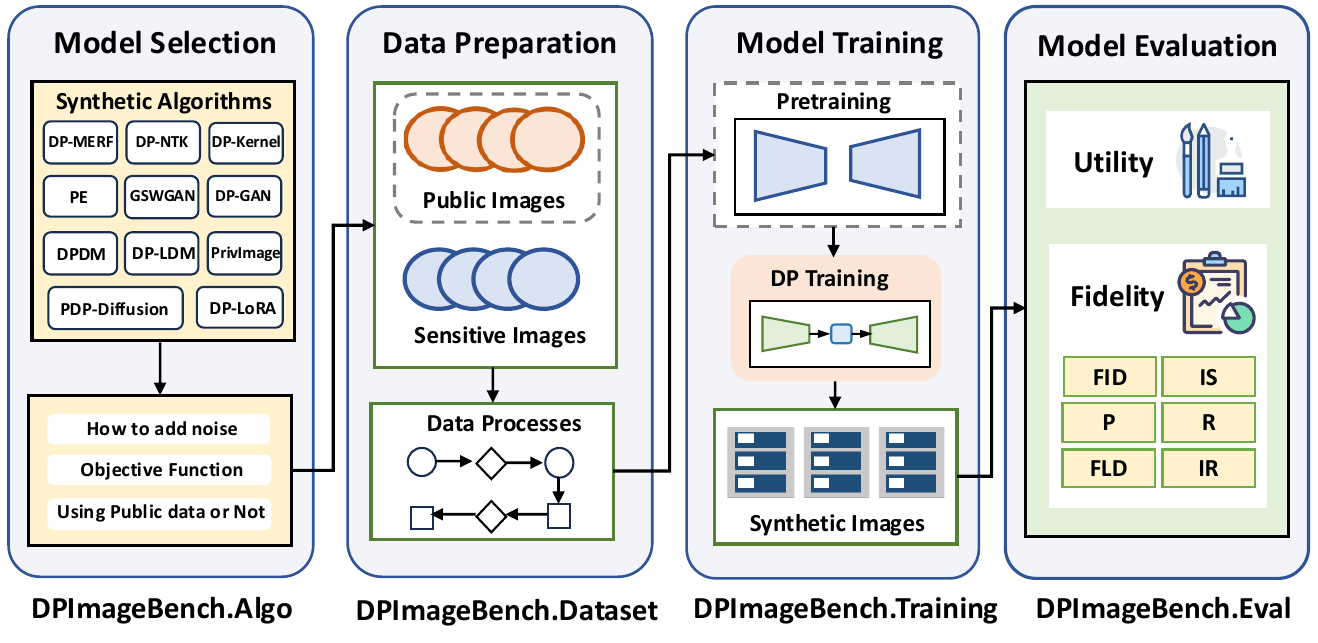}
    \caption{ Overview of \toolname{}. The operations in the gray dashed box mean using public images for pretraining, which is optional, depending on the algorithms.}
    \label{fig:dpimagebench_ovewflow}
    \vspace{-3mm}
\end{figure}

\vspace{-1.5mm}
\subsection{Image Generative Models}
\label{subsec:generative}

We introduce three image synthesizers: (1) diffusion models (DMs), {including standard DMs~\cite{ddpm} and latent DMs~\cite{labelembedding},} (2) Generative Adversarial Networks (GANs)~\cite{wgan}, and (3) foundational APIs~\cite{dpsda}. %

\noindent \textbf{Diffusion Models.} DMs~\cite{ddpm} consist of two processes: (1) The \emph{forward diffusion process} that progressively adds noise to a clean image $x_0$ until it becomes close enough to pure random noise, generating a sequence of increasingly noisy images $\{x_i\}_{i=1}^T$, where $T$ is the number of noising steps. (2) The \emph{reverse process} that progressively denoises a random noise to a clean image via a trainable neural network. In the forward diffusion process, the transition between adjacent noisy images, $p\left( {{x_t}\left| {{x_{t - 1}}} \right.} \right)$, follows a multi-dimensional Gaussian distribution,
\begin{equation}
\label{DiffusionProcess}
    p\left( {{x_t}\left| {{x_{t - 1}}} \right.} \right) = \mathcal{N}\left( {{x_t};\sqrt {1 - {\beta _t}} {x_{t - 1}},{\beta _t}\mathbb{I}} \right)
    \nonumber
\end{equation}
where ${\beta _t}$ is a hyper-parameter that defines the variance of the noise distribution at each step. We denote ${{\bar \alpha }_t} = \prod\nolimits_{s = 1}^t {\left( {1 - {\beta _s}} \right)} $. The likelihood between the clean image $x_0$ and its noisier version at step $t$ is derived as, $ p\left( {{x_t}\left| {{x_0}} \right.} \right) = \mathcal{N}\left( {{x_t};\sqrt {{{\bar \alpha }_t}} {x_0},\left( {1 - {{\bar \alpha }_t}} \right)\mathbb{I}} \right).$ 
Then, we can sample $x_t$ %
directly from $x_0$ in closed form,
\begin{equation}
    {x_t} = \sqrt {{{\bar \alpha }_t}} {x_0} + e_t\sqrt {1 - {{\bar \alpha }_t}} ,e_t\sim \mathcal{N}\left( {0,\mathbb{I}} \right)
\label{eq:dm:forward}
\end{equation}
The final objective of DMs is to train a network to predict the added noise at each step, as defined in~\cite{ddpm}.
\begin{equation}
\label{eq:L_DM}
\mathcal{L} = {\mathbb{E}_{x_0\sim D,t \sim  \text{U}\left( {1,T} \right),e_t \sim \mathcal{N}\left( {0,\mathbb{I}} \right)}}{\left\| {e - {e_\theta }\left( {{x_t},t} \right)} \right\|^2}
\end{equation}
where $D$ is the dataset of clean images and $e_\theta$ is a denoising network parameterized by $\theta$. The network $e_\theta(x_t, t)$ learns to predict the noise $e$ in any noisy image $x_t$ at step $t$. Once well-trained, $e_\theta$ can be used to denoise random Gaussian noise into a clean image.

\noindent \textbf{Latent Diffusion Models (LDMs).}
LDMs~\cite{labelembedding} are a variant of DMs that operate in a latent space instead of directly in pixel space. LDMs first use a pretrained autoencoder to map images into a lower-dimensional latent representation. The same diffusion and denoising steps discussed before proceed in this compressed latent space rather than on raw pixels. After performing the reverse diffusion steps in latent spaces, the decoder reconstructs final high-resolution images. LDMs operate in a lower-dimensional space, enabling more efficient training and sampling than standard pixel-based DM.

Stable diffusion is a representative example of LDMs that 
generate images from textual descriptions or class labels.
DP-LDM~\cite{dpldm} and DP-LoRA~\cite{dplora} leverage stable diffusion~\cite{labelembedding} as the synthesizer. 

\noindent \textbf{Generative Adversarial Nets (GAN).} GAN consists of two networks: a generator $\text{Gen}$ and a discriminator $\text{Dis}$~\cite{wgan,gan}. The $\text{Gen}$ receives a random noise vector and outputs an image. The $\text{Dis}$ receives an image and outputs a score, which indicates how real the input image is. The $\text{Gen}$ is trained to deceive the discriminator, and the $\text{Dis}$ is trained to distinguish whether its input image comes from the true dataset or the $\text{Gen}$. Both networks are trained simultaneously, constantly improving in response to each other's updates. Overall, the objective function of GAN is defined as, 
\begin{equation}
\label{eq:L_GAN}
\begin{split}
    \mathop {\min }\limits_{\text{Gen}} \mathop {\max }\limits_{\text{Dis}} V\left( {\text{Gen},\text{Dis}} \right) & = {\mathbb{E}_{x \sim \;q\left( x \right)}}\left[ {\log \text{Dis}\left( x \right)} \right] \\
     + {\mathbb{E}_{z \sim p\left( z \right)}} & \left[ {\log \left( {1 - \text{Dis}\left( {\text{Gen}\left( z \right)} \right)} \right)} \right]
\end{split}
\end{equation}
\noindent The $q(x)$ and $p(z)$ indicate the distributions of the real image and noise vector.
After the training is done, the discriminator is discarded, and the generator can be used for generation.

\noindent \textbf{Foundational APIs.} Foundation APIs are either cloud-based services (e.g., Dall-E~\cite{dalle2}) or local software libraries (e.g., Stable Diffusion~\cite{labelembedding}) that provide simple API access to pretrained models capable of generating high-quality synthetic images. The underlying model can be any generative model, including DMs and GANs.

\vspace{-1mm}
\subsection{DP Image Synthesis}
\vspace{-1mm}

Various sensitive image datasets, such as medical images~\cite{camelyon1} and face images~\cite{celeba}, face the risk of privacy leakage. Attackers may infer information about the training dataset from the synthetic images~\cite{2023whithmiadiffusion}. DP image synthesis aims to generate artificial images that closely resemble real data while ensuring the privacy protection of the original dataset. DP quantifies leakage to infer the training dataset using the synthetic
data.
They enable organizations to share and use synthetic images,
reducing privacy concerns.

\section{Overview of \toolname{}}

We introduce \toolname{}, a modular toolkit designed to evaluate DP image synthesis algorithms in terms of utility and fidelity. We envisage \toolname{} for the following four purposes. 

\begin{itemize}[leftmargin=*]
\item \toolname{} should incorporate advanced synthesis methods that are widely used and have achieved state-of-the-art performance within the community. 
\item This benchmark should enable users to apply these techniques to their own private image datasets as well as public datasets for DP image release. 
\item We intend for \toolname{} to offer straightforward implementation methods and consistent hyper-parameter settings for the studied methods, facilitating a comprehensive understanding of their strengths and weaknesses. 
\item \toolname{} should comprehensively evaluate synthetic images generated by different algorithms for comparative studies.
\end{itemize}

As presented in Figure~\ref{fig:dpimagebench_ovewflow}, we design four key modules for \toolname{}, to achieve these goals point by point:

\noindent \textbf{(1) Model Selection.} \toolname{} allows users to select among twelve distinct DP image synthesis algorithms, which are categorized into three groups as introduced in Section~\ref{sec:taxonomy}. \toolname{} also provides simple interfaces to integrate new algorithms.

\noindent\textbf{(2) Data Preparation.} This module prepares the public images and sensitive images for the following modules by partitioning and preprocessing the data, such as normalization. \toolname{} allows users to select public and sensitive datasets flexibly. %

\noindent\textbf{(3) Model Training.} This module involves (optional) synthesizer pretraining on the public dataset and the main training on the sensitive dataset. We unify the model architectures and hyper-parameters across different methods. Besides, we provide flexible configuration files, allowing users to adjust hyper-parameters according to their preferences. \toolname{} supports pretraining for all algorithms, even those not originally designed with pretraining. Details are provided in Appendix~\ref{app:pretraining}.

\noindent\textbf{(4) Evaluation.} This module assesses the synthetic images generated by algorithms on aspects of utility and fidelity. 

In the following section, we elaborate on the taxonomy method for model selection in Section~\ref{sec:taxonomy}, followed by the key design elements of \toolname{} for data preparation, model training, and evaluation in Section~\ref{sec:key_designs}.

\vspace{-1mm}
\section{Taxonomy and Method Description}
\label{sec:taxonomy}

We present a novel taxonomy of existing methods from two orthogonal dimensions: (1) The Private Data Perspective: where the noise is added, and (2) The Public Data Perspective: how public data is used to enhance DP image synthesis.
Note that this paper does not intend to benchmark every possible method but focuses only on state-of-the-art ones. 
For example, we omitted some GAN- and VAE-based methods, 
which are not state-of-the-art~\cite{li2023PrivImage,dpldm,dpdm}. %
, and we include all diffusion-based algorithms due to their exceptional performance.
That said, as \toolname{} is a general framework, users can always add more methods to it.

\vspace{-1mm}
\subsection{The Private Data Perspective}

\begin{figure}[!t]
\vspace{-1mm}
    \centering
    \setlength{\abovecaptionskip}{0pt}
    \includegraphics[width=1.0\linewidth]{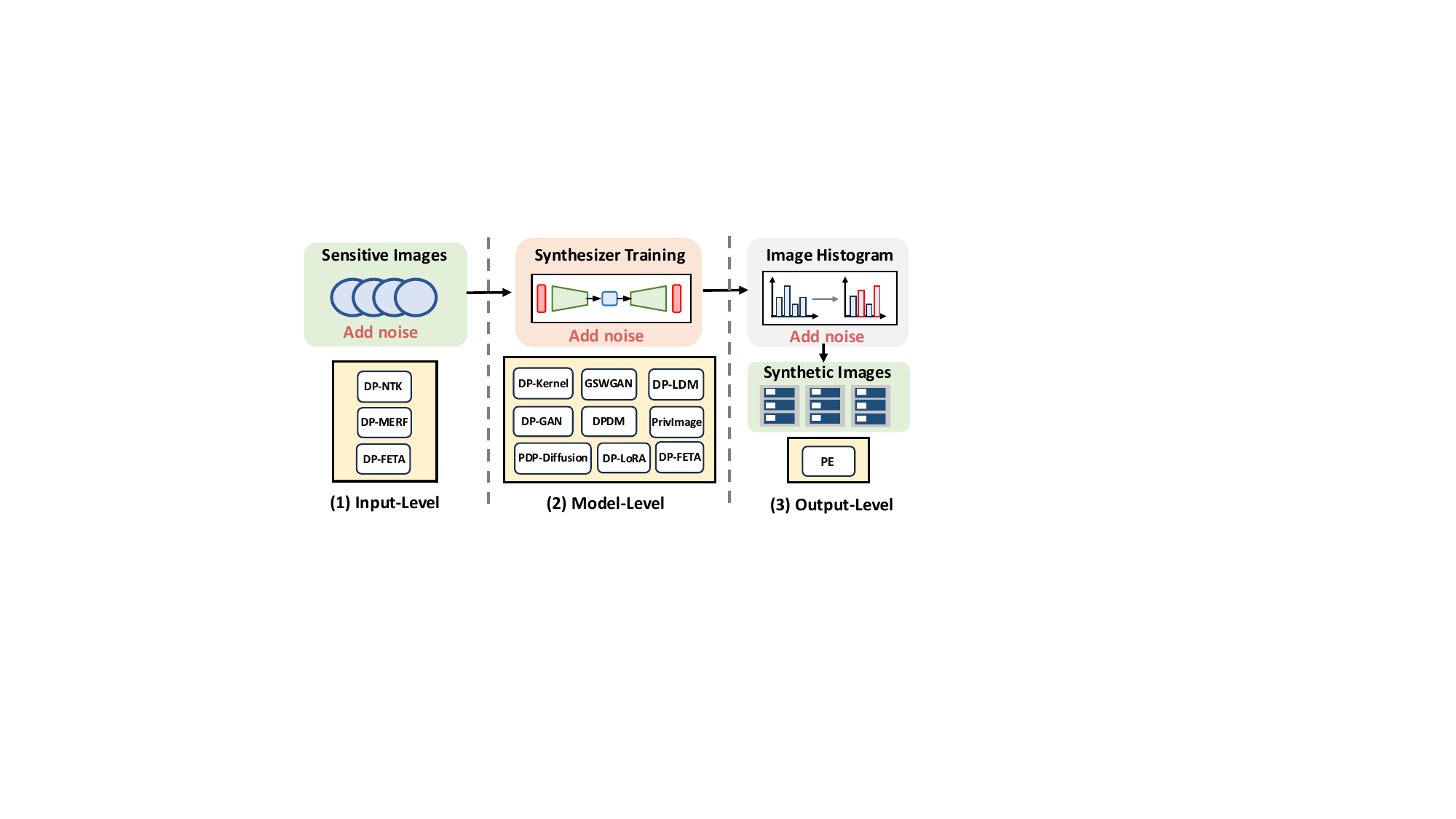}
    \caption{ The private data perspective in taxonomy categorizes the studied algorithms into three levels: input-level, model-level, and output-level.}
    \label{fig:taxonomy}
\vspace{-4mm}
\end{figure}

We formulate the pipeline of DP image synthesis as follows: public images and sensitive images are fed into the synthesizer for pretraining and finetuning. The well-trained models then generate synthetic images as output. We propose a taxonomy based on where noise is added, categorizing current methods into input-level, model-level, and output-level approaches (Figure~\ref{fig:taxonomy}).

\begin{itemize}[leftmargin=*]
\item \textit{Input-level}: DP-MERF and DP-NTK add noise to the feature embeddings of sensitive images.

\item \textit{Model-level}: GS-WGAN, DP-LDM, DP-GAN, DPDM, PrivImage, and PDP-Diffusion use DP-SGD to train the synthesizer on sensitive images while ensuring DP. DP-Kernel adds noise to the training loss of the synthesizer.
\item \textit{Output-level}: PE leverages model inference APIs to generate images, iteratively voting for the most similar API-generated images per private image to create a histogram. Noise is added to this histogram for DP, guiding pretrained models to produce images.

\item \gc{\textit{Mix-level}: Our taxonomy enables the systematic analysis of algorithms that introduce noise at various levels, such as DP-FETA, which applies noise at both the input and model levels.}
\end{itemize}

This taxonomy also reveals
the potential of combining privacy sanitization methods across various levels. For example, the finetuning-free method, PE, heavily relies on the pretrained model's ability to generate images similar to the sensitive ones. Its performance may be suboptimal if the model fails in this regard. Our taxonomy offers a clear framework that can guide PE in incorporating finetuning on sensitive images to enhance performance. 

We explain in Appendix~\ref{app:dppromise} why DP-Promise~\cite{dppromise} does not strictly satisfy DP and, therefore, is not included in our analysis.

\vspace{-1mm}
\subsubsection{Input-Level} %
These algorithms
first extract high-level statistical characteristics from images. They then add noise to these characteristics to ensure privacy. 

\noindent \textbf{DP-MERF~\cite{dp-merf}.} This method uses random Fourier feature representations~\cite{rahimi2007random} of Gaussian kernel mean embeddings as the high-level statistical characteristics of the image dataset. Kernel mean embedding (KME)~\cite{kme} maps a probability distribution into a reproducing kernel Hilbert space (RKHS)~\cite{rhks} by computing the mean of kernel evaluations. DP-MERF uses random Fourier features to approximate KME. DP-MERF adds noise, determined by the Gaussian mechanism~\cite{sgm}, to the KME of sensitive images. Then, DP-MERF trains a generator that minimizes the Maximum Mean Discrepancy (MMD)~\cite{MMD} between the noisy KME and KME of generated images.

\noindent \textbf{DP-NTK~\cite{dp-ntk}.} This method claimed that computing the KME directly from high-dimensional image datasets is challenging~\cite{dp-ntk,dp-mepf}. It proposes the use of Neural Tangent Kernels (NTKs)~\cite{ntks}, outperforming the Gaussian kernel in DP-MERF~\cite{dp-merf}. DP-NTK uses NTKs to represent images, whereby the gradient of the neural network function acts as the feature map to extract perceptual features from the original images. Computing KME from NTKs yields improved synthetic image quality.

\vspace{-1mm}
\subsubsection{Model-Level}
\label{sub:model-level}
These methods add DP noise in the training stage of generative models.

\noindent \textbf{DP-Kernel~\cite{dp-kernel}.} DP-MERF matches the KME across all data, which can lead to \textit{mode collapse}~\cite{modecollapse}, resulting in synthetic images that are overly similar. Besides, using random Fourier features to approximate KME may also cause potential errors. Instead of adding Gaussian noise to the KME of sensitive images, DP-Kernel proposes directly adding noise to the MMD loss function that measures the discrepancy between the distributions of sensitive and synthetic images in an RKHS to avoid the mentioned issues.
$$\mathcal{\Tilde{L}}_{\text{MMD}}(p,q) = \mathcal{L}_{\text{MMD}}(p,q) + \text{Gaussian noise},$$
where $p$ and $q$ mean the distribution of real and synthetic images.

\noindent \textbf{DP-GAN~\cite{dpgan}.} As described in Section~\ref{subsec:generative}, the discriminator in GANs guides the generator to generate images similar to the real images. 
DP-GAN trains the discriminator network on the sensitive images using DP-SGD, ensuring that the discriminator weights satisfy DP. As presented in Eq.~(\ref{eq:L_GAN}), the weight update of the generator solely relies on the discriminator. Due to the post-processing property of DP~\cite{dpbook}, the generator also satisfies DP.

\noindent \textbf{DPDM~\cite{dpdm}.} This paper pioneers DMs~\cite{ddpm} as DP synthesizers. It trains DMs on sensitive images using DP-SGD to ensure DP. Traditional DMs, without DP, use small batches and over-parameterized models~\cite{ddpm,edm}, but under DP, this leads to noisy gradients~\cite{dpsgd}, degrading synthesis. 
DPDM introduces the \textit{noise multiplicity}, where the image is re-used for training to learn to denoise at multiple steps. DPDM shows that this modification does not incur additional privacy costs while reducing the variance of the gradient.

\noindent \textbf{GS-WGAN~\cite{gs-wgam}.} 
Unlike DPGAN, which perturbs the entire discriminator, GS-WGAN perturbs only the discriminator’s feedback guiding the generator. It splits the generator’s gradient via the chain rule into an \textit{upstream gradient} (the output of discriminator with respect to the generated image) and a \textit{local gradient} (the generated image with respect to the generator's parameters), perturbing the upstream gradient with GM for DP, as only the discriminator accesses sensitive images. This ``privacy barrier'' prevents leakage to the generator, enabling pretraining of discriminators on sensitive images without privacy concerns to warm up the synthesizing.

\noindent \textbf{PDP-Diffusion~\cite{dp-diffusion}.} This paper first proposes the paradigm of pretraining the DP image synthesizer on a public dataset without privacy concerns. By leveraging this public dataset, the synthesizer can benefit from a broader knowledge base, which may enhance its performance during the subsequent DP-SGD~\cite{dpsgd} finetuning process on the sensitive dataset. Later methods using DMs, such as DP-LDM~\cite{dpldm} %
and PrivImage~\cite{li2023PrivImage}, are all built upon this paradigm. %

\noindent {\textbf{DP-LDM~\cite{dpldm}.} DP-LDM~\cite{dpldm} proposes using LDMs as synthesizers instead of standard DMs. It further improves efficiency by finetuning only specific components of the pretrained LDM, namely the label embedding module~\cite{labelembedding} and the attention module~\cite{vaswani2017attention}. DP-LDM reduces the number of trainable parameters and requires less noise to maintain the same privacy budget. }

\noindent {\textbf{DP-LoRA~\cite{dplora}.} Built on DP-LDM, DP-LoRA leverages Low-Rank Adapters (LoRA)~\cite{lora}, 
a parameter-efficient finetuning approach,
to train LDMs, thus reducing the size of trainable parameters and improving the privacy-utility trade-off.

This paper shows that pretraining on a dataset with a distribution similar to the sensitive dataset is more effective~\cite{yu2023selective,yin2022practical}. PrivImage thus selects pretraining data matching the sensitive dataset’s distribution, using semantics~\cite{Semantic1} to capture high-level image meaning rather than low-level features like color or texture. PrivImage defines a semantic query function on public dataset $D_p$, using image captioning~\cite{ImageCaption1,ImageCaption2} or classification, with the objective:
$${\mathcal{L}_Q} = \mathbb{E}_{\left( {x, {s}} \right) \sim {D_p}}\left[ \sum\nolimits_{i = 1}^{NS} - \left({s_i}\log {Q_i}\left( x \right) \right) \right].$$
\noindent Here, $x$ is an image, $s$ means semantic labels ($s_i = 1$ if the $i$-th label applies, else $0$), and $Q_i(x)$ is the probability of label $s_i$. PrivImage queries semantics from sensitive images; adds Gaussian noise for DP; selects public images based on noisy distributions; and fine-tunes the pretrained model on sensitive datasets using DP-SGD~\cite{dpsgd}.

\subsubsection{Output-Level} This kind of method introduces noise during the synthesis process under DP.

\noindent \textbf{Private Evolution (PE)~\cite{dpsda}.} 
This method uses foundation APIs from either black-box models like DALL-E 2~\cite{dalle2} or open-source models such as Stable Diffusion~\cite{labelembedding}. PE iteratively employs each private image to vote for the most similar API-generated images, constructing a voting histogram. DP is achieved by adding noise to this histogram. Using the noisy histogram, PE identifies images closely resembling the private ones and prompts pretrained models to generate additional variations. Consequently, PE is adaptable to any generative model capable of producing image variants (e.g., diffusion models that denoise noisy images). Notably, PE operates solely via model inference, requiring no training on sensitive images. Lin et al.~\cite{dpsda} prove that the distribution of images generated by PE can converge to that of the private images.

\subsubsection{Mix-Level}
\label{subsubsec:mix}

\gc{This taxonomy facilitates the analysis of methods that apply DP noise at multiple levels.}

\noindent \gc{\textbf{DP-FETA~\cite{dp-feta}.} This method allows synthesizers to learn image features from simple to complex. DP-FETA splits training into two stages: (1) It extracts basic features (e.g., outlines, colors) from sensitive images using ``central images''—representations of central tendency measures (e.g., mean, mode). These images are perturbed with Gaussian noise at the input level for DP, building a shortcut to facilitate early-phase synthesizer training. After the training on central images, the diffusion models can generate rough and statistically imperfect images. (2) Fine-tuning on sensitive images with DP-SGD~\cite{dpsgd} adds model-level noise to produce realistic images.}

\begin{table}[!t]
\small
    \centering
    \caption{Overview of studied algorithms.  ``\ding{51}'' means that the algorithm involves this property; ``\ding{55}'' means that it does not. DP-MERF, DP-NTK, and DP-Kernel only borrow the generator in GAN, and the critic is unnecessary because the optimization target is explicitly well-defined. Unlike DP-GAN and GS-WGAN, which rely on a critic to provide an adversarial loss to indirectly guide the generator.} 
    \vspace{-3mm}
    \label{tab:methodinfo}
    \resizebox{0.47\textwidth}{!}{
    \begin{tabular}{l|cccc}
    \toprule
    \multirow{2}{*}{Algorithm} & \multirow{2}{*}{Pretrain}  & \multicolumn{2}{c}{Privacy} & \multirow{2}{*}{Model Type}   \\
    \Xcline{3-4}{0.7pt}
     &  & Type & Noise Added & \\
    \hline
    DP-MERF & \ding{55} &  Input & Feature Embedding &  GAN   \\
    DP-NTK & \ding{55} &  Input & Feature Embedding & GAN    \\
    DP-Kernel & \ding{55} &  Model & Training Loss &GAN     \\
    PE & \ding{51}  &  Output & Histogram & API    \\
    GS-WGAN & \ding{55} &  Model & Output Gradient & GAN   \\
    DP-GAN & \ding{55} &  Model & Weight Gradient & GAN    \\
    DPDM & \ding{55}  &  Model & Weight Gradient & DM    \\
    DP-FETA & \ding{55}  &  Mix & Weight Gradient \& Images & DM    \\
    PDP-Diffusion & \ding{51} &  Model  & Weight Gradient & DM  \\
    DP-LDM &  \ding{51} &  Model & Weight Gradient & LDM     \\
    DP-LoRA &  \ding{51} &  Model & Weight Gradient & LDM    \\
    PrivImage & \ding{51} & Model & Weight Gradient & DM   \\
    \bottomrule
\end{tabular}
}
\vspace{-3mm}
\end{table}

\vspace{-1.5mm}
\subsection{The Public Data Perspective}
\label{sec:pretraining}
As shown in Table~\ref{tab:methodinfo}, pretraining synthesizers on public datasets, such as {\tt ImageNet}~\cite{imagenet}, which are openly available and do not raise privacy concerns, can enhance the utility and fidelity of DP synthetic images~\cite{li2023PrivImage,dp-diffusion,dpldm}. We believe that pretraining brings new insight into the community. Previous taxonomy lacks a systematic analysis of how public data is utilized~\cite{chen2023unified,hu2023sok}.  We fill this gap and outline potential directions as follows.

\begin{itemize}[leftmargin=*]
\item \textit{Pretrained model}: PE~\cite{dpsda} utilizes inference APIs of foundation models, which %
have been trained on public datasets.
\item \textit{Public dataset without selection}: DP-LDM and PDP-Diffusion use the whole public dataset for pretraining.
\item \textit{Public dataset with selection}: PrivImage~\cite{li2023PrivImage} shows that selecting a subset of pretraining images whose distribution closely matches the sensitive dataset yields better performance compared to using the entire public dataset.
\end{itemize}

{This taxonomy provides insights for developing new technologies. (1) Inspired by the public dataset selection approach in PrivImage, one potential avenue is to refine methods that do not consider selecting pretrained models (like PE). 
(2) Integrating the public dataset selection into approaches that have not yet been considered could be another promising direction. Section~\ref{subsec:rq3} investigates how incorporating the public dataset selection from PrivImage into methods without selection can enhance synthetic performance.
}

\vspace{-1mm}
\section{Key Designs of \toolname{}}
\label{sec:key_designs} 

\subsection{Data Preparation}
DPImageBench offers a flexible tool that allows users to select both public and sensitive datasets. This setup facilitates the pretraining of DP image synthesizers on public datasets for all studied algorithms, including those that were not initially designed to leverage public datasets for pretraining, e.g., DP-MERF~\cite{dp-merf}, DP-NTK~\cite{dp-ntk}, etc. The \texttt{README.md} file of our repository$^{\text{\ref{link:opensource}}}$ provides instructions for users to customize their dataset.

\vspace{-1mm}
\subsection{Model Training}

\subsubsection{The Failure Probability $\delta$} 
\label{subsec:delta}
For a meaningful DP guarantee, $\delta$ should satisfy that $\delta \ll \frac{1}{N}$, where $N$ presents the size of the sensitive dataset~\cite{fault_possibility}. Many previous works ignore this relationship and use $\delta = 1 \times 10^{-5}$. In \toolname{}, we set $\delta = 1 / {N \log(N)}$
following prior works~\cite{fault_possibility,yue-etal-2023-synthetic}.

\begin{table}[!t]
\small
    \centering
    \caption{Overview of evaluations for the studied algorithms. ``$\epsilon$ $(\min, \max)$'' mean the minimum and maximin values of $\epsilon$ used for DP image synthesis in their paper. }
    \label{tab:method_eval}
    \vspace{-2mm}
    \setlength{\tabcolsep}{3mm}{
    \resizebox{0.47\textwidth}{!}{
    \begin{tabular}{l|cccc}
    \toprule
    Algorithm &  Fidelity & Utility & DP testing & $\epsilon$ $(\min, \max)$ \\
    \hline
    DP-MERF~\cite{dp-merf} & \ding{55} & Acc & \ding{55} & $(0.2, 10.0)$ \\
    DP-NTK~\cite{dp-ntk} & \ding{55} & Acc & \ding{55} &  $(0.2, 10.0)$ \\
    DP-Kernel~\cite{dp-kernel} & FID & Acc & \ding{55} &  $(0.2, 1.0)$ \\
    PE~\cite{dpsda} & FID  & Acc & Nearest Sample & $(0.1,32.0)$  \\
    \hline
    GS-WGAN~\cite{gs-wgam} & FID, IS  & Acc & \ding{55} & $(0.1, 85.0)$  \\
    DP-GAN~\cite{dpgan} &   FID& Acc & \ding{55} & $(9.6,29.0)$  \\
    DPDM~\cite{dpdm}   &  FID & Acc  & \ding{55} & $(0.2, 10.0)$  \\
    DP-FETA~\cite{dp-feta}   &  FID & Acc  & \ding{55} & $(1.0, 10.0)$  \\
    PDP-Diffusion~\cite{dp-diffusion} &  FID & Acc  & \ding{55} & $(10.0, 10.0)$ \\
    DP-LDM~\cite{dpldm}  &  FID & Acc & \ding{55} & $(1.0, 10.0)$ \\
    DP-LoRA~\cite{dplora}  &  FID & Acc & \ding{55} & $(0.2, 10.0)$ \\
    DP-Promise~\cite{dppromise}   &  FID, IS & Acc & \ding{55} &  $(1.0, 10.0)$  \\
    PrivImage~\cite{li2023PrivImage} & FID & Acc & MIA &  $(1.0, 10.0)$ \\
    \bottomrule
\end{tabular}
}}
\vspace{-3mm}
\end{table}

\vspace{-1mm}
\subsubsection{Unified Framework} Inconsistent pretraining datasets and model architectures can lead to unfair comparisons. Even with identical synthesizer types, differences in hyperparameters for the synthesizer and DP-SGD—such as batch sizes, epochs, and learning rates—can significantly affect performance. \toolname{} standardizes public pretraining images and model architectures to ensure fair evaluations.

In DP-LDM and DP-LoRA, the synthesizer uses Stable Diffusion~\cite{labelembedding}, whereas methods like DPDM~\cite{dpdm} and PrivImage~\cite{li2023PrivImage} utilize standard diffusion models (DMs). As outlined in Section~\ref{subsec:generative}, LDMs integrate an autoencoder into DMs. To facilitate fair comparisons, \toolname{} provides standardized implementations for shared components of LDMs and standard DMs, including the noise prediction network structure and DP-SGD hyperparameters. Additionally, we implement DP-LDM with standard DMs—termed ``DP-LDM (SD)''—to assess how synthesizer architecture variations impact performance improvements.

\subsubsection{Extensions to New Algorithms} {\toolname{} has a modular design to make it easy to (1) add new algorithms and (2) develop new algorithms based on refining current methods (e.g., PDP-Diffision and PrivImage are both built based on DPDM).}

To add new algorithms, users simply need to implement the abstract base class with three interfaces: (1) pretraining the synthesizer with a public image dataset as input, (2) training the synthesizer with a sensitive image dataset as input, and 
(3) generating synthetic images. 
Users can flexibly design their methods inside the interfaces.
To refine existing methods, since \toolname{} isolates the parameter configurations from the algorithm implementation, users typically only need to modify the configuration files 
to adjust hyperparameters and refine algorithms.

We provide examples in the {\tt README.md} file of our repository$^{\text{\ref{link:opensource}}}$ to illustrate how to add new algorithms. 

\begin{figure}[!t]
    \centering
    \setlength{\abovecaptionskip}{0pt}
    \includegraphics[width=0.98\linewidth]{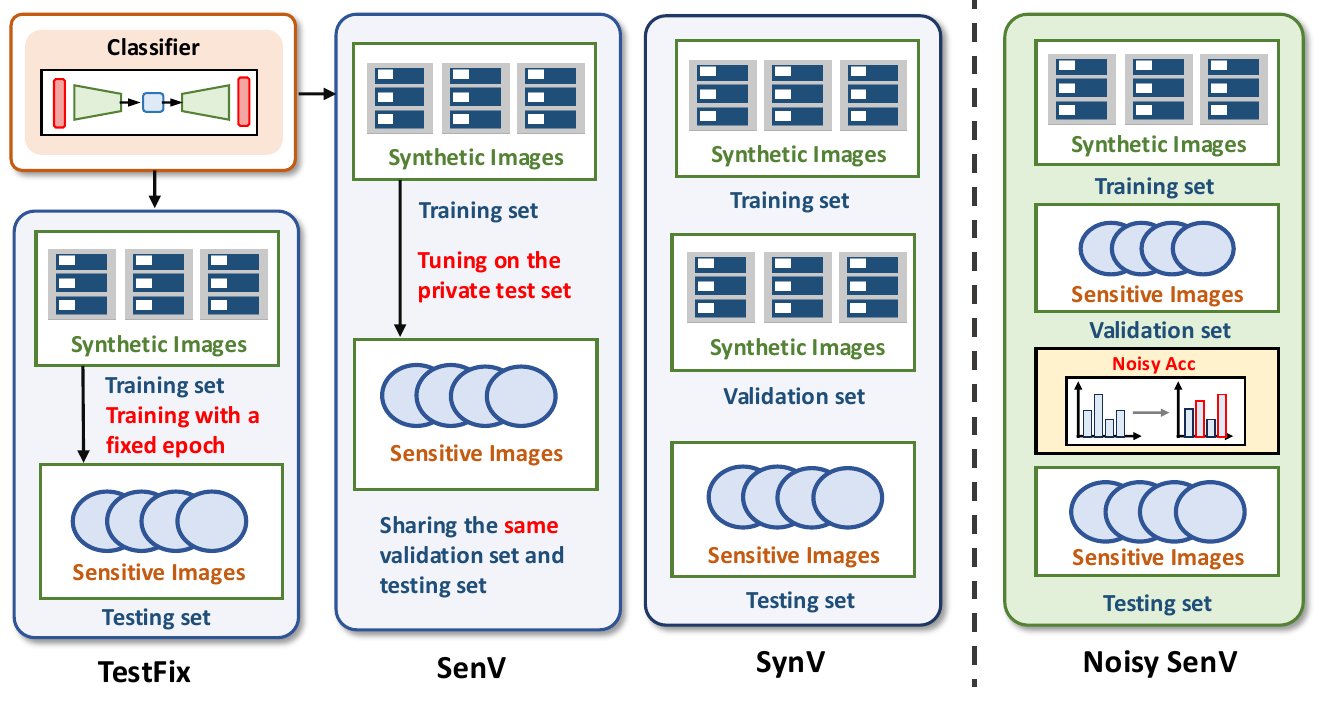}
    \caption{Comparison between current utility evaluation methods and our proposed approach.}
    \label{fig:utility}
    \vspace{-3mm}
\end{figure}

\vspace{-1.0mm}
\subsection{Evaluation}

After the evaluation is done, the results, including the synthetic images, logs, and metrics, are stored in the `exp'  folder. Please see the \texttt{README.md} file of our repository$^{\text{\ref{link:opensource}}}$ for more detail.

\begin{table}[!t]
\small
    \centering
    \caption{Overview of utility evaluations for the studied algorithms. ``\ding{55}'' indicates that it did not release codes. }
    \vspace{-2mm}
    \label{tab:method_acc_eval}
    \resizebox{0.49\textwidth}{!}{
    \begin{tabular}{lc|lc}
    \toprule
    Algorithm &  Evaluation Way & Algorithm &  Evaluation Way \\
    \hline
    DP-MERF & Test on a fixed epoch & DPDM  &  Synthetic images as val  \\
    DP-NTK & Test on a fixed epoch & PDP-Diffusion &  \ding{55} \\
    DP-Kernel & Test on a fixed epoch & DP-LDM  &  Synthetic images as val   \\
    PE & Sensitive test set as val & DP-LoRA  &  Sensitive test set as val \\
    GS-WGAN & Sensitive test set as val & DP-FETA   &  Sensitive test set as val \\
    DP-GAN & \ding{55} & PrivImage & Sensitive test set as val \\
    \bottomrule
\end{tabular}
}
\vspace{-3mm}
\end{table}

\subsubsection{Previous Synthesis Evaluations} \label{subsec:syn_eva}
Table~\ref{tab:method_eval} summarizes the evaluations of the investigated algorithms. We introduce them as follows.

\noindent \textit{Fidelity.} The Fréchet Inception Distance (FID)~\cite{fid} measures the quality and diversity of generated images by comparing feature distributions of sensitive and synthetic images in a pretrained network’s feature space.
It is used by all papers. The Inception Score (IS)~\cite{is} assesses classification confidence and class diversity~\cite{fld}, but is relevant only for natural images (e.g., ImageNet~\cite{imagenet}).

\noindent \textit{Utility.} Utility evaluates synthetic images in downstream tasks, primarily classification, where classifiers trained on synthetic images are tested for accuracy (Acc) on sensitive datasets.

\noindent \textit{Empirical DP testing.} Current DP image synthesis methods~\cite{li2023PrivImage,dpsda} propose to empirically assess privacy leakage from synthetic images. PrivImage uses MIAs~\cite{2023whithmiadiffusion} to infer sensitive images from synthetic ones, while PE~\cite{dpsda} analyzes distance distributions between synthetic and nearest sensitive images to ensure they aren’t copies.

\subsubsection{Fidelity Evaluations} 
\label{subsec:fidelity}

As discussed above, most DP image synthesis methods evaluate the fidelity of synthetic images using the FID~\cite{fid} and IS~\cite{is}. They have certain limitations, such as being insensitive to whether synthetic images memorize training images and lacking alignment with human perception~\cite{fld}. We establish evaluation from four perspectives.
\begin{itemize}[leftmargin=*] 
    \item \textit{Similarity}. Evaluates the proportion of synthetic images that resemble images in the real dataset.
    \item \textit{Diversity}. Assesses the variety in images to ensure the generated data covers the full range of the real dataset. 
    \item \textit{Novelty}. Evaluates how distinct the synthetic images are from the images in the training dataset, focusing on avoiding overfitting or memorization. 
    \item \textit{Human}. Incorporates human perception to assess the realism of synthetic images from a subjective standpoint. 
\end{itemize}

\toolname{} uses precision and recall~\cite{precision&recall}, which provide separate scores to offer a clearer understanding of whether a model excels at generating high-quality images (precision) or diverse images (recall). Moreover, \toolname{} also incorporates Feature Likelihood Divergence (FLD)~\cite{fld}, which evaluates the novelty of the synthetic dataset. From a human-level perspective, we introduce additional evaluation metrics ImageReward (IR)~\cite{xu2024imagereward}. We elaborate on the selected metrics in Appendix~\ref{supp:hyper_fidelity}.

\subsubsection{Utility Evaluations}\label{subsubsec:utility} When evaluating synthetic image utility, many studies use simple classifiers, limiting classification performance and reducing evaluation credibility. For instance, DP-MERF~\cite{dppromise} reports the best accuracy of 87.9\% on real {\tt FashionMNIST} dataset, 
which is lower than the 94.4\% achieved by Inception v3~\cite{inceptionv3}. \toolname{} trains classifiers using methods that achieve accuracy comparable to SOTA methods. We also provide options for the other classification algorithms. 

{\subsubsection{Promoting Utility Evaluations}\label{subsec:utility_error}
Current utility evaluation methods have limitations, as shown in Table~\ref{tab:method_acc_eval}. Prior studies use three approaches:
(1) \textit{Test on a fixed epoch} (TestFix):Train the classifier to a fixed epoch and test it on the sensitive set.
(2) \textit{Sensitive test set as val} (SenV): Train classifiers on the synthetic images, and use test sets without separate validation sets for \textit{tuning} classifier to select optimal parameters. 
(3) \textit{Synthetic images as val} (SynV): Use synthetic images as the validation set for classifier selection. Table~\ref{tab:utility_eval} reveals their drawbacks, detailed below,

\begin{table}[!t]
\small
    \centering
    \caption{Utility evaluation comparison. ``\ding{51}'' denotes the method includes this property; ``\ding{55}'' denotes it does not.}
    \vspace{-2mm}
    \label{tab:utility_eval}
    \resizebox{0.47\textwidth}{!}{
    \begin{tabular}{l|ccc}
    \toprule
    Methods &  Tuning for Better Results & Avoid Bias &  Satisfy DP \\
    \hline
    TestFix & \ding{55} &  \ding{55} & \ding{51} \\
    SenV & \ding{51} & \ding{55}  &  \ding{55}  \\
    SynV & \ding{51} & \ding{51}  &   \ding{51} \\
    \hline
    \rowcolor{gray0} Noisy SenV (Ours)  & \ding{51} & \ding{51}  &  \ding{51}  \\
    \bottomrule
\end{tabular}
}
\vspace{-3mm}
\end{table}

\begin{itemize}[leftmargin=*]
    \item TestFix: The optimal classifier training iterations that do not cause overfitting and underfitting depend on the data. `TestFix' trains classifiers for a predetermined number of epochs, which is not necessarily optimal. 
    \item SenV: Even worse, as Figure~\ref{fig:utility} illustrates, `SenV' leads to two key issues: (1) \textit{Bias in Test Accuracy}: Using the Test set for model selection biases results in higher accuracy. 
    (2) \textit{Privacy Risks}: Model selection based on test or validation metrics from sensitive datasets may breach DP guarantees~\cite{ponomareva2023dp}, as the test and validation sets share distributions with the training data.
    \item SynV: Due to the distribution gap between synthetic images and sensitive images, using synthetic images as the validation set to select classifiers may underestimate the utility.
\end{itemize}
}

\begin{table*}[!t]
\small
    \centering
    \caption{Acc (\%) of the classifier trained on synthetic images. We present the classifier that achieves the highest Acc on the test set of sensitive images. This table also presents relative changes in the Acc while using synthetic images as the validation set, i.e., `SynV'. `No DP' refers to the accuracy of classifiers trained directly on the sensitive dataset.}
    \vspace{-3mm}
    \label{tab:utility}
    \setlength{\tabcolsep}{5.5pt}
    \resizebox{0.99\textwidth}{!}{
    \begin{tabular}{l|cc|cc|cc|cc|cc|cc|cc}
    \toprule
    \multirow{2}{*}{Algorithm} & \multicolumn{2}{c|}{{\tt MNIST}} & \multicolumn{2}{c|}{{\tt FashionMNIST}} & \multicolumn{2}{c|}{{\tt CIFAR-10}} & \multicolumn{2}{c|}{{\tt CIFAR-100}}  & \multicolumn{2}{c|}{{\tt EuroSAT}} & \multicolumn{2}{c|}{{\tt CelebA}} & \multicolumn{2}{c}{{\tt Camelyon}} \\
    \cline{2-15}
     & $\epsilon = 1$ & $\epsilon = 10$ &  $\epsilon = 1$ & $\epsilon = 10$  &  $\epsilon = 1$ & $\epsilon = 10$  &  $\epsilon = 1$ & $\epsilon = 10$  &  $\epsilon = 1$ & $\epsilon = 10$  &  $\epsilon = 1$ & $\epsilon = 10$ &  $\epsilon = 1$ & $\epsilon = 10$ \\
    \hline
    DP-MERF & 85.1\textsubscript{\color{red} $\downarrow$10.1} & 84.6\textsubscript{\color{red} $\downarrow$5.4} & 63.9\textsubscript{\color{red} $\downarrow$2.1} & 66.8\textsubscript{\color{red} $\downarrow$4.8} & 28.1\textsubscript{\color{red} $\downarrow$2.1} & 29.1\textsubscript{\color{red} $\downarrow$5.3} & 4.3\textsubscript{\color{red} $\downarrow$1.5} & 3.6\textsubscript{\color{red} $\downarrow$1.4} & 37.3\textsubscript{\color{red} $\downarrow$0.3} & 33.7\textsubscript{\color{blue} $-$0.0} & 77.8\textsubscript{\color{red} $\downarrow$0.4} & 83.0\textsubscript{\color{red} $\downarrow$0.1} & 66.1\textsubscript{\color{red} $\downarrow$6.7} & 67.7\textsubscript{\color{red} $\downarrow$10.8} \\
    DP-NTK & 57.2\textsubscript{\color{red} $\downarrow$0.1} & 91.4\textsubscript{\color{blue}-0.0} & 66.5\textsubscript{\color{red} $\downarrow$1.3} & 73.5\textsubscript{\color{red} $\downarrow$0.4} & 17.0\textsubscript{\color{red} $\downarrow$0.6} & 25.5\textsubscript{\color{red} $\downarrow$1.2} & 3.5\textsubscript{\color{red} $\downarrow$1.7} & 3.6\textsubscript{\color{red} $\downarrow$1.8} & 23.6\textsubscript{\color{red} $\downarrow$3.2} & 38.6\textsubscript{\color{red} $\downarrow$1.9} & 61.4\textsubscript{\color{red} $\downarrow$4.9}  & 61.4\textsubscript{\color{red} $\downarrow$0.5} & 58.1\textsubscript{\color{red} $\downarrow$6.0} & 66.8\textsubscript{\color{blue}-0.0}  \\
    DP-Kernel & 92.1\textsubscript{\color{red}$\downarrow$6.8} & 92.8\textsubscript{\color{red}$\downarrow$1.7} & 68.1\textsubscript{\color{red}$\downarrow$14.5} & 70.0\textsubscript{\color{red}$\downarrow$12.8} & 25.7\textsubscript{\color{red}$\downarrow$8.5} & 26.8\textsubscript{\color{red}$\downarrow$10.1} &  5.7\textsubscript{\color{red}$\downarrow$2.8} & 5.7\textsubscript{\color{red}$\downarrow$3.8} & 48.2\textsubscript{\color{red}$\downarrow$21.2} & 47.9\textsubscript{\color{red}$\downarrow$24.9} & 83.6\textsubscript{\color{red}$\downarrow$0.5} & 83.0\textsubscript{\color{red}$\downarrow$6.4} & 73.1\textsubscript{\color{red}$\downarrow$2.5} & 73.8\textsubscript{\color{red}$\downarrow$8.0}  \\
    PE & 33.7\textsubscript{\color{red}$\downarrow$4.0} & 32.3\textsubscript{\color{red}$\downarrow$4.3} & 51.3\textsubscript{\color{red}$\downarrow$6.5} & 60.7\textsubscript{\color{red}$\downarrow$4.5} &  67.3\textsubscript{\color{red}$\downarrow$6.2} & 73.9\textsubscript{\color{red}$\downarrow$5.5}  & \colorbox{gray0}{17.9\textsubscript{\color{red}$\downarrow$1.8}} & 25.0\textsubscript{\color{red}$\downarrow$2.3} & 31.0\textsubscript{\color{red}$\downarrow$1.5} & 38.0\textsubscript{\color{red}$\downarrow$4.6} & 69.8\textsubscript{\color{red}$\downarrow$3.2} & 75.7\textsubscript{\color{red}$\downarrow$4.4} & 61.2\textsubscript{\color{red}$\downarrow$0.5} & 62.4\textsubscript{\color{red}$\downarrow$5.9} \\
    GS-WGAN & 74.5\textsubscript{\color{red}$\downarrow$2.8} & 77.8\textsubscript{\color{red}$\downarrow$0.7} & 51.4\textsubscript{\color{red}$\downarrow$3.3} & 60.2\textsubscript{\color{red}$\downarrow$3.8} & 20.4\textsubscript{\color{red}$\downarrow$3.1} &  20.4\textsubscript{\color{red}$\downarrow$3.1} & 1.7\textsubscript{\color{red}$\downarrow$0.6} & 1.8\textsubscript{\color{red}$\downarrow$0.3} & 30.5\textsubscript{\color{red}$\downarrow$2.0} & 29.8\textsubscript{\color{red}$\downarrow$2.3} & 62.2\textsubscript{\color{red}$\downarrow$3.7} & 62.3\textsubscript{\color{red}$\downarrow$2.3}  & 52.0\textsubscript{\color{red}$\downarrow$0.1} & 60.0\textsubscript{\color{red}$\downarrow$3.2} \\
    DP-GAN & 94.3\textsubscript{\color{red} $\downarrow$0.1} & 94.5\textsubscript{\color{red}$\downarrow$1.1} & 73.1\textsubscript{\color{red}$\downarrow$1.1} & 76.5\textsubscript{\color{red}$\downarrow$1.5} & 28.3\textsubscript{\color{red}$\downarrow$3.4} & 32.1\textsubscript{\color{red}$\downarrow$5.1} & 2.6\textsubscript{\color{red}$\downarrow$0.5} & 2.4\textsubscript{\color{red}$\downarrow$0.4} & 45.9\textsubscript{\color{red}$\downarrow$9.2} & 41.7\textsubscript{\color{red}$\downarrow$10.1} & 80.4\textsubscript{\color{red}$\downarrow$2.0} & 89.2\textsubscript{\color{red}$\downarrow$6.4} & 83.6\textsubscript{\color{red}$\downarrow$3.8} & 84.2\textsubscript{\color{red}$\downarrow$8.3} \\
    DPDM & 90.0\textsubscript{\color{red}$\downarrow$1.9} & 98.0\textsubscript{\color{red}$\downarrow$0.2} & 77.0\textsubscript{\color{red}$\downarrow$4.8} & 85.9\textsubscript{\color{red}$\downarrow$1.2} & 30.8\textsubscript{\color{red}$\downarrow$3.3} & 40.8\textsubscript{\color{red}$\downarrow$5.1} & 3.1\textsubscript{\color{red}$\downarrow$0.5} & 5.7\textsubscript{\color{red}$\downarrow$0.7} & 50.0\textsubscript{\color{red}$\downarrow$6.7} & 76.6\textsubscript{\color{red}$\downarrow$3.2} & 77.2\textsubscript{\color{red}$\downarrow$2.0}  & 93.2\textsubscript{\color{red}$\downarrow$2.0} & 83.3\textsubscript{\color{red}$\downarrow$6.1} & 83.6\textsubscript{\color{red}$\downarrow$6.0} \\
    DP-FETA & \colorbox{gray0}{96.2\textsubscript{\color{blue}-0.0}} & \colorbox{gray0}{98.3\textsubscript{\color{blue}-0.0}} & \colorbox{gray0}{83.3\textsubscript{\color{red} $\downarrow$0.4}} & \colorbox{gray0}{87.8\textsubscript{\color{red} $\downarrow$0.1}} & 33.5\textsubscript{\color{red} $\downarrow$3.2} & 45.3\textsubscript{\color{red} $\downarrow$1.0} & 4.8\textsubscript{\color{red} $\downarrow$0.8} & 7.4\textsubscript{\color{blue}-0.0} & 57.2\textsubscript{\color{red} $\downarrow$0.7} & 76.6\textsubscript{\color{red} $\downarrow$2.2} & 85.2\textsubscript{\color{red} $\downarrow$3.4} & \colorbox{gray0}{95.0\textsubscript{\color{red} $\downarrow$0.5}} & 84.6\textsubscript{\color{red} $\downarrow$9.9} & 86.3\textsubscript{\color{red} $\downarrow$9.9} \\
    PDP-Diffusion & 94.5\textsubscript{\color{blue} $-$0.0} & 97.7\textsubscript{\color{red} $\downarrow$0.1} & 79.0\textsubscript{\color{red} $\downarrow$0.3} & 85.4\textsubscript{\color{red} $\downarrow$1.3} & 62.3\textsubscript{\color{red} $\downarrow$1.7}  & 69.4\textsubscript{\color{red} $\downarrow$1.5} & 4.4\textsubscript{\color{red} $\downarrow$0.5} & 18.6\textsubscript{\color{red} $\downarrow$0.2} & 50.3\textsubscript{\color{red} $\downarrow$6.7} & 74.4\textsubscript{\color{blue} $-$0.0} & 91.2\textsubscript{\color{red} $\downarrow$3.1} &  94.5\textsubscript{\color{red} $\downarrow$2.0} & \colorbox{gray0}{86.0\textsubscript{\color{red} $\downarrow$2.2}} & \colorbox{gray0}{87.6\textsubscript{\color{red} $\downarrow$3.2}} \\
    DP-LDM (SD) & 81.1\textsubscript{\color{red} $\downarrow$9.8} & 95.0\textsubscript{\color{red} $\downarrow$0.1} & 76.8\textsubscript{\color{red} $\downarrow$6.7} & 82.0\textsubscript{\color{red} $\downarrow$0.3} & 65.2\textsubscript{\color{red} $\downarrow$0.2}  & 69.9\textsubscript{\color{red}$\downarrow$1.3} & 7.4\textsubscript{\color{red} $\downarrow$0.1} & 20.3\textsubscript{\color{red} $\downarrow$0.9} &  55.1\textsubscript{\color{red} $\downarrow$5.0}  & 72.6\textsubscript{\color{red} $\downarrow$1.3}   & 84.5\textsubscript{\color{red}$\downarrow$0.8} & 89.7\textsubscript{\color{red}$\downarrow$1.1}   & 84.6\textsubscript{\color{red} $\downarrow$2.6} & 85.3\textsubscript{\color{blue}$-$0.0} \\
    DP-LDM & 54.2\textsubscript{\color{red}$\downarrow$21.1} & 94.3\textsubscript{\color{red}$\downarrow$15.3} & 63.8\textsubscript{\color{red}$\downarrow$4.7} & 85.5\textsubscript{\color{red}$\downarrow$0.3} & 43.9\textsubscript{\color{red}$\downarrow$3.8} & 63.1\textsubscript{\color{red}$\downarrow$2.0} & 4.2\textsubscript{\color{red}$\downarrow$0.7} & 17.5\textsubscript{\color{blue}$-$0.0} & 67.2\textsubscript{\color{red}$\downarrow$4.1} & 79.2\textsubscript{\color{red}$\downarrow$4.4} & 86.3\textsubscript{\color{red}$\downarrow$1.4} & 92.4\textsubscript{\color{blue}$-$0.0} & 81.8\textsubscript{\color{red}$\downarrow$2.4} & 83.7\textsubscript{\color{red}$\downarrow$0.2} \\
    DP-LoRA & 76.6\textsubscript{\color{red}$\downarrow$50.8} & 97.4\textsubscript{\color{red}$\downarrow$2.9} & 65.7\textsubscript{\color{red}$\downarrow$16.8} &  84.0\textsubscript{\color{red}$\downarrow$1.5} & 64.1\textsubscript{\color{red}$\downarrow$1.8} & 77.7\textsubscript{\color{red}$\downarrow$0.6} & 5.4\textsubscript{\color{red}$\downarrow$1.0} & \colorbox{gray0}{35.2\textsubscript{\color{blue}$-$0.0}} & \colorbox{gray0}{70.9\textsubscript{\color{red}$\downarrow$2.5}} & \colorbox{gray0}{84.6\textsubscript{\color{red}$\downarrow$3.1}}  & 88.5\textsubscript{\color{red}$\downarrow$0.1} & 92.5\textsubscript{\color{red}$\downarrow$1.3} & 85.7\textsubscript{\color{red}$\downarrow$0.6} & 87.1\textsubscript{\color{red}$\downarrow$1.0} \\
    PrivImage & 95.2\textsubscript{\color{red}$\downarrow$2.2} & 97.9\textsubscript{\color{red}$\downarrow$0.1} & 81.9\textsubscript{\color{red}$\downarrow$1.6} & 86.7\textsubscript{\color{red}$\downarrow$0.4} & \colorbox{gray0}{75.0\textsubscript{\color{red}$\downarrow$0.8}} & \colorbox{gray0}{78.8\textsubscript{\color{red}$\downarrow$1.4}} & 10.2\textsubscript{\color{red}$\downarrow$0.4} & 16.5\textsubscript{\color{red}$\downarrow$0.3} & 51.0\textsubscript{\color{red}$\downarrow$1.5} & 72.1\textsubscript{\color{red}$\downarrow$0.3} & \colorbox{gray0}{92.1\textsubscript{\color{red}$\downarrow$0.4}} & 94.6\textsubscript{\color{red}$\downarrow$1.3} & 85.8\textsubscript{\color{red}$\downarrow$3.4} & 85.9\textsubscript{\color{red}$\downarrow$2.8} \\
    \hline
    No DP & \multicolumn{2}{c|}{99.7}  &  \multicolumn{2}{c|}{94.5} & \multicolumn{2}{c|}{93.1}  &  \multicolumn{2}{c|}{74.6}  &  \multicolumn{2}{c|}{97.5} & \multicolumn{2}{c|}{97.7}  & \multicolumn{2}{c}{87.7}  \\
    \bottomrule
\end{tabular}
}
\vspace{-3mm}
\end{table*}

\toolname{} presents the solutions to overcome the three issues presented in Table~\ref{tab:utility_eval} as follows.

To address the tuning and the bias issues, Figure~\ref{fig:utility} illustrates that we use the sensitive validation set (as detailed in Table~\ref{tab:datainfo}), which is distinct from the sensitive training and test set, for classifier tuning. However, the validation set shares the same distribution as the sensitive training data, potentially still violating DP~\cite{ponomareva2023dp}.

To satisfy DP, we use the ``report noisy max'' method~\cite{report-noisy-max}. We select the best classifier checkpoint that has the most correct samples. To satisfy $\epsilon$-DP, we add the Laplace noise from $\text{Lap}\left(1/\epsilon\right)$ to their correct number and select the best classifier checkpoint based on the noisy numbers. The best classifier is used to calculate the test set accuracy. %
Since the training and validation sets are two disjoint sets, according to the \textit{parallel composition theorem}~\cite{PrivacyIntegratedQueries}, the ultimate privacy guarantee depends only on the worst of the guarantees of each analysis, $\text{max}\{\epsilon_t,\epsilon_v\}$, where $\epsilon_v$ is the privacy costs of the validation set and $\epsilon_t$ is the privacy cost of training the synthesizer on the training set. \toolname{} sets $\epsilon_t = \epsilon_v$ without additional privacy cost. Thus, the final privacy cost is still $\epsilon_t$~\cite{dpbook}. We abbreviate this method as `Noisy SenV'. 

\toolname{} provides `SynV' and `Noisy SenV', the two methods that satisfy DP, avoid bias, and support tuning, for evaluation.

\vspace{-0.5mm}
\subsubsection{Empirical DP Testing} 
Empirical DP testing experimentally audits an algorithm’s privacy by running it on specially constructed data to assess DP parameters and measure potential exposure of personal information. The goal is to determine whether the system effectively obscures sensitive details and adheres to DP~\cite{dpauditing}.

We chose not to perform empirical DP testing for the following reasons.
(1) \textit{Lack of suitable methods:} Most existing research focuses on auditing DP in \textit{classification} models~\cite{dpauditing,tightDPauditing}, models trained on non-image data (e.g., tabular data~\cite{dptabularauditing}), or the DP-SGD algorithm~\cite{steinke2023privacy,jagielski2020auditing}. Applying these methods to our problem is not straightforward, as we focus on DP \textit{image generation}, and some of the approaches we evaluate, such as PE, do not use DP-SGD.
(2) \textit{Theoretical guarantee:} DP synthesis algorithms already have the rigorous theoretical privacy guarantee that the sensitive information leaked from the output data is bounded.
If the goal is empirical privacy, we do not even need DP in the first place. 

That said, we believe empirical DP testing is still an important direction that can flag the errors of DP implementation. We call for future research on DP auditing for \textit{image synthesis}.  

\vspace{-2mm}
\section{Experimental Setup}
Appendices~\ref{supp:imple} and \ref{supp:hyper_algo} introduce algorithm implementations and hyper-parameter settings.  Appendix~\ref{supp:privacy_cost_hyper} discusses privacy costs for hyper-parameter tuning.

\noindent \textbf{Investigated Datasets.} Drawing on prior work~\cite{li2023PrivImage,dpsda,dp-diffusion,dppromise,dpdm}, we categorize datasets into \textit{sensitive datasets} requiring protection and \textit{public datasets} that can be shared without privacy concerns. Appendix~\ref{app:dataset} Table~\ref{tab:datainfo} summarizes the datasets, noting that some public datasets like {\tt LAION}~\cite{LAION} raise privacy and copyright issues, making them unsuitable for pretraining~\cite{privacyissueoflaion}. 

We select seven widely studied sensitive datasets: (1) {\tt MNIST}~\cite{mnist}, (2) {\tt FashionMNIST}~\cite{fmnist}, (3) {\tt CIFAR-10}~\cite{cifar10}, (4) {\tt CIFAR-100}~\cite{cifar10}, (5) {\tt EuroSAT}~\cite{eurosat}, (6) {\tt CelebA}~\cite{celeba}, and (7) {\tt Camelyon}~\cite{camelyon1}, commonly used in prior studies~\cite{li2023PrivImage,dpsda,dpldm}. The {\tt ImageNet ISLVRC2012}~\cite{imagenet} and {\tt Places365}~\cite{places365} datasets are treated as public datasets. For {\tt CelebA}~\cite{celeba}, we focus on gender classification tasks following\cite{li2023PrivImage}. {\tt MNIST} and {\tt FashionMNIST} consist of grayscale images, while the others are in color, with further details in Appendix~\ref{app:dataset}.

\noindent \textbf{Implementation of Evaluation Metrics.}
This section describes the metrics for evaluating synthetic images. We generate 60,000 synthetic images per sensitive dataset for evaluation. For metrics like FID, IS, Precision, and Recall, we use the pretrained Inception v3~\cite{inceptionv3} from Github to extract features or classify images.\footnote{\url{https://github.com/mseitzer/pytorch-fid/releases/download/fid_weights/pt_inception-2015-12-05-6726825d.pth}} Hyper-parameter settings for utility and fidelity metrics are detailed in Appendices~\ref{supp:hyper_class} and \ref{supp:hyper_fidelity}. We assess the utility of synthetic images in image classification tasks using \toolname{}, which employs three classification algorithms—ResNet~\cite{resnet}, WideResNet~\cite{zagoruyko2017wideresidualnetworks}, and ResNeXt~\cite{resnext}—trained on synthetic images, with accuracy evaluated on the sensitive test set.

\begin{table*}[!t]
\small
    \centering
    \caption{
    Acc (\%) of the classifier trained on synthetic images. We use `Noisy SenV' for utility evaluation, as introduced in Section~\ref{subsubsec:utility}. Relative changes indicate differences from the highest sensitive test accuracy reported in Table~\ref{tab:utility}.}
    \vspace{-3mm}
    \label{tab:utility_val}
    \setlength{\tabcolsep}{4.5pt}
    \resizebox{0.99\textwidth}{!}{
    \begin{tabular}{l|cc|cc|cc|cc|cc|cc|cc}
    \toprule
    \multirow{2}{*}{Algorithm} & \multicolumn{2}{c|}{{\tt MNIST}} & \multicolumn{2}{c|}{{\tt FashionMNIST}} & \multicolumn{2}{c|}{{\tt CIFAR-10}} & \multicolumn{2}{c|}{{\tt CIFAR-100}}  & \multicolumn{2}{c|}{{\tt EuroSAT}} & \multicolumn{2}{c|}{{\tt CelebA}} & \multicolumn{2}{c}{{\tt Camelyon}} \\
    \cline{2-15}
     & $\epsilon = 1$ & $\epsilon = 10$ &  $\epsilon = 1$ & $\epsilon = 10$  &  $\epsilon = 1$ & $\epsilon = 10$  &  $\epsilon = 1$ & $\epsilon = 10$  &  $\epsilon = 1$ & $\epsilon = 10$  &  $\epsilon = 1$ & $\epsilon = 10$ &  $\epsilon = 1$ & $\epsilon = 10$ \\
    \hline
    DP-MERF & 80.3\textsubscript{\color{red}$\downarrow$4.8} & 81.3\textsubscript{\color{red}$\downarrow$3.3} & 62.2\textsubscript{\color{red}$\downarrow$1.7} & 62.2\textsubscript{\color{red}$\downarrow$4.6} & 27.2\textsubscript{\color{red}$\downarrow$0.9} & 29.0\textsubscript{\color{red}$\downarrow$0.1} & 3.4\textsubscript{\color{red}$\downarrow$0.9} & 4.2\textsubscript{\color{brown}$\uparrow$0.6} & 34.9\textsubscript{\color{red}$\downarrow$2.4} & 28.0\textsubscript{\color{red}$\downarrow$5.7} & 81.0\textsubscript{\color{brown}$\uparrow$3.2}  & 81.2\textsubscript{\color{red}$\downarrow$1.8} & 60.4\textsubscript{\color{red}$\downarrow$5.7}  & 58.3\textsubscript{\color{red}$\downarrow$9.4} \\
    DP-NTK & 50.0\textsubscript{\color{red}$\downarrow$0.2} & 91.3\textsubscript{\color{red}$\downarrow$0.1} & 64.4\textsubscript{\color{red}$\downarrow$2.1} & 76.3\textsubscript{\color{brown}$\uparrow$1.8} & 17.0\textsubscript{\color{blue}$-$0.0} & 28.2\textsubscript{\color{brown}$\uparrow$1.4} & 1.8\textsubscript{\color{red}$\downarrow$1.7} & 2.1\textsubscript{\color{red}$\downarrow$1.5}  & 22.8\textsubscript{\color{red}$\downarrow$0.8} & 30.8\textsubscript{\color{red}$\downarrow$7.8} & 61.2\textsubscript{\color{red}$\downarrow$0.2} & 64.2\textsubscript{\color{brown}$\uparrow$2.8} & 53.1\textsubscript{\color{red}$\downarrow$5.0} & 64.1\textsubscript{\color{red}$\downarrow$2.7} \\
    DP-Kernel & 94.0\textsubscript{\color{brown}$\uparrow$1.9} & 93.6\textsubscript{\color{brown}$\uparrow$0.8} & 68.4\textsubscript{\color{brown}$\uparrow$0.4} & 70.0\textsubscript{\color{blue}$-$0.0} & 26.4\textsubscript{\color{brown}$\uparrow$0.7}  & 25.1\textsubscript{\color{red}$\downarrow$1.4} & 6.0\textsubscript{\color{brown}$\uparrow$0.3} & 6.1\textsubscript{\color{red}$\downarrow$1.9} & 50.2\textsubscript{\color{brown}$\uparrow$2.0} & 50.2\textsubscript{\color{brown}$\uparrow$2.3} & \colorbox{gray0}{83.0\textsubscript{\color{red}$\downarrow$0.6}} & 83.7\textsubscript{\color{red}$\downarrow$0.6} & 68.0\textsubscript{\color{red}$\downarrow$5.1} & 68.7\textsubscript{\color{red}$\downarrow$5.1} \\
    PE & 27.9\textsubscript{\color{red}$\downarrow$5.8} & 32.7\textsubscript{\color{brown}$\uparrow$0.4} & 47.9\textsubscript{\color{red}$\downarrow$3.4} & 57.8\textsubscript{\color{red}$\downarrow$2.9} & 64.6\textsubscript{\color{red}$\downarrow$2.7} & 75.3\textsubscript{\color{brown}$\uparrow$1.4} & \colorbox{gray0}{15.4\textsubscript{\color{brown}$\uparrow$2.5}} & 24.9\textsubscript{\color{red}$\downarrow$0.1} & 32.1\textsubscript{\color{brown}$\uparrow$1.2} & 36.8\textsubscript{\color{red}$\downarrow$1.2}  & 70.5\textsubscript{\color{brown}$\uparrow$0.7} & 74.2\textsubscript{\color{red}$\downarrow$1.5} & 63.3\textsubscript{\color{brown}$\uparrow$2.1} & 64.9\textsubscript{\color{brown}$\uparrow$2.5} \\
    GS-WGAN & 72.4\textsubscript{\color{red}$\downarrow$ 2.1} & 75.3\textsubscript{\color{red}$\downarrow$2.8}  & 52.7\textsubscript{\color{brown}$\uparrow$1.3} & 56.7\textsubscript{\color{red}$\downarrow$ 3.5} & 20.4\textsubscript{\color{blue}$-$0.0} & 21.3\textsubscript{\color{brown}$\uparrow$0.9} & 1.3\textsubscript{\color{red}$\downarrow$ 0.4} & 1.6\textsubscript{\color{red}$\downarrow$ 0.2} & 28.6\textsubscript{\color{red}$\downarrow$1.9} & 29.7\textsubscript{\color{red}$\downarrow$0.1} & 61.4\textsubscript{\color{red}$\downarrow$0.8} & 61.5\textsubscript{\color{red}$\downarrow$0.8} & 52.1\textsubscript{\color{brown}$\uparrow$0.1} & 58.9\textsubscript{\color{red}$\downarrow$1.1} \\
    DP-GAN & 92.4\textsubscript{\color{red}$\downarrow$1.9} & 92.7\textsubscript{\color{red}$\downarrow$1.8} & 71.8\textsubscript{\color{red}$\downarrow$1.3} & 70.3\textsubscript{\color{red}$\downarrow$6.2} & 26.2\textsubscript{\color{red}$\downarrow$2.1} & 30.5\textsubscript{\color{red}$\downarrow$1.6} & 2.0\textsubscript{\color{red}$\downarrow$0.6} & 1.7\textsubscript{\color{red}$\downarrow$0.7} & 39.4\textsubscript{\color{red}$\downarrow$6.5} & 38.2\textsubscript{\color{red}$\downarrow$3.5} & 77.9\textsubscript{\color{red}$\downarrow$2.5} & 89.2\textsubscript{\color{blue}$-$0.0} & 83.2\textsubscript{\color{red}$\downarrow$0.4} & 79.6\textsubscript{\color{red}$\downarrow$4.6} \\
    DPDM & 89.2\textsubscript{\color{red}$\downarrow$0.8} & 97.7\textsubscript{\color{red}$\downarrow$0.3} & 76.4\textsubscript{\color{red}$\downarrow$0.6} & 85.6\textsubscript{\color{red}$\downarrow$0.3} & 28.9\textsubscript{\color{red}$\downarrow$1.9} & 36.8\textsubscript{\color{red}$\downarrow$2.0} & 2.4\textsubscript{\color{red}$\downarrow$0.7} &  6.4\textsubscript{\color{brown}$\uparrow$1.7} & 48.8\textsubscript{\color{red}$\downarrow$1.2} & 72.8\textsubscript{\color{red}$\downarrow$3.8}  & 74.5\textsubscript{\color{red}$\downarrow$2.7} & 91.8\textsubscript{\color{red}$\downarrow$1.4} & 80.6\textsubscript{\color{red}$\downarrow$2.7} & 79.5\textsubscript{\color{red}$\downarrow$4.1} \\
    DP-FETA & \colorbox{gray0}{95.6\textsubscript{\color{red}$\downarrow$0.6}} & \colorbox{gray0}{98.1\textsubscript{\color{red}$\downarrow$0.2}} & \colorbox{gray0}{81.7\textsubscript{\color{red}$\downarrow$1.6}} & \colorbox{gray0}{87.3\textsubscript{\color{red}$\downarrow$0.5}} & 35.1\textsubscript{\color{brown}$\uparrow$1.7} & 43.3\textsubscript{\color{red}$\downarrow$2.0} & 5.0\textsubscript{\color{brown}$\uparrow$0.2} & 8.1\textsubscript{\color{brown}$\uparrow$0.7} & 58.0\textsubscript{\color{brown}$\uparrow$0.8} & 75.0\textsubscript{\color{red}$\downarrow$1.6} & 82.3\textsubscript{\color{red}$\downarrow$2.9} & \colorbox{gray0}{94.2\textsubscript{\color{red}$\downarrow$0.8}} & 77.3\textsubscript{\color{red}$\downarrow$7.3} & 82.9\textsubscript{\color{red}$\downarrow$3.4} \\
    PDP-Diffusion & 94.5\textsubscript{\color{blue}$-$0.0} & 97.4\textsubscript{\color{red}$\downarrow$0.3} & 79.2\textsubscript{\color{brown}$\uparrow$0.2} & 85.4\textsubscript{\color{blue}$-$0.0} & 59.3\textsubscript{\color{red} $\downarrow$3.0} & 70.1\textsubscript{\color{brown}$\uparrow$0.7} & 3.9\textsubscript{\color{red}$\downarrow$0.5} & 17.3\textsubscript{\color{red}$\downarrow$1.3} & 46.6\textsubscript{\color{red}$\downarrow$3.7} & 73.8\textsubscript{\color{red}$\downarrow$0.6} & 89.4\textsubscript{\color{red}$\downarrow$1.8} & 94.0\textsubscript{\color{red}$\downarrow$0.5} & \colorbox{gray0}{85.2\textsubscript{\color{red} $\downarrow$0.8}} & 84.8\textsubscript{\color{red} $\downarrow$2.8} \\
    DP-LDM (SD) & 78.8\textsubscript{\color{red}$\downarrow$2.3} & 94.4\textsubscript{\color{red}$\downarrow$0.6} & 75.9\textsubscript{\color{red}$\downarrow$0.9} & 81.6\textsubscript{\color{red} $\downarrow$0.4} & 63.0\textsubscript{\color{red}$\downarrow$2.2} & 69.9\textsubscript{\color{blue}$-$0.0}  &  6.0\textsubscript{\color{red}$\downarrow$1.4} & 19.9\textsubscript{\color{red}$\downarrow$0.4}  & 48.1\textsubscript{\color{red}$\downarrow$7.0}  & 73.1\textsubscript{\color{brown}$\uparrow$0.5} & 84.4\textsubscript{\color{red}$\downarrow$0.1} & 89.1\textsubscript{\color{red}$\downarrow$0.6} & 84.0\textsubscript{\color{red}$\downarrow$0.4} & 84.7\textsubscript{\color{red}$\downarrow$0.6} \\
    DP-LDM & 44.2\textsubscript{\color{red}$\downarrow$10.0} & 95.5\textsubscript{\color{brown}$\uparrow$1.2} & 62.3\textsubscript{\color{red}$\downarrow$1.5} & 86.3\textsubscript{\color{brown}$\uparrow$0.8} & 41.7\textsubscript{\color{red}$\downarrow$2.2} & 64.8\textsubscript{\color{brown}$\uparrow$1.7} & 3.5\textsubscript{\color{red}$\downarrow$0.7} & 16.9\textsubscript{\color{red}$\downarrow$0.6} & 65.1\textsubscript{\color{red}$\downarrow$2.1} & 78.8\textsubscript{\color{red}$\downarrow$0.4} & 85.8\textsubscript{\color{red}$\downarrow$0.5} & 92.4\textsubscript{\color{blue}$-$0.0} & 81.9\textsubscript{\color{brown}$\uparrow$0.1} & 84.7\textsubscript{\color{brown}$\uparrow$1.0} \\
    DP-LoRA & 82.2\textsubscript{\color{brown}$\uparrow$5.6} & 97.1\textsubscript{\color{red}$\downarrow$0.3} & 63.5\textsubscript{\color{red}$\downarrow$2.2} & 83.8\textsubscript{\color{red}$\downarrow$0.2} & 64.6\textsubscript{\color{brown}$\uparrow$0.5} & 77.2\textsubscript{\color{red}$\downarrow$0.5} & 4.3\textsubscript{\color{red}$\downarrow$1.1} & \colorbox{gray0}{33.2\textsubscript{\color{red}$\downarrow$2.0}} & \colorbox{gray0}{70.7\textsubscript{\color{red}$\downarrow$0.2}} & \colorbox{gray0}{83.6\textsubscript{\color{red}$\downarrow$1.0}} & 87.0\textsubscript{\color{red}$\downarrow$1.5} & 92.0\textsubscript{\color{red}$\downarrow$0.5} & 84.1\textsubscript{\color{red}$\downarrow$1.6} & \colorbox{gray0}{87.0\textsubscript{\color{red}$\downarrow$0.1}}  \\
    PrivImage & 94.0\textsubscript{\color{red}$\downarrow$1.2} & 97.8\textsubscript{\color{red}$\downarrow$0.1} & 79.9\textsubscript{\color{red}$\downarrow$2.0} & 87.1\textsubscript{\color{brown}$\uparrow$0.4} & \colorbox{gray0}{74.5\textsubscript{\color{red}$\downarrow$0.5}} & \colorbox{gray0}{78.4\textsubscript{\color{red}$\downarrow$0.4}} & 9.9\textsubscript{\color{red}$\downarrow$0.3}  & 15.8\textsubscript{\color{red}$\downarrow$0.7} & 47.3\textsubscript{\color{red}$\downarrow$3.7} & 71.0\textsubscript{\color{red}$\downarrow$1.3} & 90.8\textsubscript{\color{red}$\downarrow$1.3} & 92.0\textsubscript{\color{red}$\downarrow$2.6} & 82.8\textsubscript{\color{red}$\downarrow$3.0} & 82.9\textsubscript{\color{red}$\downarrow$3.0} \\
    \bottomrule
\end{tabular}}
\vspace{-1mm}
\end{table*}

\begin{table*}[!t]
\small
    \centering
    \caption{ Fidelity evaluations of synthetic images under $\epsilon = 10$ and $\delta = 1 / {N \log(N)}$, where $N$ is the size of the sensitive dataset. Due to the space limitation, we report the results of the {\tt Camelyon} dataset in Table~\ref{table:camelyon_fidelity} of the Appendix. Note that IS is more informative in natural image datasets such as {\tt CIFAR-10} and {\tt CIFAR-100}; we include it in all datasets for completeness.}
    \vspace{-3mm}
    \label{tab:fidelity}
    \setlength{\tabcolsep}{3.8pt}
    \resizebox{0.99\textwidth}{!}{
    \begin{tabular}{l|cccccc|cccccc|cccccc}
    \toprule
    \multirow{2}{*}{Algorithm} & \multicolumn{6}{c|}{{\tt MNIST}} & \multicolumn{6}{c|}{{\tt FashionMNIST}} & \multicolumn{6}{c}{{\tt CIFAR-10}} \\
    \cline{2-19}
     & FID{\color{red}$\downarrow$} & IS{\color{blue}$\uparrow$} & Precision{\color{blue}$\uparrow$} & Recall{\color{blue}$\uparrow$} & FLD{\color{red}$\downarrow$} & IR{\color{blue}$\uparrow$} &  FID{\color{red}$\downarrow$} & IS{\color{blue}$\uparrow$} & Precision{\color{blue}$\uparrow$} & Recall{\color{blue}$\uparrow$} & FLD{\color{red}$\downarrow$} & IR{\color{blue}$\uparrow$} & FID{\color{red}$\downarrow$} & IS{\color{blue}$\uparrow$} & Precision{\color{blue}$\uparrow$} & Recall{\color{blue}$\uparrow$} & FLD{\color{red}$\downarrow$} & IR{\color{blue}$\uparrow$}\\
    \hline
    DP-MERF &  106.3 & 2.64 & 0.03 & 0.03 & 34.9 & -2.23  &  106.4 & 2.86 & 0.08 & 0.01 & 29.2 & -2.05 & 214.1 & 3.06 & 0.25 & 0.00 & 32.1 & -2.27 \\
    DP-NTK & 69.2 & 2.18 & 0.09 & 0.08 & 25.5 & -2.21 & 120.5 & 3.05 & 0.04 & 0.00 & 36.4 & -2.19 & 346.9 & 1.59 & 0.01 & 0.00 & 41.7 & -2.27 \\
    DP-Kernel & 38.9 & 2.19 & 0.24 & 0.02 & 17.8 & -2.18 & 74.2 & 3.45 & 0.23 & 0.01 & 21.3 & -1.95 & 161.4 & 3.69 & 0.18 & 0.00 & 27.2 & -2.27 \\
    PE &  45.3 & 2.79 & 0.08 & 0.28  & 25.6 & -2.21   & 23.1  & \colorbox{gray0}{5.37} & 0.15 & 0.53 & 16.1 & -1.90 &  \colorbox{gray0}{9.2} & \colorbox{gray0}{14.97} & 0.59 & 0.54 & \colorbox{gray0}{2.7} & \colorbox{gray0}{-1.31} \\
    GS-WGAN & 47.7 & 2.37 & 0.13 & 0.01  & 25.4 & -2.14 & 97.2 & 2.95 & 0.17 & 0.00 & 28.1 & -1.95 & 194.4 & 2.34 & 0.17 & 0.00 & 31.1 & -2.28 \\
    DP-GAN & 30.3 & 2.06 & 0.19 & 0.22 & 15.0 & -2.16 & 76.9 & 3.60 & 0.17 & 0.02 & 23.9 & -1.94 & 138.7 & 2.65 & \colorbox{gray0}{0.67} & 0.01 & 22.5 & -2.28 \\
    DPDM &  4.4 & 2.07 & 0.63 & 0.73 & 3.3 & -2.00 & 17.1 & 3.92 & 0.54 & 0.38 & 6.6 & -1.63 & 110.1 & 3.12 & 0.59 & 0.04 & 19.4 & -2.21 \\
    DP-FETA & 3.4 & 2.09 & \colorbox{gray0}{0.66} & 0.77 & \colorbox{gray0}{2.7} & \colorbox{gray0}{-1.99} & 13.2 & 3.93 & \colorbox{gray0}{0.60} & 0.43 & 4.6 & \colorbox{gray0}{-1.61} & 95.3 & 3.52 & 0.60 & 0.07 & 18.1 & -2.20 \\
    PDP-Diffusion & 3.8 & \colorbox{gray0}{2.87} & 0.61 & 0.82 & 3.4 & -2.02 & 6.2 & 4.23 & 0.53 & 0.71 & 4.9 & -1.66 & 18.4 & 8.11 & 0.50 & 0.66 & 7.2 & -2.09 \\
    DP-LDM (SD) & 18.7 & 2.23 & 0.21 & 0.77 & 12.2 & -2.13  &  20.1 & 4.33 & 0.32 & 0.71  & 11.7  & -1.87 & 19.8 & 8.14 & 0.48 & \colorbox{gray0}{0.68} & 9.0 & -2.13 \\
    DP-LDM  & 99.1 & 1.85 & 0.00 & 0.03 & 32.1 &  -2.18 & 53.2 & 3.74 & 0.28 & 0.40 & 15.4 & -1.87 & 47.4 & 6.40 & 0.49 & 0.49& 14.1 & -2.23 \\
    DP-LoRA & 95.4 & 1.85 & 0.00 & 0.03 & 32.1 & -2.18  &  43.7 & 3.98 & 0.26 & 0.53 & 14.8 & -1.89 & 27.8 & 7.64 & 0.49 & 0.64 & 9.3 & -2.18 \\
    PrivImage &  \colorbox{gray0}{2.3} & 2.16 & 0.62 &  \colorbox{gray0}{0.83} & 2.8 & -2.01  & \colorbox{gray0}{5.3} & 4.29 & 0.56 & \colorbox{gray0}{0.72} & \colorbox{gray0}{4.3} & -1.64 & 13.1 & 8.41 & 0.56 & 0.63 & 5.1 & -1.92 \\
    \bottomrule
    \toprule
    \multirow{2}{*}{Algorithm} & \multicolumn{6}{c|}{{\tt CIFAR-100}} & \multicolumn{6}{c|}{{\tt EuroSAT}} & \multicolumn{6}{c}{{\tt CelebA}} \\
    \cline{2-19}
     & FID{\color{red}$\downarrow$} & IS{\color{blue}$\uparrow$} & Precision{\color{blue}$\uparrow$} & Recall{\color{blue}$\uparrow$} & FLD{\color{red}$\downarrow$} & IR{\color{blue}$\uparrow$} &  FID{\color{red}$\downarrow$} & IS{\color{blue}$\uparrow$} & Precision{\color{blue}$\uparrow$} & Recall{\color{blue}$\uparrow$} & FLD{\color{red}$\downarrow$} & IR{\color{blue}$\uparrow$} & FID{\color{red}$\downarrow$} & IS{\color{blue}$\uparrow$} & Precision{\color{blue}$\uparrow$} & Recall{\color{blue}$\uparrow$} & FLD{\color{red}$\downarrow$} & IR{\color{blue}$\uparrow$} \\
    \hline
    DP-MERF &  197.5 & 2.93 & 0.32 & 0.00 & 25.2 & -2.26 & 182.2 & 2.20 & 0.05 & 0.00 & 34.6 & -2.26 & 147.9 & 3.61 & 0.08 & 0.00 & 17.1 & -2.20 \\
    DP-NTK & 398.1 & 1.28 & 0.13 & 0.00 & 43.0 & -2.28 & 257.6 & 1.95 & 0.06 & 0.00  & 40.3 & -2.27 & 227.8 & 3.21 & 0.05 & 0.00 & 30.6 & -2.23 \\
    DP-Kernel & 224.1 & 2.43 & 0.12  &  0.00 & 26.1 & -2.27 & 181.6 & 2.52 & 0.05 & 0.01 & 32.2 & -2.23 & 128.8 & 2.90 & 0.11 & 0.00 & 14.4 & -1.72 \\
    PE & \colorbox{gray0}{9.6} & \colorbox{gray0}{19.65} & 0.58 & 0.54 & \colorbox{gray0}{3.7} & \colorbox{gray0}{-2.22} & 48.1 & \colorbox{gray0}{8.16} & 0.08 & \colorbox{gray0}{0.69} & 16.7 & -2.10 & 22.0 & \colorbox{gray0}{3.87} & 0.28 & 0.52 & 5.1 & -0.96 \\
    GS-WGAN &  246.4 & 1.87 & 0.44 & 0.00 & 26.1 & -2.28 & 236.3 & 1.75 & 0.02 & 0.0 & 45.0 & -2.28 & 290.0 & 1.66 & 0.03 & 0.00 & 43.2 & -2.27 \\
    DP-GAN & 181.0 & 2.34 & 0.45 &  0.00 & 23.2 & -2.28 & 222.4 & 1.95 & 0.03 & 0.00 & 47.0 & -2.27 & 31.7 & 2.28 & 0.62 & 0.05 & 3.9 & \colorbox{gray0}{-0.69} \\
    DPDM & 130.2 & 2.73 & \colorbox{gray0}{0.68} & 0.01 & 19.0  &  -2.23 & 168.8 & 1.67 & 0.61 & 0.10 & 20.3 & -1.57 & 28.8 & 2.23 & 0.60 & 0.15 & 4.5 & -1.38 \\
    DP-FETA & 114.0 & 3.30 & 0.63 & 0.02 & 18.6 & \colorbox{gray0}{-2.22} &   137.0 & 1.81 & \colorbox{gray0}{0.68} & 0.03 & 14.6 & \colorbox{gray0}{-1.53} & 24.8 & 2.30 & \colorbox{gray0}{0.63} & 0.17 & 3.5 & -1.26 \\
    PDP-Diffusion & 19.9 & 8.28 & 0.53 & \colorbox{gray0}{0.64} & 9.1 & -2.25 & \colorbox{gray0}{24.2} & 3.39 & 0.57 & 0.61 & 4.8 & -1.62  & \colorbox{gray0}{8.1} & 2.73 & 0.53 & 0.61 & \colorbox{gray0}{2.2} & -1.09 \\
    DP-LDM (SD) & 19.8 & 8.39 & 0.53 & 0.63 & 8.94 & -2.25 & 29.7 & 3.39 & 0.49 & 0.60 & 6.7 & -1.68 & 24.1 & 3.15 & 0.38 & \colorbox{gray0}{0.64} & 5.1 & -1.49 \\
    DP-LDM & 51.5 & 6.08 & 0.57 & 0.39 & 12.4 & -2.26 & 69.6 & 3.06 & 0.42 & 0.27 & 15.3 & -1.96 &  40.4 & 3.02 & 0.40 & 0.40 & 6.7 & -1.65 \\
    DP-LoRA & 29.1 & 7.71 & 0.54 & 0.57 & 9.25 & -2.26 & 50.9 & 3.23 & 0.46 & 0.36 & 12.8 & -1.93 & 32.2 &3.03  & 0.38 & 0.52 & 6.0 & -1.62 \\
    PrivImage & 21.4 & 8.08  & 0.54 & 0.62 & 8.9 & -2.24 & \colorbox{gray0}{24.2} & 3.75 & 0.49 & 0.66 & 5.4 & -1.69 & 11.3 & 2.88 & 0.50 & 0.61 & 2.9 & -1.25 \\
    \bottomrule
\end{tabular}
}
\vspace{-2mm}
\end{table*}

\vspace{-1mm}
\section{EMPIRICAL EVALUATIONS}
\label{sec:eval}
\vspace{-1mm}

Using our proposed evaluation framework and metrics, we conduct a series of comprehensive experiments to address the following four research questions (RQs):

\begin{itemize}[leftmargin=*]
\item \textbf{RQ1:} How do the various studied methods perform regarding both the utility and fidelity of synthetic images?
\item \textbf{RQ2:} How do model architecture and privacy budget affect the performance of studied algorithms?
\item \textbf{RQ3:} How does pretraining with public datasets affect the performance of synthetic images?
\item \gc{\textbf{RQ4:} Can we combine various methods to further enhance synthetic performance?}
\end{itemize}

We provide the computational resources and runtime analysis of studied methods in Table~\ref{tab:computationalResource} of Appendix~\ref{suppsub:computation_resources}.

\vspace{-2mm}
\subsection{The Quality of Synthetic Images (RQ1)}
\label{sec:quality}
\noindent \textbf{Experiment Design.} 
To address RQ1, this section evaluates the utility and fidelity of synthetic images generated by the studied methods under privacy budgets $\epsilon = \{1, 10\}$, with visualizations for intuitive insights. The synthesizer sizes for GAN-based and diffusion-based methods are 5.4M and 3.8M, respectively. For GAN-based methods, we use BigGAN~\cite{biggan} with a generator size of 3.8M and a discriminator size of 1.6M, aligning the generator sizes with diffusion-based methods. For diffusion-based methods, we adopt the model size settings from DPDM~\cite{dpdm} and PrivImage~\cite{li2023PrivImage}. In latent diffusion-based methods (e.g., DP-LDM and DP-LoRA), as described in Section~\ref{subsec:generative}, an autoencoder is integrated into the diffusion process, with DMs sized at 3.8M and the autoencoder following the original implementations in DP-LDM~\cite{dpldm} and DP-LoRA~\cite{dplora} (see Appendix~\ref{supp:gen_network} for network architecture details). 

\begin{figure*}[!t]
    \centering
    \hspace{-0.2cm}
    \subfigure[GAN-Based Synthesizers.]{
        \includegraphics[width=3.44 in]{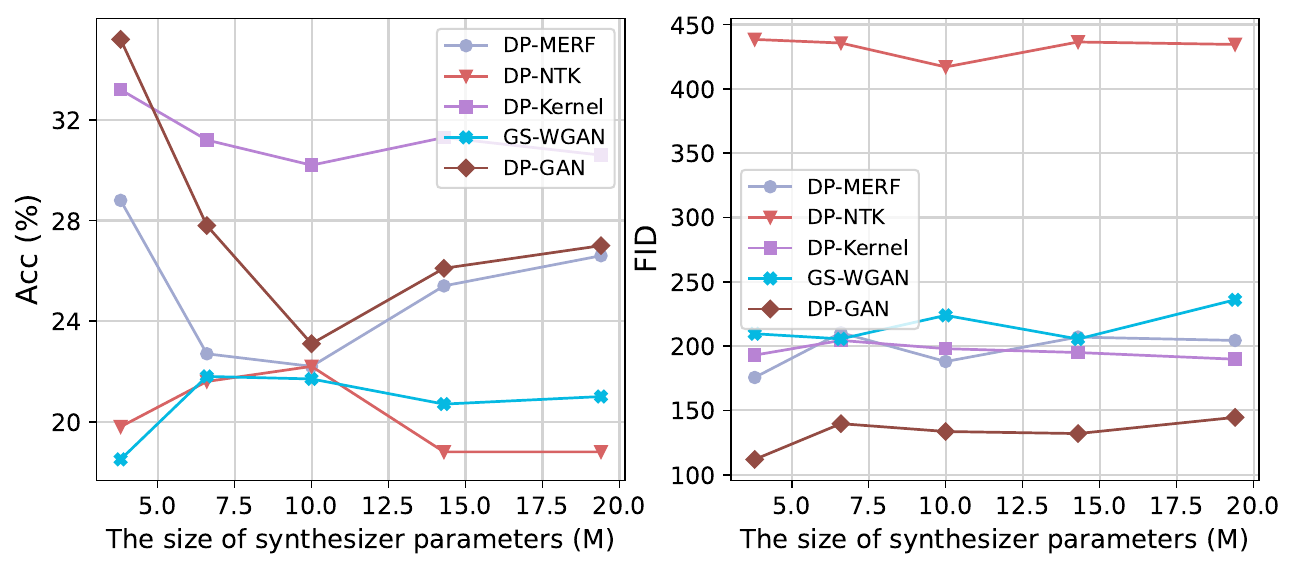}    
    }
    \hspace{-0.5cm}
    \subfigure[Diffusion-based Synthesizers.]{
        \includegraphics[width=3.44 in]{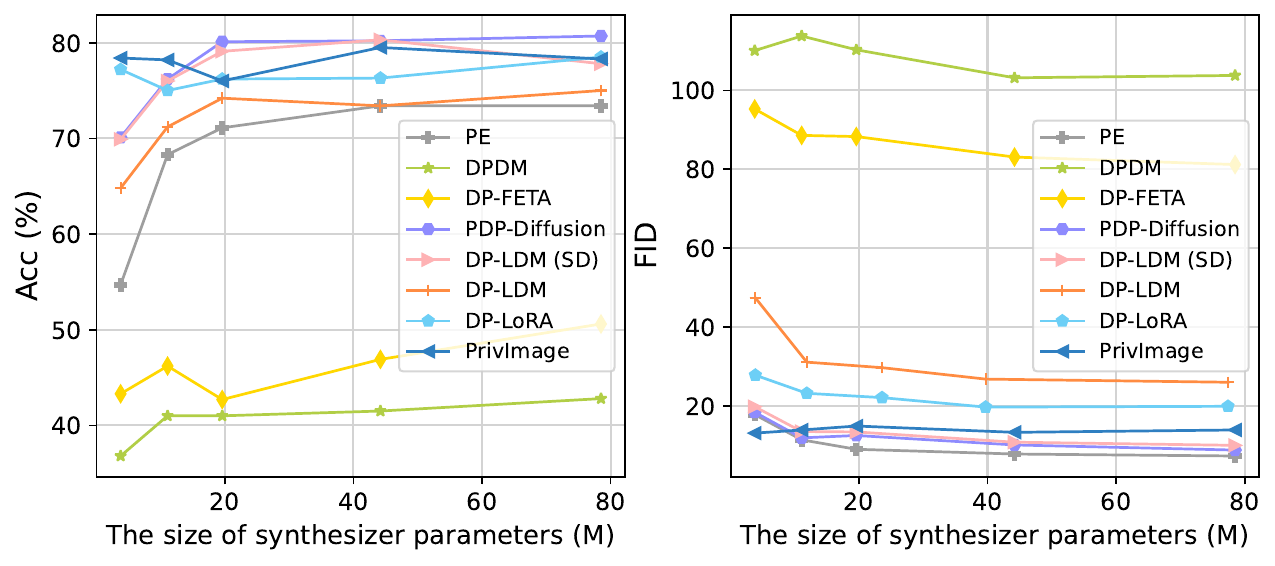}
    }
    \vspace{-4.5mm}
    \caption{The Acc and FID of synthetic images generated by the studied algorithms for {\tt CIFAR-10} under privacy budgets $\epsilon = 10$.}
    \label{fig:model_size_change}
    \vspace{-3mm}
\end{figure*}

High-dimensional sensitive datasets with resolutions above 
$32 \times 32$ are resized to  $32 \times 32$ for fair comparison, following prior work~\cite{dp-diffusion,li2023PrivImage,dpdm}. For {\tt CelebA}, we also tested higher resolutions (
$64 \times 64$ and $128 \times 128$). Unless specified, utility evaluations in this paper use the `Noisy SenV' method.

\noindent \textbf{Result Analysis.}
Tables~\ref{tab:utility} and \ref{tab:utility_val} present the Acc of classifiers trained on synthetic images, with the best score per column highlighted in a \colorbox{gray0}{gray} box. Table~\ref{tab:fidelity} reports fidelity results under $\epsilon=10$. Note that our experimental settings differ from the original papers in model architecture, failure probability $\delta$, and evaluation methods, which may cause performance variations compared to the reported results. Figure~\ref{fig:eps10_visual} in Appendix~\ref{suppsubsec:visualize} provides examples of synthetic images. These findings reveal the following key takeaways:

\noindent {\textbf{Directly tuning classifiers using the sensitive dataset overestimates the utility of synthetic images.} For instance, in Table~\ref{tab:utility}, DP-Kernel on {\tt EuroSAT} shows an accuracy drop of 21.2\% and 24.9\% under $\epsilon = \{1,10\}$ when using the synthetic dataset as the validation set. {Tables~\ref{tab:utility} and~\ref{tab:utility_val} present the average results for Noisy SenV' and SynV,' showing a decrease of 1.33\% and 4.48\% in accuracy under $\epsilon=1$, and 1.12\% and 3.00\% under $\epsilon=10$, compared to the `SenV' evaluation method.} { As $\epsilon$ increases, the quality of synthetic images improves, leading to reduced utility evaluation errors. These results also indicate that the `Noisy Senv' method achieves better utility evaluation than the `SynV.' }

\begin{table}[!t]
\small
    \centering
    \caption{The Acc and FID of synthetic images using {\tt CelebA} under $\epsilon=10$, with varying image resolutions. The best and second-best results are highlighted in bold and underlined.}
    \vspace{-3mm}
    \label{tab:resulotion}
    \setlength{\tabcolsep}{3.6mm}{
    \resizebox{0.48\textwidth}{!}{
    \begin{tabular}{l|cc|cc|cc}
    \toprule
    \multirow{1}{*}{Algorithm} & \multicolumn{2}{c|}{{\tt $32 \times 32$}} & \multicolumn{2}{c|}{{\tt $64 \times 64$}}  & \multicolumn{2}{c}{{\tt $128 \times 128$}} \\
    \cline{2-7}
     ({\tt CelebA}) & Acc  & FID & Acc & FID  & Acc & FID \\
    \hline
    DP-MERF & 81.2 & 147.9 & 61.7  & 381.2 & 61.4 & 311.2 \\
    DP-NTK & 64.2 & 346.6 & 64.2 & 299.9 & 62.9 & 272.1 \\
    DP-Kernel & 83.7 & 128.8 & 73.5 & 272.4 & 53.7 & 359.2 \\
    PE & 74.2 & 22.0 & 70.2 & \underline{26.8} & 68.9 & \textbf{59.4} \\
    GS-WGAN & 61.5 & 290.0 & 61.8 & 433.6 & 61.4 & 383.2 \\
    DP-GAN & 89.2 & 31.7 & 61.8 & 395.0 & 39.9 & 320.2 \\
    DPDM   & 91.8 & 28.8 & 78.7 &  106.7 & 71.1 & 210.8 \\
    DP-FETA   & \textbf{94.2} & 24.8 & 82.6 & 89.4 & 69.3 & 173.6 \\
    PDP-Diffusion & \underline{94.0} & \textbf{8.1} & \textbf{92.7} & 91.7 & 74.1 & 121.1 \\
    DP-LDM (SD)  & 89.1 & 24.1 & 84.3 & 117.8  & 67.3 & 188.6 \\
    DP-LDM  & 92.4 & 40.4 & 90.2 & 110.7 & \textbf{83.6} & 156.1 \\
    DP-LoRA  & 92.0 & 32.2 & 82.6 & 99.2 & \underline{78.0} & 169.9 \\
    PrivImage & 92.0 & \underline{11.3} &  \underline{91.3} & \textbf{23.6} & 61.2 & \underline{106.3} \\
    \hline
    \rowcolor{gray0} \textbf{Average} & 84.0 & 87.5 & 76.6 & 188.4 & 65.6 & 218.4 \\
    \bottomrule
\end{tabular}
}}
\vspace{-3mm}
\end{table}

\noindent {\textbf{Our fidelity and utility evaluation provides a more comprehensive view of the quality of 
synthetic images.}
IS evaluates the practical quality of synthetic images but struggles with unnatural datasets like {\tt MNIST}, {\tt FashionMNIST}, and {\tt EuroSAT}~\cite{fld}. In Table~\ref{tab:fidelity}, PE achieves IS scores of 2.79, 5.37, 14.97, 19.65, 8.16, 3.87, and 4.58 for the seven sensitive datasets; six out of seven datasets achieve the highest IS across all methods. However, despite the high practical quality of these images, they are not of good utility, achieving Acc scores of 32.7\%, 57.8\%, 75.3\%, 24.9\%, 36.8\%, 74.2\%, and 64.9\%. Five out of seven of them are lower than the best methods. 

{For DP-LDM and DP-LoRA for {\tt MNIST}, the Acc under $\epsilon=10$ is 95.5\% and 97.1\%, but the FID is 99.1 and 95.4. The synthetic images retain essential information for downstream models but may lack realistic textures, colors, and fine details, as these methods prioritize class-discriminative features like shapes over photorealism. In their original implementations, DP-LDM and DP-LoRA resize {\tt MNIST} images to $32 \times 32$, whereas the original size of {\tt MNIST} image is $28 \times 28$. This resizing results in better FID scores reported in their paper compared to \toolname{}. When resizing to $32 \times 32$, DP-LDM and DP-LoRA achieve FID scores of 13.4 and 12.1, similar to what was reported in their papers. Precision and Recall are decoupled from FID, meaning that having one value high and the other low does not necessarily improve the FID for synthetic images. For example, in the case of DPGAN with {\tt CIFAR-10}, the precision is 0.67, the best among all methods. However, its recall is as low as 0.01, and the FID is as high as 138.7.} FLD considers the novelty of synthetic images, and better FLD scores generally indicate that most other fidelity metrics are also favorable. The human perspective metric, IR, is consistently negative due to the low $32 \times 32$ resolution of synthetic images compared to $1024 \times 1024$ in paper~\cite{xu2024imagereward}.
Further fidelity evaluation details for {\tt Camelyon} are in Appendix~\ref{suppsubsec:camelyon}.

\begin{figure*}[!t]
    \centering
    \setlength{\abovecaptionskip}{0pt}
    \includegraphics[width=\textwidth]{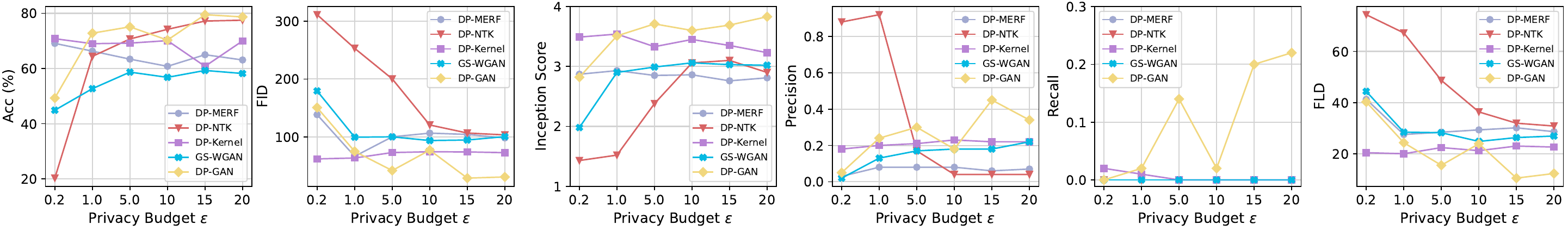}
    \caption{The utility and fidelity of synthetic images generated by the five GAN-based methods for {\tt FashionMNIST} dataset under privacy budgets $\epsilon = \{0.2, 1.0, 5.0, 10.0, 15.0, 20.0\}$ and a fixed $\delta = 1/N \log(N)$. Due to the space limitation, Figure~\ref{fig:change_eps_diffusion} in the Appendix presents the results for the remaining eight diffusion-based methods. }
    \label{fig:change_eps}
    \vspace{-3mm}
\end{figure*}

\begin{figure*}[!t]
    \centering
    \hspace{-0.2cm}
    \subfigure[With and without pretraining.]{
        \includegraphics[width=3.44 in]{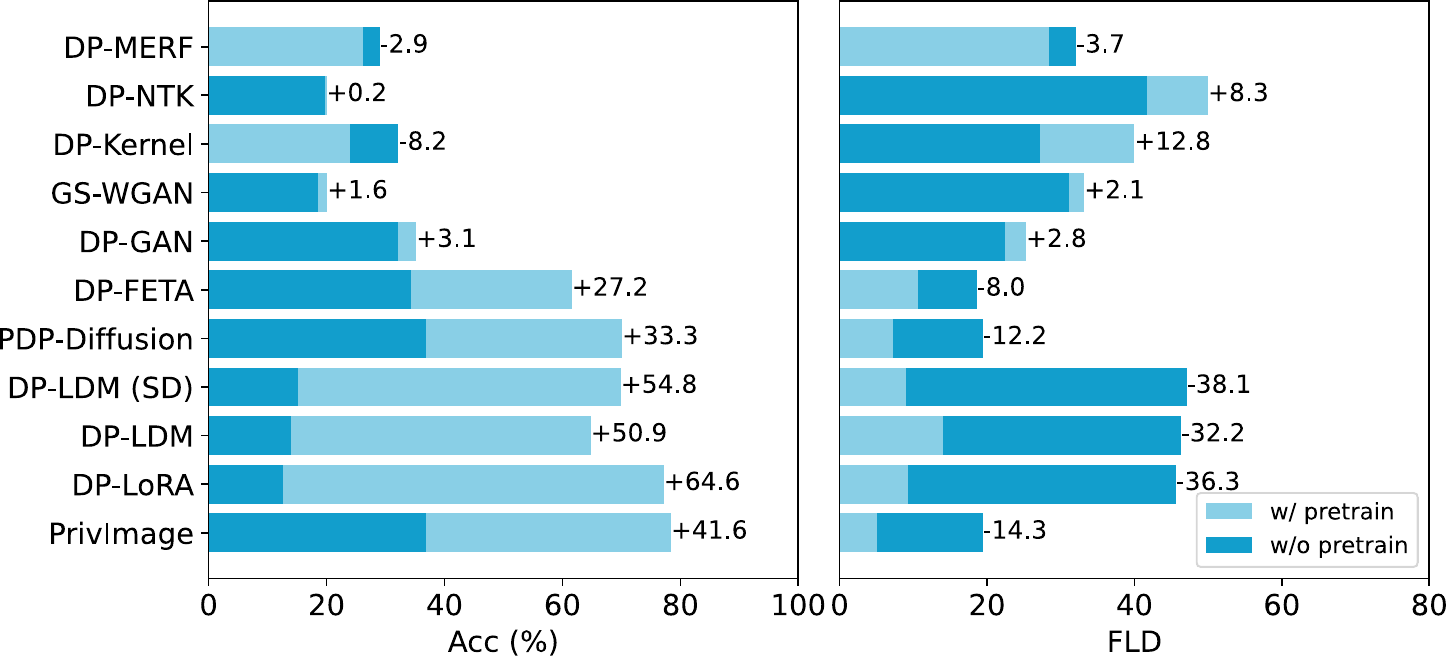}    
    }
    \hspace{-0.3cm}
    \subfigure[Conditional and unconditional pretraining.]{
        \includegraphics[width=3.44 in]{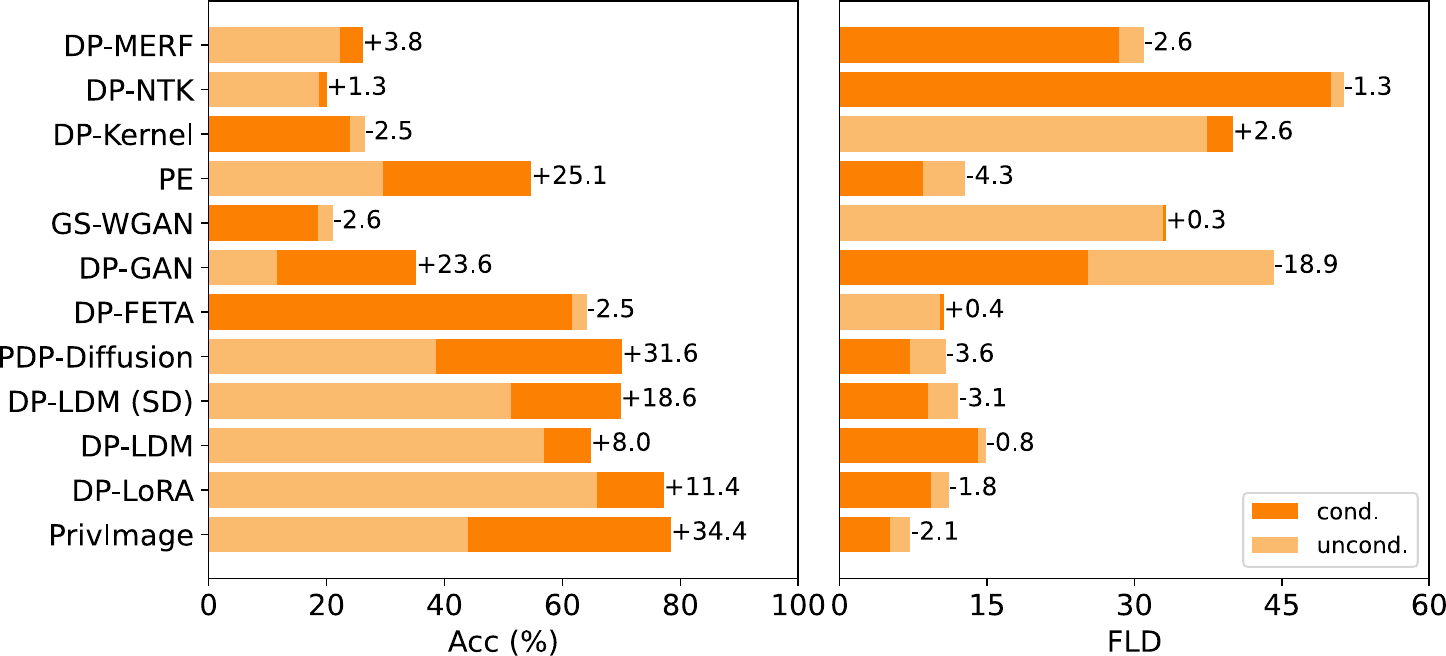}
    }
    \vspace{-3.5mm}
    \caption{The Acc and FLD evaluations of synthetic {\tt CIFAR-10} images under $\epsilon = 10$ and $\delta = 1/N \log(N)$. The labels ``w/ pretrain'' and ``w/o pretrain'' denote algorithms with and without pretraining, while ``cond.'' and ``uncond.'' indicate conditional and unconditional pretraining. Relative changes reflect ``w/ pretrain'' minus ``w/o pretrain'' and ``cond.'' minus ``uncond.''.}
    \label{fig:cifar10-convsuncon}
    \vspace{-3mm}
\end{figure*}

\noindent {\textbf{Current methods still face challenges in high-resolution DP image synthesis.} Table~\ref{tab:resulotion} shows the Acc and FID of synthetic {\tt CelebA} with varying resolutions. As resolutions of synthetic images increases from $32\times 32$ to $128\times 128$, average accuracy and FID drop from 84.0 and 87.5 to 65.6 and 218.4, respectively, indicating that current methods struggle with high-resolution synthesis. However, PE is the least affected by the increased resolutions in synthetic images, as the synthetic images of PE only depend on the capability of foundational APIs. Other methods, such as PrivImage and DP-LDM, which train synthesizers using DP-SGD, fail in high-resolution image synthesis. High-resolution images contain more complex spatial structures, requiring more precise gradients to capture meaningful patterns. However, DP-SGD injects noise into gradients, which disrupts these updates. As image resolution increases, the number of model parameters and/or the number of required model updates also grows, making it harder for DP-SGD to converge. 
We use a default synthesizer size of 3.8M; larger sizes may enhance performance.}

\vspace{-1mm}
\subsection{Model Size and Privacy Budget (RQ2)} 
\label{subsec:rq2}

\noindent \textbf{Experiment Design.} 
We evaluate the utility and fidelity of synthetic images across six privacy budgets $\epsilon = \{0.2, 1.0, 5.0, 10, 15, 20\}$. We further evaluate the performance of DP image synthesis methods with varying synthesizer sizes under a certain privacy budget $\epsilon=10$. For diffusion-based synthesizers, we consider sizes of $\{3.8\text{M}, 10.0\text{M}, 20.0\text{M}, 40.0\text{M}, 80.0\text{M}\}$, while for the model size of GAN-based synthesizers, we use $\{3.8\text{M}, 6.6\text{M}, 10.0\text{M}, 14.3\text{M}, 19.4\text{M}\}$.

\noindent \textbf{Result Analysis.} 
We summarize the key takeaways from this section as follows:

\noindent {\textbf{Adding noise on low-dimensional features is less sensitive to privacy budgets.}}
Figure~\ref{fig:change_eps} shows that adding noise to high-dimensional features, like weight gradients, generally yields superior synthesis performance compared to adding noise to low-dimensional features, such as adding noise to training loss in DP-Kernel, especially at large privacy budgets like $\epsilon = 10$ and $20$. However, methods that add noise to high-dimensional features are more sensitive to the privacy budget. At very low privacy budgets, such as $\epsilon = 0.2$, methods that add noise to low-dimensional features outperform, with DP-Kernel and DP-MERF achieving the highest accuracy of 71.2\% and 70.6\%, due to their lower sensitivity and smaller noise scale.

\begin{figure}[!t]
    \centering
    \setlength{\abovecaptionskip}{0pt}
    \includegraphics[width=0.98\linewidth]{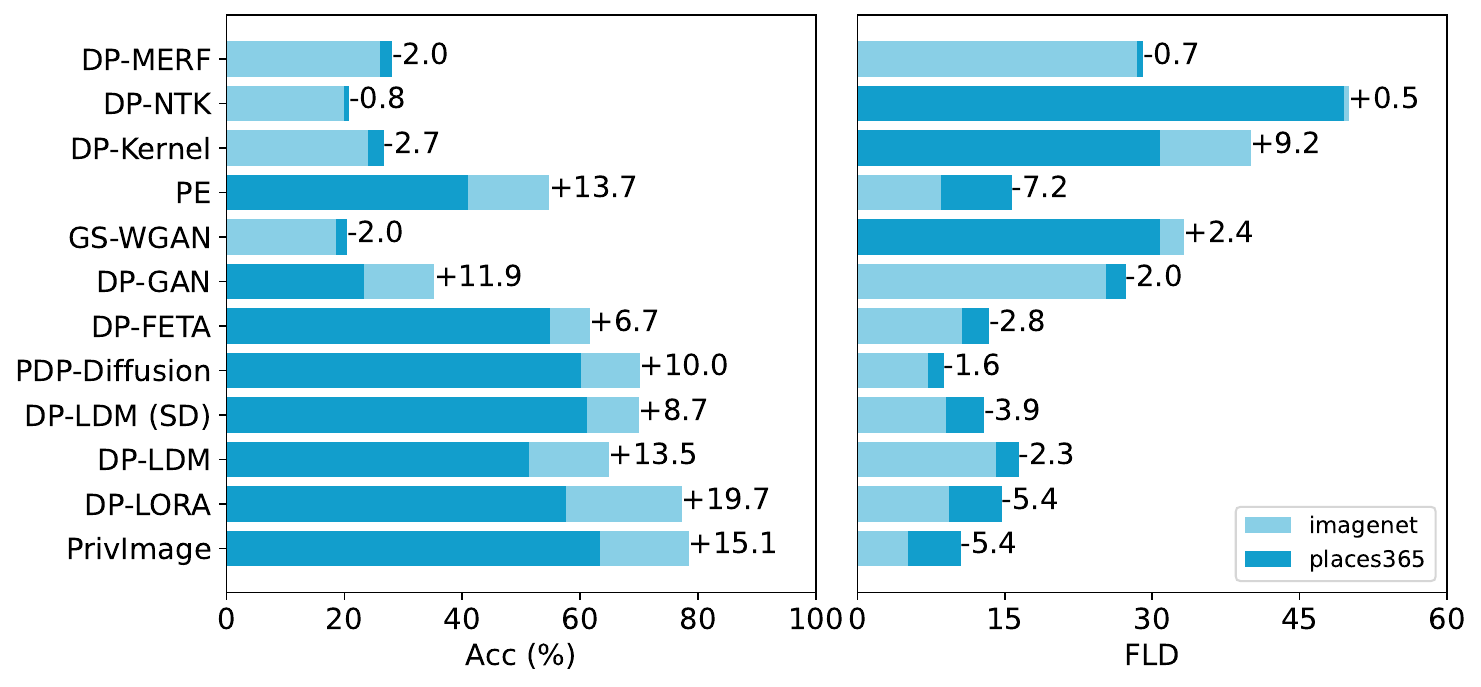}
    \caption{The Acc and FLD evaluations of synthetic images under $\epsilon = 10$ and $\delta = 1 / N \log(N)$ for the sensitive dataset {\tt CIFAR-10}, with public datasets {\tt ImageNet} and {\tt Places365}.}
    \label{fig:chang_public_images}
    \vspace{-4mm}
\end{figure}

\noindent {\textbf{Larger synthesizer models do not always yield better performance.}}
GANs often face unstable training~\cite{wgan,lin2021spectral} and mode collapse~\cite{lin2018pacgan,goodfellow2016nips}, with DP further complicating training of larger models. Figure~\ref{fig:model_size_change} shows that scaling up model sizes does not always benefit GAN-based methods. Similarly, in diffusion-based methods like PrivImage, accuracy drops from 78.2\% to 76.0\% as synthesizer size increases from 11.1M to 19.6M. Larger synthesizers, despite better learning capacity, suffer from increased DP noise under the same privacy budget, potentially degrading performance. Li et al.~\cite{li2023PrivImage} noted that when synthesizers reach their generalization limit, DP exacerbates performance decline in larger models. Increasing the batch size in DP-SGD can mitigate DP noise~\cite{de2022unlocking}, as shown in Table~\ref{tab:batchsize} (Appendix~\ref{suppsubsec:batch}), where PrivImage achieves better Acc with a batch size of 16,384 (used in PrivImage~\cite{li2023PrivImage}) compared to 4,096, especially for large model sizes.

\vspace{-2mm}
\subsection{Leveraging Public Image Datasets (RQ3)}
\label{subsec:rq3}

\noindent \textbf{Experiment Design.} This RQ examines how pretraining on public image datasets benefits DP image synthesis under a privacy budget $\epsilon = 10$ from three perspectives as follows.
\begin{itemize}[leftmargin=*]
    \item We compare the performance of DP image synthesizers with and without pretraining on a public dataset. 
    \item We compare the synthesis performance between unconditional and conditional pretraining. Details of the pretraining methods are provided in Appendix~\ref{app:pretraining}.
    \item We compare the performance of synthesizers with different pretraining datasets, {\tt ImageNet} and {\tt Places365}.
\end{itemize}

Appendix~\ref{app:pretraining} introduces the default pretraining strategy. As noted in Section~\ref{sub:model-level}, GS-WGAN pretrains synthesizers using sensitive datasets, unlike PrivImage~\cite{li2023PrivImage}, which rely on external public datasets. PrivImage and PDP-Diffusion, built on DPDM, use a distinct pretraining method, reverting to DPDM without pretraining.

\noindent \textbf{Result Analysis.} The takeaways from this section are as follows:

\noindent {\textbf{Pretraining is more beneficial for diffusion-based methods than for GAN-based methods.
}}
Figure~\ref{fig:cifar10-convsuncon} compares the Acc and FLD of synthetic {\tt CIFAR-10} images with and without pretraining, using both conditional and unconditional methods. Lower values are shown on the left bar and higher on the right. It reveals that pretraining consistently enhances utility and fidelity in diffusion-based methods, but its effect on GAN-based methods varies. For DP-Kernel, the Acc shows an 8.2\% reduction while pretraining the synthesizer. GAN-based models have limited ability to produce high-quality images, so pretraining has less effect on their performance than it does for diffusion-based methods. 

\noindent {\textbf{Conditional pretraining outperforms unconditional pretraining.}}
In Figure~\ref{fig:cifar10-convsuncon}, using unconditional pretraining for PrivImage results in a 34.4\% drop in accuracy and a 2.1 increase in FLD. The reason is that using labels for conditional pretraining helps establish a stronger connection between the labels and images. When using entire public datasets for pretraining, the presence of various labels unrelated to the sensitive images can cause the model to learn redundant label-to-image mappings. This distraction compromises its ability to learn the meaningful features necessary for accurately synthesizing sensitive images. Table~\ref{tab:fine_tuning_pre_training_steps} of Appendix~\ref{suppsubsec:pretraining} shows that excessive pretraining may reduce the synthetic performance.

\noindent {\textbf{A closer distribution match between pretraining and sensitive datasets enhances DP image synthesis.}}
PrivImage, which uses a public image subset closely matching the sensitive images’ distribution, achieves the highest utility on {\tt CIFAR-10} in Table~\ref{tab:utility_val}. Figure~\ref{fig:chang_public_images} compares the Acc and FLD for the sensitive datasets {\tt CIFAR-10} when pretraining with public datasets {\tt ImageNet} and {\tt Places365}. Appendix~\ref{suppsubsec:pretraining} of Figure~\ref{fig:chang_public_images_fashion} shows the results of {\tt FashionMNIST}. The average accuracy scores for {\tt CIFAR-10} using {\tt ImageNet} and {\tt Places365} as the pretraining datasets are 48.3 and 42.3, respectively.
Actually, {\tt ImageNet}’s distribution is closer to {\tt CIFAR-10} than {\tt Places365} because its \textit{object-focused} categories align more with {\tt CIFAR-10}’s content~\cite{imagenet}. {\tt Places365}, focused on scenes~\cite{places365}, introduces less relevant background information, making its distribution less similar to {\tt CIFAR-10}’s object-centric data.
For GAN-based methods, pretraining offers minimal improvement, with performance varying slightly depending on whether {\tt ImageNet} or {\tt Places365} is used. In contrast, for diffusion-based methods, pretraining enhances performance, as shown in Figure~\ref{fig:cifar10-convsuncon}.

\vspace{-2mm}
\subsection{\gc{Combining Algorithmic Improvement (RQ4)}}
\label{subsec:combining_improvement}

\noindent \textbf{Experiment Design.} 
We create four subsets from {\tt ImageNet} to compare synthesis performance during pretraining: (1) 5\% randomly selected images; (2) 10\% randomly selected images; (3) 5\% selected via PrivImage’s distribution matching to align with the sensitive dataset’s semantics; and (4) 5\% from PrivImage plus an additional 5\% randomly selected. PrivImage is equivalent to PDP-Diffusion with 5\% selected pretraining data. %

\noindent \textbf{Result Analysis.} \gc{Table~\ref{tab:select_public_dataset} presents the results of using just 5\% of a public dataset, aligned with the categories of {\tt CIFAR-10}. Table~\ref{tab:select_public_dataset_whole} in Appendix shows the results of FLD. This table highlights the following key takeaway:}

\noindent \gc{\textbf{Integrating the pretraining dataset selection proposed in PrivImage can enhance synthetic performance.}} The pretraining dataset selection improves the similarity between the public and 
sensitive dataset, boosting accuracy by 8.8\% (78.9\%–70.1\%) for PDP-Diffusion and 4.9\% (74.8\%–69.9\%) for DP-LDM (SD).
Compared to randomly selecting images from the public dataset, there are performance decreases in Acc when using the entire public dataset for pretraining. The improvement for GAN-based methods is minimal because GANs struggle more than DMs to learn the distribution of the public dataset. These results indicate that the relationship between the distributions of public and sensitive datasets plays a crucial role in improving synthetic image performance, which aligns with findings in other data domains~\cite{yin2022practical}. This
experiments also present efforts that incorporate the strength of selecting public datasets as proposed taxonomy in Section~\ref{sec:taxonomy} to enhance methods. 

\begin{table}[!t]
\setlength{\tabcolsep}{4.5pt}
\small
    \centering
    \caption{The Acc of synthetic {\tt CIFAR-10} images under $\epsilon = 10$. `5\%' means matching categories from {\tt CIFAR10} dataset in {\tt ImageNet} (50 out of the 1000 available categories) using PrivImage. `5\% rand.' refers to 5\% categories randomly selected. }
    \vspace{-3.5mm}
    \setlength{\tabcolsep}{2.3mm}{
    \resizebox{0.47\textwidth}{!}{
    \begin{tabular}{l|cc|cc|c}
    \toprule
    \multirow{1}{*}{\textbf{Acc (\%)}} & 5\% & 5\% (rand.) & 5\%+5\% (rand.) & 10\% (rand.) & Whole \\
    \midrule
    DP-MERF & 22.1 & 25.1 & 18.1 & 23.7 & 26.1  \\
    DP-NTK & 22.0 & 22.3 & 19.8 & 20.0  & 20.0 \\
    DP-Kernel & 26.7 & 28.3 & 30.1 & 29.2 & 24.0 \\
    DP-GAN & 24.9 & 18.6 & 21.0 & 21.1 & 28.1 \\
    \hline
    PE & 67.7 & 52.6 & 39.4 & 37.9 & 54.7 \\
    DP-FETA & 62.2 & 49.2 & 65.0 & 51.5 &  61.4 \\
    PDP-Diffusion & 78.9 & 53.8 & 73.7 & 59.5 &  70.1\\
    DP-LDM (SD) & 74.8 & 55.2 & 73.3 & 57.8 & 69.9  \\
    DP-LDM & 68.3 & 55.2 & 62.9 & 58.2 &  64.8 \\
    DP-LoRA & 78.2 & 70.5 & 75.9 & 69.7 &  77.2 \\
    \bottomrule
\end{tabular}}
\label{tab:select_public_dataset}
}
\vspace{-3mm}
\end{table}

\gc{Appendix~\ref{suppsubsec:combining_feta} presents the results of combining the DP shortcut proposed in DP-FETA~\cite{dp-feta} with other methods, showing that algorithmic combinations do not always enhance synthetic performance. We hypothesize that the DP shortcut may disrupt the pretraining process, whereas pretraining dataset selection does not affect other algorithmic components. We need further research to effectively integrate the strengths of diverse algorithms.}

\noindent \textbf{Privacy Analysis.} Since both the data selection of PrivImage and central image constructions of DP-FETA will introduce additional privacy costs. We use Renyi DP~\cite{rdp} to compose privacy costs.

\vspace{-2mm}
\section{Related Work}
\label{sec:related}
Several studies provide benchmarks for DP in machine learning~\cite{tao2021benchmarking,hu2023sok,Dpmlbench,pate-bench}. Du et al.~\cite{symster} and Tao et al.~\cite{tao2021benchmarking} focus on DP tabular data synthesis, while Hu et al.~\cite{hu2023sok} cover various data types (e.g., tabular, graph) but overlook DP image synthesizers with pretraining~\cite{li2023PrivImage,dppromise} and foundational APIs~\cite{dpsda}. \toolname{} is the first benchmark to include state-of-the-art synthesizers (e.g., DMs and APIs) and pretraining for DP image synthesis.

\vspace{-1mm}
\section{Conclusions}
\label{sec:con}

DP image synthesis faces challenges from complex codebases, inconsistent hyper-parameters, diverse architectures, and lacking standardized evaluation metrics. Advancing these algorithms requires understanding existing methods and fair comparisons. This paper introduces \toolname{}, a benchmark for DP image synthesis evaluation. First, for \textit{methods}, we analyze twelve prominent approaches, detailing their architectures, pretraining, and privacy mechanisms, using a novel taxonomy to explore noise addition for DP and combined methods. Second, for \textit{evaluation}, we use seven private and two public datasets with seven fidelity and utility metrics, revealing that selecting classifiers based on the highest accuracy on sensitive test sets violates DP and inflates utility scores—issues \toolname{} addresses. Third, the \textit{platform} provides a standardized interface for the unified implementation of current and future methods. Our analysis yields several key insights. The pretraining on public image datasets does not always improve performance; the distributional similarity between pretraining and sensitive images is key and can limit benefits. Besides, current methods still face challenges in high-resolution DP image synthesis. These findings highlight potential research for optimizing pretraining strategies. 

\bibliographystyle{ACM-Reference-Format}
\bibliography{bib}

\vspace{3mm}
\begin{center}
    \Large{\textbf{Appendix}}
\end{center}

\appendix

\setcounter{section}{0}
\renewcommand\thesection{\Alph{section}}

\section{Privacy Notations}
\label{app:privacy-notations}

As shown in Table~\ref{tab:notion}, different definitions of neighboring datasets (i.e., replace-one, add-or-remove-one) have been used in prior work. In this section, we show that the methods that used the replace-one definition---DP-MERF, DP-NTK, and DP-Kernel---can all be adapted to the add-or-remove-one definition without affecting their $(\epsilon,\delta)$ DP parameters.

Our key idea is to show that the \emph{sensitivity} of the function these methods privatize remains the same when switching from replace-one to add-or-remove-one while the dataset size $m$ is unknown. 
For all these three methods, the function to be privatized is the \emph{mean (vector or kernel) feature embedding of the private dataset}. For simplicity, we will use DP-MERF as an example, and the proof for the other two methods is similar. Let $m$ be the number of private samples, and $\hat{\phi}(x)$ be the feature of sample $x$. Let the two neighboring datasets (under the replace-one definition) be $\mathcal{D}=\{x_1,...,x_m\}$ and $\mathcal{D}'=\{x_1',...,x_m'\}$ where they only differ on the $i$-th sample: $x_i\not=x_i'$.  Eq. (9) and (10) in DP-MERF \cite{dp-merf} show that the sensitivity of the mean feature embedding is $\frac{2}{m}$:

\begin{table*}[!t]
\small
    \centering
    \caption{
    Acc (\%) of the classifier trained on synthetic images. We adjust the DP notion of DP-MERF, DP-NTK, and DP-Kernel from ``Replace-one'' to ``Add-or-remove-one.'' }
    \vspace{-3mm}
    \label{tab:utility_1_m}
    \setlength{\tabcolsep}{7.5pt}
    \resizebox{0.99\textwidth}{!}{
    \begin{tabular}{l|cc|cc|cc|cc|cc|cc|cc}
    \toprule
    \multirow{2}{*}{Algorithm} & \multicolumn{2}{c|}{{\tt MNIST}} & \multicolumn{2}{c|}{{\tt FashionMNIST}} & \multicolumn{2}{c|}{{\tt CIFAR-10}} & \multicolumn{2}{c|}{{\tt CIFAR-100}}  & \multicolumn{2}{c|}{{\tt EuroSAT}} & \multicolumn{2}{c|}{{\tt CelebA}} & \multicolumn{2}{c}{{\tt Camelyon}} \\
    \cline{2-15}
     & $\epsilon = 1$ & $\epsilon = 10$ &  $\epsilon = 1$ & $\epsilon = 10$  &  $\epsilon = 1$ & $\epsilon = 10$  &  $\epsilon = 1$ & $\epsilon = 10$  &  $\epsilon = 1$ & $\epsilon = 10$  &  $\epsilon = 1$ & $\epsilon = 10$ &  $\epsilon = 1$ & $\epsilon = 10$ \\
    \hline
    DP-MERF & 85.0 & 82.5 & 63.7 & 61.8 & 22.8 & 25.3 & 1.8 & 1.8 &  27.8 & 39.2 & 79.2 &  76.3 & 67.9 & 67.1 \\
    DP-NTK & 62.0 & 84.0 & 59.7 & 75.4 & 17.2 & 23.5 & 1.3 & 4.2 & 25.4 & 31.6 & 61.8 & 61.7 & 55.0 & 51.5 \\
    DP-Kernel & 95.7 & 94.3 & 76.7 & 78.2 &  30.4 & 28.0 & 6.7 & 6.9 & 51.0 &  50.0 & 81.2 & 84.9 & 69.8 & 72.1 \\
    \bottomrule
\end{tabular}}
\end{table*}

\begin{table*}[!t]
\small
    \centering
    \caption{ Fidelity evaluations of synthetic images. We adjust the DP notion of DP-MERF, DP-NTK, and DP-Kernel from ``Replace-one'' to ``Add-or-remove-one.''}
    \vspace{-3mm}
    \label{tab:fidelity-2-m}
    \setlength{\tabcolsep}{3.8pt}
    \resizebox{0.99\textwidth}{!}{
    \begin{tabular}{l|cccccc|cccccc|cccccc}
    \toprule
    \multirow{2}{*}{Algorithm} & \multicolumn{6}{c|}{{\tt MNIST}} & \multicolumn{6}{c|}{{\tt FashionMNIST}} & \multicolumn{6}{c}{{\tt CIFAR-10}} \\
    \cline{2-19}
     & FID{\color{red}$\downarrow$} & IS{\color{blue}$\uparrow$} & Precision{\color{blue}$\uparrow$} & Recall{\color{blue}$\uparrow$} & FLD{\color{red}$\downarrow$} & IR{\color{blue}$\uparrow$} &  FID{\color{red}$\downarrow$} & IS{\color{blue}$\uparrow$} & Precision{\color{blue}$\uparrow$} & Recall{\color{blue}$\uparrow$} & FLD{\color{red}$\downarrow$} & IR{\color{blue}$\uparrow$} & FID{\color{red}$\downarrow$} & IS{\color{blue}$\uparrow$} & Precision{\color{blue}$\uparrow$} & Recall{\color{blue}$\uparrow$} & FLD{\color{red}$\downarrow$} & IR{\color{blue}$\uparrow$}\\
    \hline
    DP-MERF & 72.8 & 2.35 & 0.03 & 0.23 & 27.4 & -2.25 & 78.8 & 3.04 & 0.15 & 0.00 & 23.9 & -1.94 & 181.8 & 3.41 & 0.26 & 0.00 &  27.5 & -2.28 \\
    DP-NTK & 31.4 & 2.16 & 0.25 & 0.34 & 15.8 & -2.14 & 90.0 & 3.41 & 0.05 & 0.03 & 26.9 & -2.02 & 142.0 & 3.33 & 0.32 & 0.00 & 25.9 & -2.28 \\
    DP-Kernel & 43.9 & 2.22 & 0.11 & 0.01 & 22.4 & -2.15 & 64.2 & 2.82 & 0.19 & 0.00 &  17.3 & -1.93 & 201.4 & 2.90 & 0.31 & 0.00 & 30.4 & -2.28 \\
    \bottomrule
    \toprule
    \multirow{2}{*}{Algorithm} & \multicolumn{6}{c|}{{\tt CIFAR-100}} & \multicolumn{6}{c|}{{\tt EuroSAT}} & \multicolumn{6}{c}{{\tt CelebA}} \\
    \cline{2-19}
     & FID{\color{red}$\downarrow$} & IS{\color{blue}$\uparrow$} & Precision{\color{blue}$\uparrow$} & Recall{\color{blue}$\uparrow$} & FLD{\color{red}$\downarrow$} & IR{\color{blue}$\uparrow$} &  FID{\color{red}$\downarrow$} & IS{\color{blue}$\uparrow$} & Precision{\color{blue}$\uparrow$} & Recall{\color{blue}$\uparrow$} & FLD{\color{red}$\downarrow$} & IR{\color{blue}$\uparrow$} & FID{\color{red}$\downarrow$} & IS{\color{blue}$\uparrow$} & Precision{\color{blue}$\uparrow$} & Recall{\color{blue}$\uparrow$} & FLD{\color{red}$\downarrow$} & IR{\color{blue}$\uparrow$} \\
    \hline
    DP-MERF & 185.6 & 3.02 & 0.36 & 0.00 &  24.6 & -2.28 & 211.9 & 2.14 & 0.12 & 0.00 & 34.8 & -2.27 & 129.0 & 2.76 & 0.13 & 0.00 & 15.0 & -2.01 \\
    DP-NTK & 440.2 & 1.26 & 0.00 & 0.00 & 45.4 & -2.28 & 156.6 & 2.57 & 0.10 & 0.00 & 29.8 & -2.21 & 200.0 & 3.01 & 0.05 & 0.00 & 33.7 & -2.00\\
    DP-Kernel & 159.2 & 2.95 & 0.34 & 0.00 & 23.0 & -2.28 & 163.4 & 2.42 & 0.11 & 0.00 & 30.1 & -2.25 & 113.9  & 2.45  & 0.13 &  0.00 & 14.4 & -1.83 \\
    \bottomrule
\end{tabular}
}
\end{table*}

\begin{align*}
    \max_{\mathcal{D},\mathcal{D'}} \left\lVert \frac{1}{m} \sum_{i=1}^m \hat{\phi}(x_i) 
  - \frac{1}{m} \sum_{i=1}^m \hat{\phi}(x_i') 
 \right\lVert \leq \frac{2}{m}.
\end{align*}
After we switch the definition to add-or-remove-one, without loss of generality, let us assume that $\mathcal{D}=\{x_1,...,x_{m-1}\}$ and $\mathcal{D'}=\{x_1',...,x_m'\}$, where $x_i=x_i', i \in [1,...,m-1]
$. In this case,  the sensitivity is:
\begin{align*}
    &\max_{\mathcal{D},\mathcal{D'}} \left\lVert \frac{1}{m-1} \sum_{i=1}^{m-1} \hat{\phi}(x_i) 
  - \frac{1}{m} \sum_{i=1}^m \hat{\phi}(x_i') 
 \right\lVert \\
 =&\max_{\mathcal{D},\mathcal{D'}}\left\lVert \frac{1}{m(m-1)}\sum_{i=1}^{m-1}\hat{\phi}(x_i)-\frac{1}{m} \hat{\phi}(x_m) \right\lVert\\
 \leq&\frac{m-1}{m(m-1)}+\frac{1}{m} =\frac{2}{m}.
\end{align*}
In addition, this bound is tight. For example, when $\hat{\phi}(x_1)=...=\hat{\phi}(x_{m-1})=1$ and $\hat{\phi}(x_m)=-1$, the upper bound is achieved.

Previous works~\cite{dpsgd,kulesza2024mean} assume a known dataset size $m$ and overlook the privacy budget required for dataset size estimation. Similarly, DPImageBench methods using DP-SGD, including PrivImage~\cite{li2023PrivImage}, DP-LoRA~\cite{dplora}, and DP-FETA~\cite{dp-feta}, also assume $m$ is known. Then, the sensitivity of the function is defined as follows,
\begin{align*}
    &\max_{\mathcal{D},\mathcal{D'}} \left\lVert \frac{1}{m} \sum_{i=1}^{m-1} \hat{\phi}(x_i) 
  - \frac{1}{m} \sum_{i=1}^m \hat{\phi}(x_i') 
 \right\lVert \\
 =&\max_{\mathcal{D},\mathcal{D'}}\left\lVert \frac{1}{m} \hat{\phi}(x_m) \right\lVert \leq \frac{1}{m}.
\end{align*}
\noindent This equation shows that the sensitivity is halved when the dataset size $m$ is known. We adjust the sensitivity for DP-MERF, DP-NTK, and DP-Kernel, and present the results in Table~\ref{tab:utility_1_m} and Table~\ref{tab:fidelity-2-m}. The reduced sensitivity can reduce the noise scale under a certain privacy budget.  14 out of 18 entries in Table~\ref{tab:fidelity} surpass the results in Table~\ref{tab:utility_val} in terms of FID. In Table~\ref{tab:utility_val}, we observe that the improvement in Acc is not consistent when compared to the results in Table~\ref{tab:utility_val} (27 out of 42 entries in Table~\ref{tab:utility_1_m} exceed the results in Table~\ref{tab:utility_val}). This inconsistency arises because the quality of the synthetic images is not sufficiently high, leading to instability in the Acc.

\begin{table}[!t]
\small
    \centering
    \caption{Privacy notion of studied algorithms.}
    \vspace{-2mm}
    \label{tab:notion}
    \setlength{\tabcolsep}{7.8mm}{
    \resizebox{0.39\textwidth}{!}{
    \begin{tabular}{lc}
    \toprule
    {Algorithm} &  {Privacy notion} \\
    \hline
    DP-MERF~\cite{dp-merf} &  Replace-one \\
    DP-NTK~\cite{dp-ntk} &  Replace-one \\
    DP-Kernel~\cite{dp-kernel} &  Replace-one \\
    PE~\cite{dpsda} & Add-or-remove-one \\
    GS-WGAN~\cite{gs-wgam} &  Both \\
    DP-GAN~\cite{dpgan} &  Add-or-remove-one \\
    DPDM~\cite{dpdm} &  Add-or-remove-one \\
    DP-FETA~\cite{dp-feta} &  Add-or-remove-one \\
    PDP-Diffusion~\cite{dp-diffusion} &  Add-or-remove-one \\
    DP-LDM~\cite{dpldm} &  Add-or-remove-one \\
    DP-LoRA~\cite{dplora} &  Add-or-remove-one \\
    PrivImage~\cite{li2023PrivImage} &  Add-or-remove-one \\
    \bottomrule
\end{tabular}
}}
\vspace{-4mm}
\end{table}

\section{DP Image Synthesizer Details}
\label{supp:dp_synth_details}

\subsection{Implementations}
\label{supp:imple}

In all experiments, we used default hyper-parameter settings and made minimal modifications to public implementations whenever possible. PE utilizes discrete-step diffusion models, whereas DPDM, PDP-Diffusion, and PrivImage use continuous-step diffusion models. Discrete-step diffusion models proceed with fixed discrete time increments, whereas continuous-step diffusion models model diffusion as a continuous-time process using differential equations. We list the code bases of the investigated methods below. 

\begin{itemize}[leftmargin=*]
    \item DP-MERF: \url{https://github.com/ParkLabML/DP-MERF}
    \item DP-Kernel: \url{https://github.com/dihjiang/DP-kernel}
    \item DP-NTK: \url{https://github.com/Justinyangjy/DP-NTK}
    \item PE: \url{https://github.com/microsoft/DPSDA}
    \item GS-WGAN: \url{https://github.com/DingfanChen/GS-WGAN}
    \item DP-GAN: \url{https://github.com/illidanlab/dpgan}
    \item DPDM: \url{https://github.com/nv-tlabs/DPDM}
    \item DP-FETA: \url{https://github.com/SunnierLee/DP-Pretrain}
    \item DP-LDM: \url{https://github.com/SaiyueLyu/DP-LDM}
    \item DP-LoRA: \url{https://github.com/EzzzLi/DP-LORA}
    \item PrivImage: \url{https://github.com/SunnierLee/DP-ImaGen}
\end{itemize}

Noted that PDP-Diffusion did not release their source code. Instead, we refer to the implementation provided by the PrivImage repository, which uses PDP-Diffusion as a baseline. Additionally, the original DP-LDM repository implements a LDM. For a fair comparison, we use a standard diffusion model as the synthesizer to implement DP-LDM (SD), built upon the DPDM repository.

\subsection{Hyper-parameter Settings}
\label{supp:hyper}

\subsubsection{Investigated Synthesis Algorithms}\label{supp:hyper_algo} This section elaborates on the hyper-parameter settings for DP image synthesis methods. Table~\ref{tab:param_lr_train} presents the learning rate and batch size, which are all the same on different datasets for each method. Table~\ref{tab:param_iter_train} presents the number of iterations for finetuning on sensitive datasets. The number of iterations is: $(\text{epoch} \times \text{size of sensitive data}) / (\text{batch size})$.

\noindent \textbf{DP-MERF.} The dimension of random Fourier features is 10,000, and the sigma for these features is 105.

\noindent \textbf{DP-NTK.} For the NTK model, We use fully-connected 1-hidden-layer and 2-hidden-layer neural network with the width \{800\} and \{32, 256\} for RGB and gray scale images.

\noindent \textbf{DP-Kernel.} The sigma used to calculated the kernel is set as \{1, 2, 4, 8, 16\}.

\noindent \textbf{PE.} Refer to the practical adoption in the original paper~\cite{dpsda},  we use the pretrained model ``imagenet64\_cond\_270M\_250K.pt''  as the foundational API\footnote{\url{https://github.com/openai/improved-diffusion}}. For generating $128 \times 128$ images, we use the pretrained model ``128x128\_diffusion.pt''  as the foundational API.\footnote{\url{https://github.com/openai/guided-diffusion}} These models are only trained on {\tt ImageNet}. 
In principle, PE can use any model. For comparison fairness, in the model size experiments in Section \ref{subsec:rq2} and all experiments in Sections \ref{subsec:rq3} and \ref{subsec:combining_improvement}, we use the APIs from the PDP-Diffusion pretrained by us in the same experiments.

\begin{table}[!t]
\small
    \centering
    \caption{Learning rate and batch size for training models. 
    }
    \label{tab:param_lr_train}
    \setlength{\tabcolsep}{5.0mm}{
    \resizebox{0.40\textwidth}{!}{
    \begin{tabular}{l|cc}
    \toprule
    Method & Learning Rate & Batch Size\\
    \hline
    DP-MERF & $1 \times 10^{-2}$ & 100 \\
    DP-NTK & $1 \times 10^{-2}$ & 4,096\\
    DP-Kernel & $3 \times 10^{-4}$ & 60\\
    PE & - & -\\
    GS-WGAN & $1 \times 10^{-4}$ & 32 \\
    DP-GAN & $3 \times 10^{-4}$ & 4,096\\
    DPDM & $3 \times 10^{-4}$ & 4,096\\
    DP-FETA & $3 \times 10^{-4}$ & 4,096 \\
    PDP-Diffusion & $3 \times 10^{-4}$ & 4,096\\
    DP-LDM (SD) & $3 \times 10^{-4}$ & 4,096\\
    DP-LDM & $2 \times 10^{-3}$ & 4,096\\
    DP-LoRA & $2 \times 10^{-3}$ & 4,096\\
    PrivImage & $3 \times 10^{-4}$ & 4,096\\
    \bottomrule
\end{tabular}
}}
\vspace{-3mm}
\end{table}

\noindent \textbf{GS-WGAN.} 
We use 1,000 discriminators and 1 generator. The clip bound of the gradient is set as 2.0.

\noindent \textbf{DP-GAN.} We use BigGAN~\cite{biggan} as the synthesizer in our implementation of DP-GAN, resulting in improved synthesis performance compared to the original DP-GAN~\cite{dpgan}.

\noindent \textbf{DPDM.} We use continuous-time DMs following their code. We use the noise multiplicity proposed in their paper, and the multiplicity number is set as 32 for all datasets.

\noindent \textbf{DP-FETA.} We first pretrain diffusion models on the central images and then fine-tune the models on the sensitive dataset like DPDM. We query 10 central images for {\tt EuroSAT}, 50 central images for {\tt MNIST}, {\tt F-MNIST}, and {\tt CIFAR-10}, and 500 central images for {\tt Celeba} and {\tt Camelyon}.

\noindent \textbf{PDP-Diffusion.} We first pretrain diffusion models on the public dataset, and then fine-tune the models on the sensitive dataset like DPDM. 

\noindent \textbf{DP-LDM (SD).} The pretraining stage is the same as PDP-Diffusion. During the finetuning, we only set the label embedding layer and attention modules trainable instead of all the modules in the diffusion model. 

\noindent \textbf{DP-LDM.}  During pretraining, we first pretrain a Variational AutoEncoder (VAE) on {\tt ImageNet}, and then pretrain DM on the latent space of VAE. The finetuning stage is the same as DP-LDM (SD).

\begin{table}[!t]
\small
    \centering
    \caption{The number of iterations for training models. 
    }
    \label{tab:param_iter_train}
    \setlength{\tabcolsep}{2.0mm}{
    \resizebox{0.48\textwidth}{!}{
    \begin{tabular}{l|ccccc}
    \toprule
    \multirow{2}{*}{Method} & {\tt MNIST} \& & {\tt CIFAR-10} \& & \multirow{2}{*}{{\tt EuroSAT}} & \multirow{2}{*}{{\tt CelebA}} & \multirow{2}{*}{{\tt Camelyon}}\\
    & {\tt F-MNIST} & {\tt CIFAR-100} & & &\\
    \hline
    DP-MERF & 3,000 & 2,500 & 1,150 & 8,138 & 15,122 \\
    DP-NTK & 2,000 & 2,000 & 2,000 & 2,000 & 2,000\\
    DP-Kernel & 200,002 & 200,002 & 200,002 & 200,002 & 200,002\\
    PE & 8 & 8 & 8 & 8 & 8\\
    GS-WGAN & 2,000 & 2,000 & 2,000 & 2,000 & 2,000\\
    DP-GAN & 2,196 & 1,830 & 842 & 3,974 & 3,692\\
    DPDM & 2,196 & 1,830 & 842 & 3,974 & 3,692\\
    DP-FETA & 2,196 & 1,830 & 842 & 3,974 & 3,692\\
    PDP-Diffusion & 2,196 & 1,830 & 842 & 3,974 & 3,692\\
    DP-LDM (SD) & 2,196 & 1,830 & 842 & 3,974 & 3,692\\
    DP-LDM & 2,196 & 1,830 & 842 & 3,974 & 3,692\\
    DP-LoRA & 2,196 & 1,830 & 842 & 3,974 & 3,692\\
    PrivImage & 2,196 & 1,830 & 842 & 3,974 & 3,692\\
    \bottomrule
\end{tabular}
}}
\vspace{-1mm}
\end{table}

\noindent \textbf{DP-LoRA.} The pretraining stage is the same as DP-LDM. During finetuning, we use the LoRA~\cite{lora} technique to finetune the label embedding layer and attention modules of DM.

\noindent \textbf{PrivImage.} We implement the semantic query function with ResNet-50, and its training epoch is set as 500. We select 5\% public data for all sensitive datasets.

\subsubsection{Network Architecture}\label{supp:gen_network} For a fair comparison, we implement the GAN-based methods using the same network architecture of BigGAN generator~\cite{biggan} except for GS-WGAN for its poor performance. Specifically, the generator has a noise vector dimension of 60, a shared feature dimension of 128, and a convolutional feature dimension of 60. For DPDM, PDP-Diffusion, DP-LDM (SD), and PrivImage, we use the same UNet as the generator with a feature dimension of 32, a feature multiplicity of $[2,4]$, an attention resolution of $[16]$. For DP-LDM and DP-LoRA, we use the same UNet as the generator with a feature dimension of 32, a feature multiplicity of $[1,2,4]$, and an attention resolution of $[4,8,16]$. The autoencoder of VAE has a feature dimension of 128, a feature multiplicity of $[1,2]$, and an attention resolution of $[8,16]$.

\begin{figure*}[!t]
    \centering
    \hspace{-0.2cm}
    \subfigure[With and without pretraining.]{
        \includegraphics[width=3.44 in]{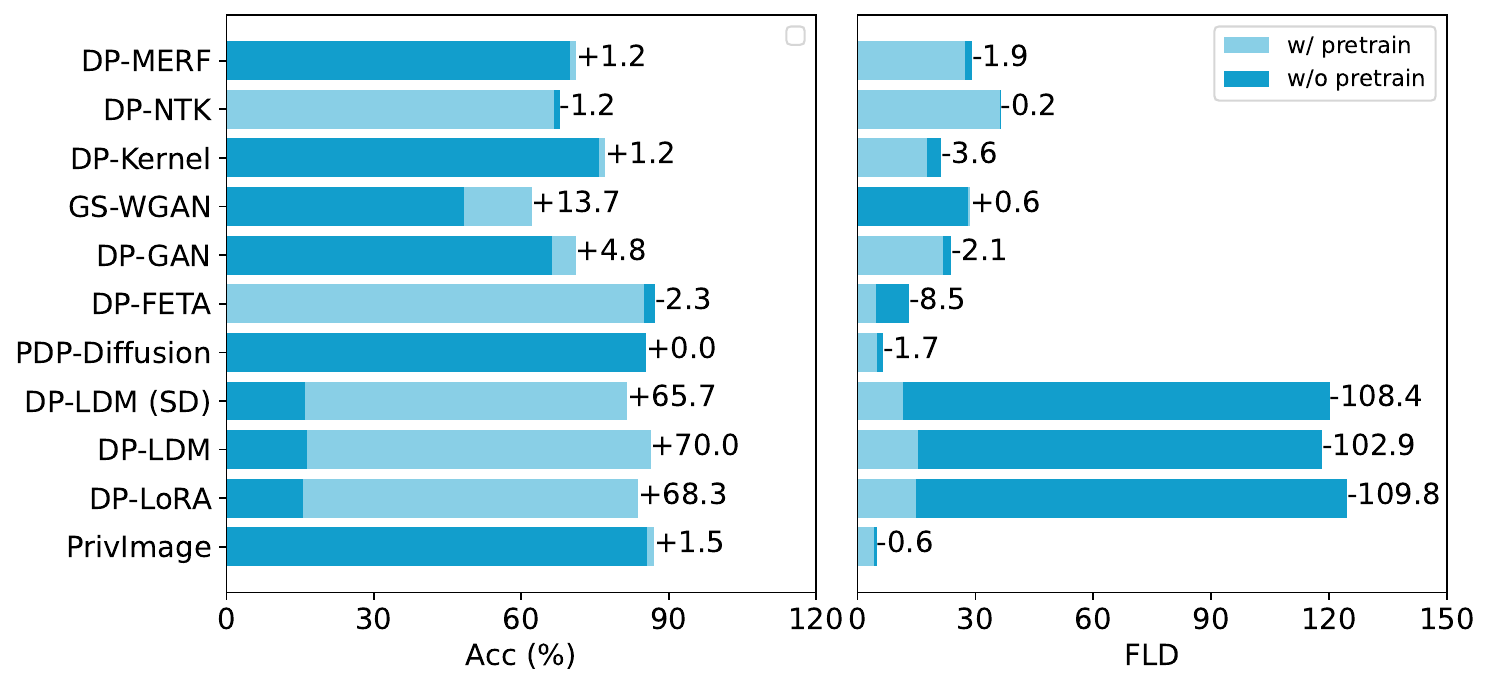}    
    }
    \hspace{-0.2cm}
    \subfigure[Conditional and unconditional pretraining.]{
        \includegraphics[width=3.44 in]{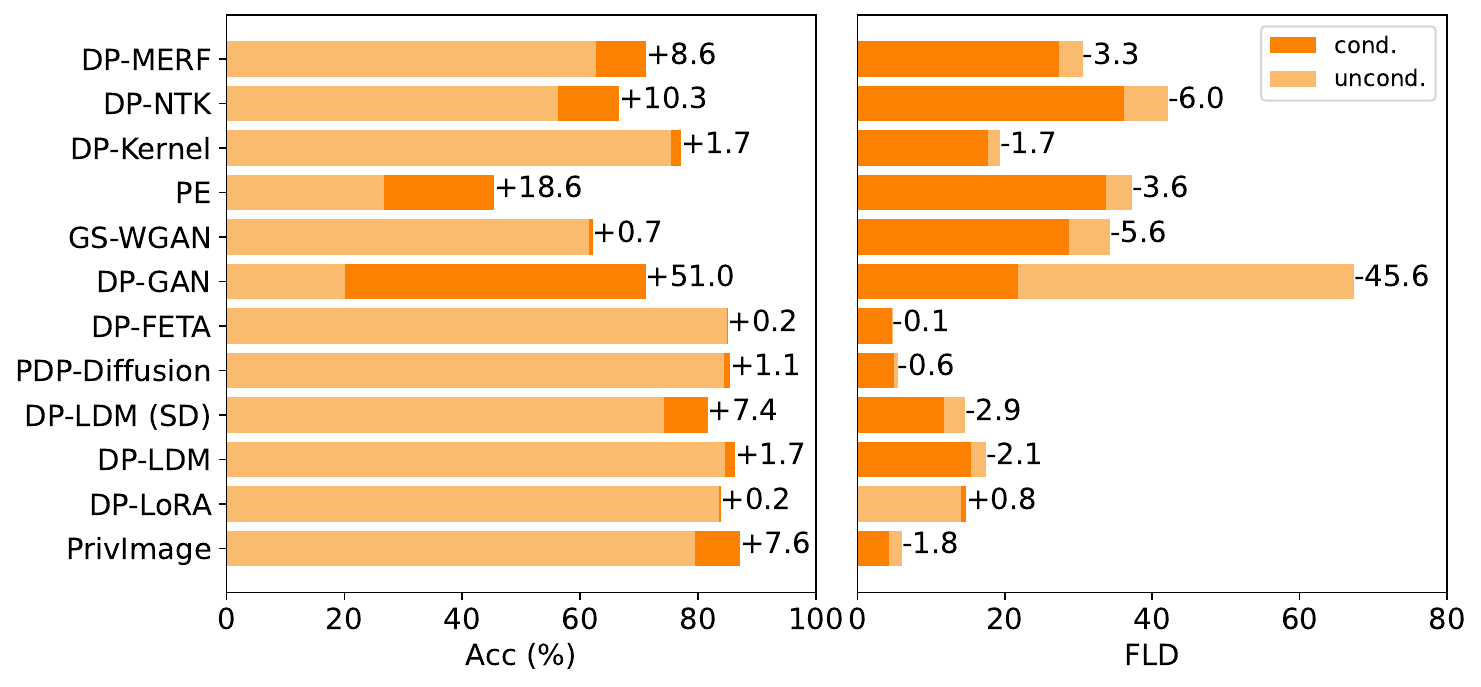}
    }
    \caption{The Acc and FLD evaluations of synthetic images under $\epsilon = 10$ and $\delta = 1/N \log(N)$ for the {\tt FashionMNIST} dataset. The labels `w/ pretrain' and `w/o pretrain' refer to the studied algorithms with and without pretraining, respectively. `cond.' and `uncond.' indicate the use of conditional and unconditional pretraining. The relative changes indicate the results of w/ pretrain' minus that of `w/o pretrain', and the result of `cond.' minus that of `uncond.'.}
    \label{fig:Fmnist-convsuncon}
\end{figure*}

\begin{table}[!t]
\small
    \centering
    \caption{The hyper-parameters for pretraining DP image synthesizers on {\tt ImageNet}. Since PrivImage only uses 5\% public images for pretraining, we multiply its epoch by 20 to achieve the same training iterations as PDP-Diffusion.}
    \label{tab:param_iter_pre}
    \setlength{\tabcolsep}{2.0mm}{
    \resizebox{0.48\textwidth}{!}{
    \begin{tabular}{l|cccc}
    \toprule
    Method & Iterations & Epoch & Learning Rate & Batch Size\\
    \hline
    DP-MERF & 12,811 & 1 & $1 \times 10^{-2}$ & 100\\
    DP-NTK & 4,000 & 16 & $1 \times 10^{-2}$ & 5,000\\
    DP-Kernel & 20,018 & 1 & $5 \times 10^{-5}$ & 64\\
    PE & - & - & - & -\\
    GS-WGAN & 40,036 & 1 & $1 \times 10^{-4}$ & 32\\
    DP-GAN & 200,182 & 160 & $3 \times 10^{-4}$ & 1,024\\
    DPDM & 200,182 & 160 & $3 \times 10^{-4}$ & 1,024\\
    DP-FETA & 2,000 & 2,000 & $3 \times 10^{-4}$ & 50 \\
    PDP-Diffusion & 200,182  & 160 & $3 \times 10^{-4}$ & 1,024\\
    DP-LDM (SD) & 200,182  & 160 & $3 \times 10^{-4}$ & 1,024\\
    DP-LDM & 200,182 & 160 & $2 \times 10^{-3}$ & 1,024\\
    DP-LoRA & 200,182 & 160 & $2 \times 10^{-3}$ & 1,024\\
    PrivImage & 200,182 & 3,200 & $3 \times 10^{-4}$ & 1,024\\
    \bottomrule
\end{tabular}
}}
\end{table}

\subsubsection{Pretraining}
\label{app:pretraining}
 There are two widely used pretraining paradigms: conditional and unconditional pretraining. We describe them as follows.
\begin{itemize}[leftmargin=*]
    \item Conditional: Given the number of sensitive categories, $N_{c}$, the public images must be labeled with one of these categories. For PrivImage, labeling is performed using a semantic query function. For other methods, we randomly assign one of the $N_{c}$ categories to the public data.
    \item Unconditional: The public data remains unlabeled, and we set the label input as a zero vector.
\end{itemize}

\begin{table*}[!t]
\small
    \centering
    \caption{Summary of investigated datasets used for evaluation. Images from the {\tt ImageNet} dataset have varying dimensions but are typically resized to 256$\times$256 for generation tasks.}
    \vspace{-2mm}
    \label{tab:datainfo}
    \setlength{\tabcolsep}{2.9mm}{
    \resizebox{1.0\textwidth}{!}{
    \begin{tabular}{l|lcccccc|c}
    \toprule
    Dataset & Size & Dimension & Category  & Training & Validation & Test & Task  & Effect  \\
    \hline
    {\tt MNIST}~\cite{mnist} & 70,000 & 28$\times$28$\times$1 & 10 & 55,000 & 5,000 & 10,000 & Classification  & \multirow{7}{*}{Sensitive}  \\
    {\tt FashionMNIST}~\cite{fmnist} & 70,000 & 28$\times$28$\times$1 & 10 & 55,000 & 5,000 & 10,000 & Classification  &  \\
    {\tt CIFAR-10}~\cite{cifar10} & 60,000 & 32$\times$32$\times$3 & 10 & 45,000 & 5,000 & 10,000 & Classification &  \\
    {\tt CIFAR-100}~\cite{cifar10} & 60,000 & 32$\times$32$\times$3 & 100 & 45,000 & 5,000 & 10,000  & Classification & \\
    {\tt EuroSAT}~\cite{eurosat} & 27,000 & 64$\times$64$\times$3 & 10 & 21,000 & 2,000 & 4,000 & Classification & \\
    {\tt CelebA}~\cite{celeba} & 182,732 & 128$\times$128$\times$3 & 2 & 145,064 & 17,706 & 19,962 & Classification & \\
    {\tt Camelyon}~\cite{camelyon1} & 337,340 & 64$\times$64$\times$3 & 2 & 269,538 & 32,898  & 34,904 & Classification & \\
    \hline
    {\tt ImageNet ISLVRC2012}~\cite{imagenet} & 1,431,167  & 256$\times$256$\times$3 & 1,000 & 1,281,167 & 50,000& 100,000 & Pretraining & \multirow{2}{*}{Public}  \\
    {\tt Places365}~\cite{places365} & 2,141,660  & 256$\times$256$\times$3 & 365 & 1,803,460 & 17,800 & 320,400 & Pretraining &   \\
    \bottomrule
\end{tabular}
}}
\vspace{-1mm}
\end{table*}

\noindent We base our implementations on the practical code available in the repositories provided by the original papers. In the default setting of \toolname{}, we use conditional pretraining. Table~\ref{tab:param_iter_pre} presents the training iterations used for pretraining DP synthesizers across different methods. For methods that do not use a public dataset, we use their default training parameters for pretraining. For methods using a public dataset, we pretrain these methods for 160 epochs with batch size 1,024. Therefore, the pretrain iterations are $1,281,167$ (number of {\tt ImageNet} images) $/ 1024 \approx 200,182$. An epoch refers to one complete pass through the entire dataset, while an iteration represents a single training step on a mini-batch. Consequently, for different datasets, it is expected to have a varying number of iterations per epoch due to differences in dataset size. Table~\ref{tab:param_iter_pre} presents the hyper-parameters used for pretraining.

\subsubsection{Classifier for Utility Evaluation}\label{supp:hyper_class} In utility evaluation, the hyper-parameter settings for the classifiers—ResNet, WideResNet (WRN), and ResNeXt—are adjusted depending on whether the dataset used is {\tt CIFAR} (including {\tt CIFAR-10} and {\tt CIFAR-100}) or another dataset. For the {\tt CIFAR} dataset, models are trained with a batch size of 126 for up to 200 epochs, utilizing the SGD optimizer with a learning rate of 0.1, momentum of 0.9, and weight decay of 5e-4. The learning rate is controlled using a StepLR scheduler with a step size of 60 and a gamma of 0.2. Specifically, WRN is configured with a depth of 28 and a widen factor of 10, ResNet uses a depth of 164 with BasicBlock modules, and ResNeXt uses a depth of 28, a cardinality of 8, and a widen factor of 10, each with a dropout rate of 0.3. For other datasets, the models are trained with a batch size of 256 for 50 epochs, using the Adam optimizer with a learning rate of 0.01. In these cases, the StepLR scheduler has a step size of 20 and a gamma of 0.2. WRN maintains a dropout rate of 0.3, with a depth of 40 and a widened factor of 4, while ResNet is adjusted to a depth of 110. Across all models, an exponential moving average with a decay rate of 0.9999 is used to enhance training stability.

\subsubsection{Metrics for Fidelity Evaluation}
\label{supp:hyper_fidelity}

We briefly outline the selected fidelity metrics as follows.

\noindent \textbf{FID:} FID~\cite{fid} first uses Inception-v3~\cite{inceptionv3} to extract features from synthetic and sensitive images. It estimates multivariate Gaussian distributions from these features and calculates the FID score by measuring the Fréchet distance~\cite{fid} between the two distributions. A lower FID indicates that the generated images are more similar to the real dataset.

\noindent \textbf{IS:} The IS uses the Inception v3~\cite{inceptionv3} network to classify each generated image into predefined classes. 
A higher IS indicates that the synthetic images are both realistic (the model classifies them with high confidence) and diverse (the images span various classes). 

\noindent \textbf{Precision and Recall:} We begin by extracting features from both sensitive and synthetic images using the Inception v3. Next, for each synthetic image, we identify the closest sensitive image in the feature space. If the distance between the synthetic image and the sensitive image falls below a specified threshold, we classify it as a ``true positive.'' Precision is then calculated as $\frac{\text{Number of True Positives}}{\text{Total Generated Images}}$. For recall, we locate the nearest synthetic image for each sensitive image. Here, a ``True Positive'' is defined as a pair where the distance between the images is below the threshold, and Recall is calculated as $\frac{\text{Number of True Positives}}{\text{Total Real Images}}$. Achieving higher Precision and Recall is preferable, although there is often a trade-off between them~\cite{precision&recall}.

\begin{table}[!t]
\small
    \centering
    \caption{
    Fidelity evaluations for the {\tt Camelyon} dataset.}
    \label{table:camelyon_fidelity}
    \setlength{\tabcolsep}{6.1pt}
    \resizebox{0.47\textwidth}{!}{
    \begin{tabular}{l|cccccc}
    \toprule
    \multirow{2}{*}{Algorithm} & \multicolumn{6}{c}{{\tt Camelyon}} \\
    \cline{2-7}
     & FID{\color{red}$\downarrow$} & IS{\color{blue}$\uparrow$} & Precision{\color{blue}$\uparrow$} & Recall{\color{blue}$\uparrow$} & FLD{\color{red}$\downarrow$} & IR{\color{blue}$\uparrow$} \\
    \hline
    DP-MERF& 251.6 & 2.45 & 0.12 & 0.00 & 50.5 & -2.16  \\
    DP-NTK & 234.5 & 1.67 & 0.00  & 0.00  & 52.7 & -2.25  \\
    DP-Kernel & 217.3 & 3.21 & 0.01 &  0.01 & 38.0 & -2.21  \\
    PE & 69.1 & \colorbox{gray0}{4.58} & 0.01 & \colorbox{gray0}{0.77}  & 13.6 &  -2.11 \\
    GS-WGAN & 291.8 & 1.45 & 0.04 & 0.01  & 64.2 & -2.27  \\
    DP-GAN & 66.9 & 1.84 & 0.37 &  0.14 & 3.01 & \colorbox{gray0}{-1.67}  \\
    DPDM & 29.2 & 1.65 & \colorbox{gray0}{0.74} & 0.29  & -2.71 & -1.87  \\
    DP-FETA & 27.8 & 1.69 & 0.72 & 0.31  & -4.4 & -1.84 \\
    PDP-Diffusion & 6.1 & 2.09 & 0.62 & 0.69 & \colorbox{gray0}{-5.9} & -1.82  \\
    DP-LDM (SD) & 15.3 & 2.03 & 0.46 & 0.68 & -4.22 & -1.92  \\
    DP-LDM & 45.4 & 1.88 & 0.32 & 0.43 & -0.21 & -1.91 \\
    DP-LoRA & 36.8 & 1.94  & 0.35  & 0.51  & -5.09  &  -1.91 \\
    PrivImage & 10.1 & 2.17 & 0.50 & 0.71 & -4.76 & -1.88  \\
    \bottomrule
\end{tabular}
}
\vspace{-3mm}
\end{table}

\noindent \textbf{FLD:} We refer to the repository released by Jiralerspong et al.~\cite{fld}. A lower FLD value is better, with negative values indicating the synthetic dataset outperforms the original.

\noindent \textbf{IR:} IR is the state-of-the-art method for assessing synthetic images from a human perspective. We use the image labels as the text description. A higher IR value indicates that the synthetic images align more with human preferences.

We refer to the open-source repository\footnote{\url{https://github.com/marcojira/FLD}} for implementing FID, IS, Precision, Recall, and FLD metrics. For the human-perspective metric, ImageReward, we refer to the implementation provided at the official repository\footnote{\url{https://github.com/THUDM/ImageReward}} of the original paper. Since ImageReward requires the prompt of each image and all the sensitive datasets we investigated do not contain the prompts, we manually design the prompt templates based on the category of the images. For {\tt MNIST}, the prompt is ``A grayscale image of a handwritten digit [label]''. For {\tt F-MNIST}, the prompt is ``A grayscale image of a/an  [label]''. For {\tt CIFAR-10} and {\tt CIFAR-100}, the prompt is ``An image of a/an  [label]''. For {\tt EuroSAT}, the prompt is ``A remote sensing image of a/an  [label]''. For {\tt CelebA}, the prompt is ``An image of a  [label] face''. For {\tt Camelyon}, the prompt is ``A normal lymph node image'' and ``A lymph node histopathology image''.  Notably, all experiments are conducted using the default hyper-parameters provided in the respective repositories.

\subsubsection{Privacy Cost for Hyper-Parameter Tuning}
\label{supp:privacy_cost_hyper}
Current DP image synthesis methods overlook the privacy cost associated with hyper-parameter tuning~\cite{dpsda,dp-diffusion,li2023PrivImage,dpdm}. \toolname{} follows them and similarly does not account for this cost. To mitigate the impact of neglecting privacy loss during hyper-parameter tuning, we should minimize the hyper-parameter tuning process as much as possible. Therefore, we follow the default hyper-parameters the same as that released in the repository as introduced in Section~\ref{supp:dp_synth_details}. Besides, different synthesis methods often use the same value for shared hyper-parameters when applied to the same synthesis task. The results could be further improved with optimized hyper-parameters beyond those we used.

\vspace{-1mm}
\section{Details Descriptions of Datasets}
\label{app:dataset}
We evaluate \toolname{} on seven widely studied sensitive datasets, including two gray and five colorful datasets: (1) {\tt MNIST}~\cite{mnist}, (2) {\tt FashionMNIST}~\cite{fmnist}, (3) {\tt CIFAR-10}~\cite{cifar10}, (4) {\tt CIFAR-100}~\cite{cifar10}, (5) {\tt EuroSAT}~\cite{eurosat}, (6) {\tt CelebA}~\cite{celeba},  and (7) {\tt Camelyon}~\cite{camelyon1}. Besides, the {\tt ImageNet ISLVRC2012}~\cite{imagenet} and {\tt Places365}~\cite{places365} datasets are considered a public dataset. This section introduces the investigated dataset as follows.

\begin{table}[!t]
\setlength{\tabcolsep}{4.5pt}
\small
\centering
\caption{GPU memory usage and runtime analysis of studied methods in \toolname{} for the {\tt CIFAR-10}.}
\setlength{\tabcolsep}{2.3mm}{
\resizebox{0.48\textwidth}{!}{
\begin{tabular}{l|l|ccc}
\toprule
\textbf{Algorithm} & \textbf{Stage} & \textbf{Memory} & \textbf{Runtime} & \textbf{Max Memory} \\
\midrule
\multirow{3}{*}{DP-MERF} & Pretrain & 0 GB & 0 H & \multirow{3}{*}{18.6 GB} \\
                         & Fine-tune & 3.5 GB & 0.02 H & \\
                         & Synthesis & 18.6 GB & 0.03 H & \\
\hline
\multirow{3}{*}{DP-NTK}  & Pretrain & 0 GB & 0 H & \multirow{3}{*}{21.3 GB} \\
                         & Fine-tune & 4.8 GB & 7.75 H & \\
                         & Synthesis & 19.1 GB & 0.05 H & \\
\hline
\multirow{3}{*}{DP-Kernel} & Pretrain & 0 GB & 0 H & \multirow{3}{*}{22.5 GB} \\
                           & Fine-tune & 8.8 GB & 3.6 H & \\
                           & Synthesis & 19.0 GB & 0.05 H & \\
\hline
\multirow{3}{*}{PE}        & Pretrain & 0 GB & 0 H & \multirow{3}{*}{183.0 GB} \\
                           & Fine-tune & 0 GB & 0 H & \\
                           & Synthesis & 183.0 GB & 12.0 H & \\
\hline
\multirow{3}{*}{GS-WGAN}   & Pretrain & 21.6 GB & 18.3 H & \multirow{3}{*}{21.6 GB} \\
                           & Fine-tune & 21.6 GB & 0.6 H & \\
                           & Synthesis & 20.0 GB & 0.01 H & \\
\hline
\multirow{3}{*}{DP-GAN}    & Pretrain & 0 GB & 0 H & \multirow{3}{*}{46.1 GB} \\
                           & Fine-tune & 46.1 GB & 3.25 H & \\
                           & Synthesis & 20.1 GB & 0.03 H & \\
\hline
\multirow{3}{*}{DPDM}      & Pretrain & 0 GB & 0 H & \multirow{3}{*}{96.3 GB}  \\
                           & Fine-tune & 96.3 GB & 20.5 H &  \\
                           & Synthesis & 32.1 GB & 0.17 H &  \\
\hline
\multirow{3}{*}{DP-FETA}      & Pretrain & 3.4 GB & 7.2 H &  \multirow{3}{*}{96.3 GB}\\
                           & Fine-tune & 96.3 GB & 20.5 H \\
                           & Synthesis & 32.1 GB & 0.17 H &  \\
\hline
\multirow{3}{*}{PDP-Diffusion} & Pretrain & 34.3 GB & 32.0 H & \multirow{3}{*}{96.3 GB} \\
                               & Fine-tune & 96.3 GB & 20.5 H &  \\
                               & Synthesis & 32.1 GB & 0.17 H &  \\
\hline
\multirow{3}{*}{DP-LDM (SD)}        & Pretrain &  34.3 GB & 32.0 H & \multirow{3}{*}{96.3 GB} \\
                               & Fine-tune & 60.0 GB & 9.1 H &  \\
                               & Synthesis & 32.1 GB & 0.17 H &  \\
\hline
\multirow{3}{*}{DP-LDM}        & Pretrain & 40.0 GB & 75.0 H & \multirow{3}{*}{40.0 GB} \\
                               & Fine-tune & 20.8 GB & 2.5 H &  \\
                               & Synthesis & 5.8 GB & 0.5 H &  \\
\hline
\multirow{3}{*}{DP-LORA}        & Pretrain & 40.0 GB & 75.0 H & \multirow{3}{*}{40.0 GB} \\
                               & Fine-tune & 35.6 GB & 20.5 H &  \\
                               & Synthesis & 5.8 GB & 0.5 H &  \\
\hline
\multirow{3}{*}{PrivImage}     & Pretrain &  34.3 GB & 32.0 H & \multirow{3}{*}{96.3 GB} \\
                               & Fine-tune & 96.3 GB & 20.5 H &  \\
                               & Synthesis & 32.1 GB & 0.17 H &  \\
\bottomrule
\end{tabular}}}
\label{tab:computationalResource}
\vspace{-3mm}
\end{table}

\begin{itemize}[leftmargin=*]
\item {\tt MNIST}: The MNIST (Modified National Institute of Standards and Technology) has 70,000 grayscale images of handwritten digits (0-9), with 60,000 images for training and 10,000 for testing. Each image is $28\times 28$ pixels. 
\item {\tt FashionMNIST}: This dataset contains 70,000 grayscale images of fashion items divided into 10 categories (including t-shirts, trousers, shoes, and bags). Each image is $28\times 28$ pixels in size, similar to the original MNIST dataset, but featuring clothing and accessories.
\item {\tt CIFAR-10}: CIFAR-10 contains 60,000 color images, each with a resolution of $32\times 32$ pixels, categorized into 10 different classes: airplanes, cars, birds, cats, deer, dogs, frogs, horses, ships, and trucks.
\item {\tt CIFAR-100}: It contains 60,000 color images, each with a resolution of $32\times 32$ pixels. It is organized into 100 classes. These classes are further grouped into 20 superclasses, each comprising several related subclasses (the superclasses ``animals'' include subclasses like ``beaver.'').
\item {\tt EuroSAT}: This dataset comprises diverse satellite images optimized for land use and land cover classification tasks. EuroSAT contains 10 distinct classes, including forests, rivers, and urban environments. EuroSAT features over 27,000 labeled and georeferenced images, each measuring $64 \times 64$ pixels, capturing a broad spectrum of geographical and environmental scenarios across Europe.
\item {\tt CelebA}: The CelebA (CelebFaces Attributes) dataset contains over 200,000 color celebrity images, each with a resolution of $128\times 128$ pixels and annotated with 40 binary attributes, including features such as ``smiling,'' ``young,'' ``wearing glasses,'' ``blond hair,'' and more.
\item {\tt Camelyon}: The Camelyon dataset is a large-scale medical imaging resource aimed at advancing research in digital pathology, particularly for the detection of metastatic cancer in lymph node tissue sections. It consists of high-resolution histopathological images, known as whole-slide images, of lymph node samples. Each image in the Camelyon dataset is meticulously annotated by expert pathologists, marking the regions where cancer is present.
\item {\tt ImageNet ISLVRC2012}: ImageNet Large Scale Visual Recognition Challenge (ISLVRC) 2012 contains over 14 million images labeled across 1,000 object categories, including animals, vehicles, and others. Each image has a resolution of 256x256 pixels. We use a subset of the whole ImageNet dataset for pretraining like previous works~\cite{dp-diffusion,li2023PrivImage}. This subset of the dataset is specifically used for the ImageNet Large Scale Visual Recognition Challenge.
\item {\tt Places365}: The Places365 dataset is a large-scale scene recognition dataset designed to support visual recognition research. It contains over 1.8 million images categorized into 365 scene classes, such as beaches, kitchens, and libraries, providing a diverse collection of places encountered in everyday environments.
\end{itemize}

\section{Additional Experiments Analysis}
\label{app:add_exp_analysis}

\begin{figure*}[!t]
\vspace{-0.5mm}
    \centering
    \setlength{\abovecaptionskip}{0pt}
    \includegraphics[width=1.0\linewidth]{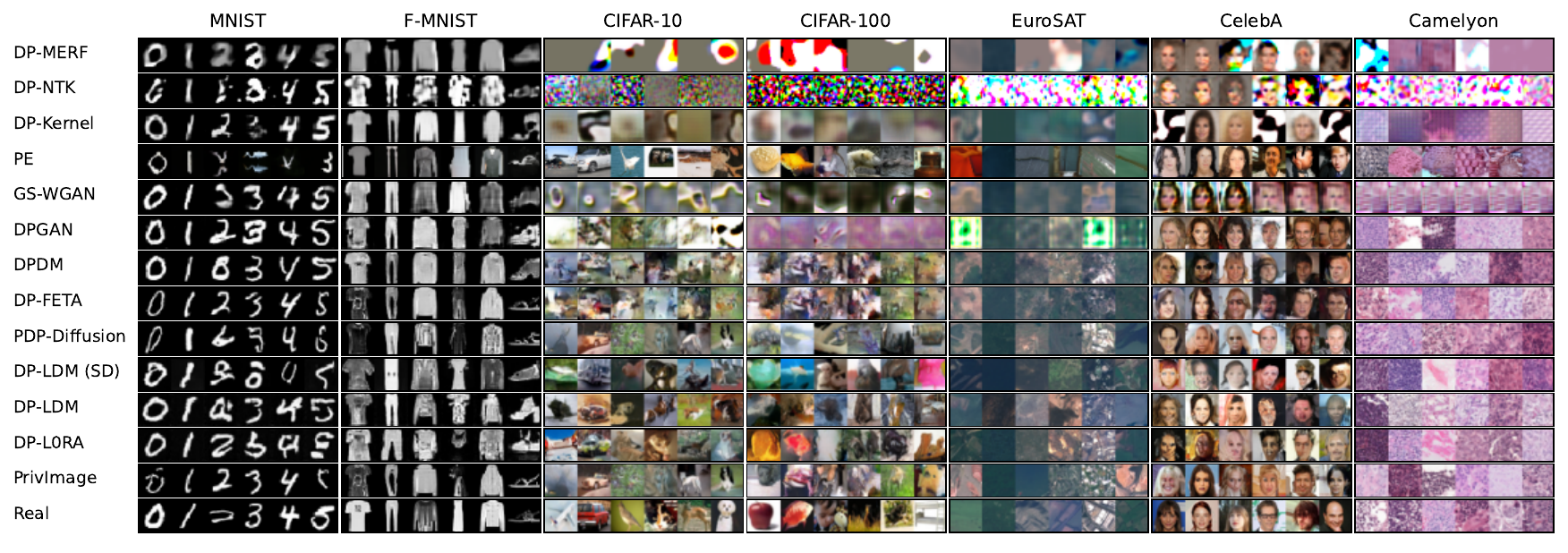}
    \caption{The examples of synthetic images under $\epsilon = 10$ and $\delta = 1 / {N \log(N)}$. The last row of images are real image samples from each sensitive image dataset.}
    \label{fig:eps10_visual}
\end{figure*}

\subsection{Fidelity Evaluations of {\tt Camelyon} Dataset}
\label{suppsubsec:camelyon}
Table~\ref{table:camelyon_fidelity} shows the fidelity results for various methods on the {\tt Camelyon} dataset. PE achieves the highest IS score of 4.58 among the studied methods. However, the downstream task accuracy achieved by PE is only 64.9\%. These results indicate that although PE can generate images with high practical quality, they are not similar to sensitive images. The precision and recall values of 0.01 and 0.77 further support this perspective. Besides, the FLD of synthetic images achieved by PDP-Diffusion and PrivImage are both negative, i.e., -5.9 and -5.09, suggesting that, according to this metric, the quality of the synthetic images surpasses that of the original images. Additionally, the accuracy achieved by PDP-Diffusion and PrivImage is 84.8\% and 87.0\%, respectively, which closely approaches the accuracy obtained on the original dataset, i.e., 87.7\%.

\subsection{Computational Resources}
\label{suppsub:computation_resources}
All methods are implemented on a server equipped with four NVIDIA GeForce A6000 Ada GPUs and 512GB of memory. The synthesizer sizes for the GAN-based and diffusion-based methods are 5.6M and 3.8M. For GAN, the generator size is 3.8M, and
the discriminator size is 1.8M. Consistently shared hyper-parameters are used across all methods, as outlined in Appendix~\ref{supp:hyper_algo}.

\begin{table}[!t]
\setlength{\tabcolsep}{4.5pt}
\small
    \centering
    \caption{The Acc of synthetic images generated by PrivImage under $\epsilon = 10$, using the sensitive dataset {\tt CIFAR-10}.}
    \vspace{-3mm}
    \setlength{\tabcolsep}{2.3mm}{
    \resizebox{0.48\textwidth}{!}{
    \begin{tabular}{c|cccc}
    \toprule
    \textbf{Batch size} & $M=3.8$M & $M=11.1$M & $M=19.6$M & $M=44.2$M  \\
    \midrule
    4,096 & 78.4 & 78.2 & 76.0 & 79.5  \\
    16,384 & 78.5 & 78.6 & 78.8 & 79.5 \\
    \bottomrule
\end{tabular}}}
\label{tab:batchsize}
\vspace{-2mm}
\end{table}

\begin{figure*}[!t]
    \centering
    \setlength{\abovecaptionskip}{0pt}
    \includegraphics[width=\textwidth]{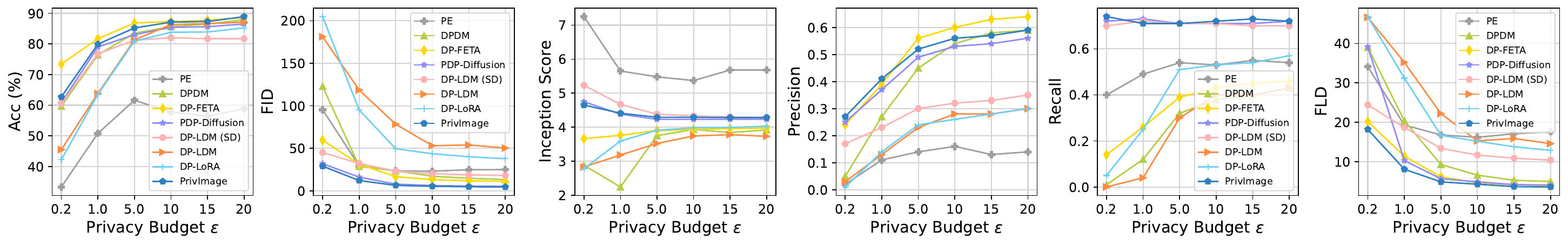}
    \caption{The utility and fidelity of synthetic images generated by the eight diffusion-based investigated methods for {\tt FashionMNIST} dataset under privacy budgets $\epsilon = \{0.2, 1.0, 5.0, 10.0, 15.0, 20.0\}$ and a fixed $\delta = 1/N \log(N)$.}
    \label{fig:change_eps_diffusion}
    \vspace{-1mm}
\end{figure*}

Table~\ref{tab:computationalResource} provides an analysis of GPU memory usage and runtime for the methods studied in \toolname{} using the {\tt CIFAR-10} dataset. Overall, the diffusion-based synthesizer requires more GPU resources and has a longer runtime compared to the GAN-based methods. In fact, the computation resource cost heavily depends on multiple factors, like the size of the synthesizer parameters, the batch size, etc. We recommend using the largest possible batch size for finetuning the synthesizers, while it needs more computation resources. However, with the same privacy budget, using a larger batch size results in a smaller scale of DP noise being added during finetuning for each individual image~\cite{dpbook}, which can lead to improved synthesis results.

\subsection{Visualize Results}
\label{suppsubsec:visualize}
Figure~\ref{fig:eps10_visual} illustrates the examples of synthetic images under $\epsilon=10$.

\subsection{Impact on Batch Size of DP-SGD}
\label{suppsubsec:batch}
Table~\ref{tab:batchsize} presents the accuracy (Acc) of synthetic images generated by PrivImage under a privacy budget of $\epsilon=10$, using the sensitive dataset {\tt CIFAR-10}. The table examines the effect of batch sizes—4,096 and 16,384 (the latter being the default batch size in PrivImage~\cite{li2023PrivImage})—across four synthesizer model sizes of 3.8M, 11.1M, 19.6M, and 44.2M parameters. For the batch size of 4,096, accuracy varies from 76.0\% (at 19.6M) to 79.5\% (at $M=$44.2M), indicating a slight upward trend as model size increases, though a dip occurs at 19.6M. In contrast, with the larger batch size of 16,384, accuracy consistently improves across model sizes, ranging from 78.5\% (at $M=$3.8M) to 79.5\% (at $M=$44.2M), with a peak of 78.8\% at $M=$19.6M. These findings suggest that larger batch sizes in DP-SGD significantly enhance performance, particularly for larger model sizes, likely because they reduce the noise scale in gradient updates, leading to more stable training and better convergence under differential privacy constraints.

\begin{figure}[!t]
    \centering
    \setlength{\abovecaptionskip}{0pt}
    \includegraphics[width=0.98\linewidth]{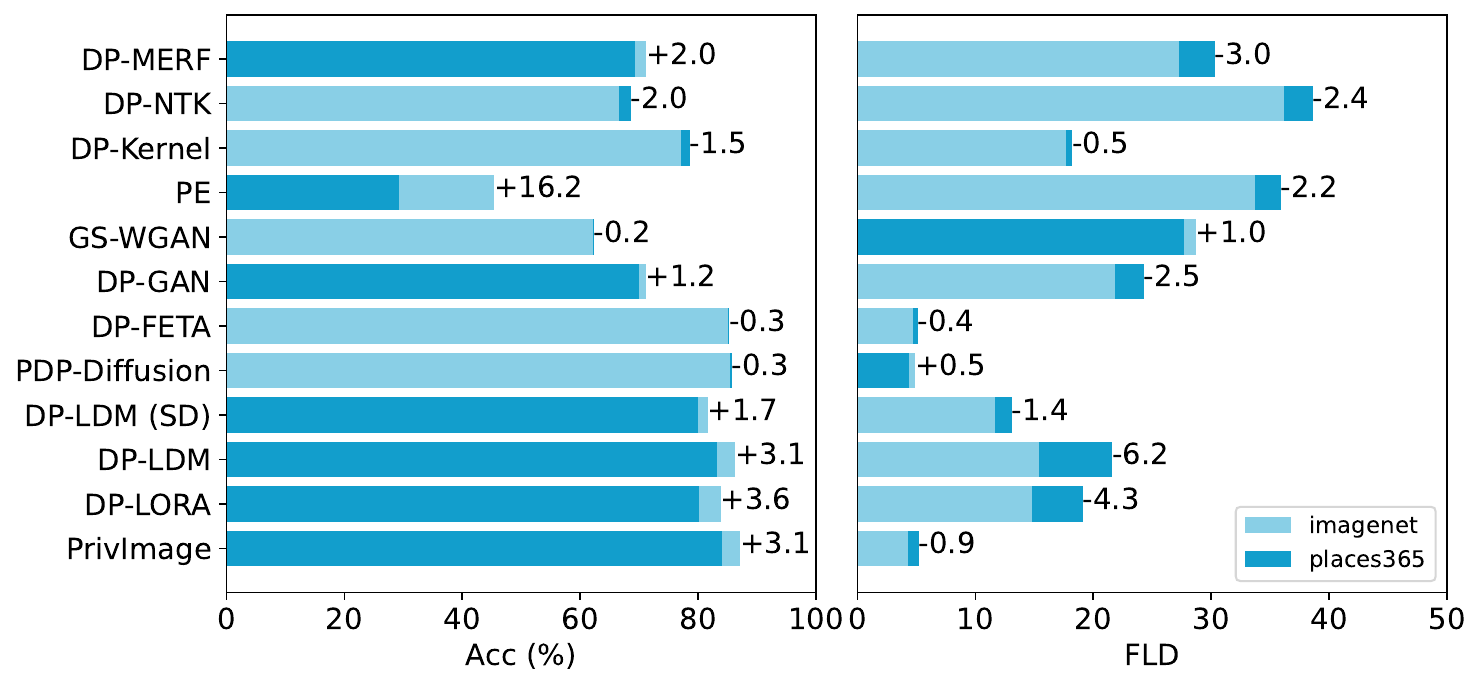}
    \caption{The Acc and FLD evaluations of synthetic images under $\epsilon = 10$ and $\delta = 1 / N \log(N)$ for the sensitive dataset {\tt FashionMNIST}, with public datasets {\tt ImageNet} and {\tt Places365}.}
    \label{fig:chang_public_images_fashion}
    \vspace{-2mm}
\end{figure}

\subsection{The Impact of Pretraining}
\label{suppsubsec:pretraining}
Figure~\ref{fig:Fmnist-convsuncon} presents the Acc and FLD evaluations of synthetic images under $\epsilon = 10$ for the {\tt FashionMNIST} dataset.

Table~\ref{tab:fine_tuning_pre_training_steps} presents the accuracy (Acc) and Fréchet Latent Distance (FLD) of synthetic images generated by PDP-Diffusion and PrivImage under a privacy budget of $\epsilon = 10$. The sensitive dataset {\tt MNIST} and public dataset {\tt ImageNet} are used, with pretraining iterations ranging from 25K to 400K. Results show that both methods achieve high Acc (above 97\%), with PrivImage generally exhibiting lower FLD values, indicating better distributional similarity to the sensitive data across varying pretraining steps.

Figure~\ref{fig:chang_public_images_fashion} compares the Acc and FLD for the sensitive datasets {\tt FashionMNIST} when pretraining with public datasets {\tt ImageNet} and {\tt Places365}. For {\tt FashionMNIST}, the average accuracy scores with {\tt ImageNet} and {\tt Places365} are 77.2 and 76.1. 
Pretraining on {\tt ImageNet} consistently outperforms {\tt Places365}, except in the case of PDP-Diffusion trained on {\tt FashionMNIST}, where the difference is marginal (0.3\%).

\begin{table}[!t]
\setlength{\tabcolsep}{4.5pt}
\small
    \centering
    \caption{The Acc and FLD of synthetic images generated by PDP-Diffusion and PrivImage under $\epsilon = 10$, using the sensitive dataset {\tt MNIST} and public dataset {\tt ImageNet} with different pretraining iterations.}
    \vspace{-1mm}
    \setlength{\tabcolsep}{3.4mm}{
    \resizebox{0.47\textwidth}{!}{
    \begin{tabular}{l|ccccc}
    \toprule
    PDP-Diffusion & 25K & 50K & 100K & 200K & 400K \\
    \midrule
    Acc (\%) & 97.4 & 97.6 & 97.4 & 97.4 &  97.1 \\
    FLD & 2.48 & 2.56 & 2.83 & 3.40 & 3.00 \\
    \toprule
    PrivImage & 25K & 50K & 100K & 200K & 400K \\
    \midrule
    Acc (\%) & 97.8 & 97.4 & 97.5 & 97.8 & 97.2  \\
    FLD & 2.23 & 2.48 & 2.62 & 2.80 & 2.90  \\
    \bottomrule
\end{tabular}}}
\label{tab:fine_tuning_pre_training_steps}
\vspace{-1mm}
\end{table}

\subsection{Combining DP-FETA with other methods}
\label{suppsubsec:combining_feta}

We incorporate a shortcut process into the DP training, extending it to other methods. Specifically, we extract central images from sensitive datasets and train synthesizers on these images prior to training on the sensitive images. Central images are perturbed with Gaussian noise for DP. PE is a fine-tuning-free method, so we exclude it from this analysis. Table~\ref{tab:w/o_dpfeta} shows the results of integrating the DP training shortcut into other methods.

\begin{table}[!t]
\small
    \centering
    \caption{\gc{The Acc and FLD of synthetic images using {\tt FashionMNIST} ($\epsilon=10$). These methods integrate shortcuts for DP training proposed in DP-FETA.}}    
    \label{tab:w/o_dpfeta}
    \setlength{\tabcolsep}{3.0mm}{
    \resizebox{0.48\textwidth}{!}{
    \begin{tabular}{l|cc|cc}
    \toprule
    \multirow{1}{*}{Algorithm} & \multicolumn{2}{c|}{With DP Training Shortcuts} & \multicolumn{2}{c}{Original}   \\
    \cline{2-5}
     ({\tt CIFAR-10}) & Acc{\color{red}$\uparrow$} (\%)  & FLD{\color{blue}$\downarrow$} & Acc{\color{red}$\uparrow$} (\%) & FLD{\color{blue}$\downarrow$}   \\
    \hline
    DP-MERF & 57.7  &  24.5 & 62.2  & 29.2  \\
    DP-NTK & 70.6 & 31.0 & 76.3 & 36.4 \\
    DP-Kernel & 78.0 & 18.5 & 70.0 & 21.3 \\
    GS-WGAN & 58.9 & 27.2 & 56.7 & 28.1 \\
    DP-GAN & 55.4 & 22.8  & 70.3 & 23.9 \\
    DPDM   & \textbf{87.3}  & \underline{4.6} & 85.6 & 6.6   \\
    \hline
    PDP-Diffusion & 85.0 & 4.7  & 85.4 & \underline{4.9} \\
    DP-LDM (SD)  & 80.6 & 13.3  & 81.6 & 11.7 \\
    DP-LDM  & 85.3 & 18.8  & \underline{86.3} & 15.4  \\
    DP-LoRA  & 86.3 & 14.2  & 83.8 & 14.8  \\
    PrivImage & \underline{86.4} & \textbf{4.6}  &  \textbf{87.1} & \textbf{4.3} \\
    \hline
    \rowcolor{gray0} \textbf{Average} & 75.6 & 16.7 &  76.8 & 17.9 \\
    \bottomrule
\end{tabular}
}}
\vspace{-1mm}
\end{table}

\noindent \gc{\textbf{Building DP training shortcut may damage the synthetic performance.} Table~\ref{tab:w/o_dpfeta} shows that incorporating a DP training shortcut may impair the synthetic performance of methods with pretraining. For instance, the Acc and FLD of DP-LDM (SD) decline from 81.6\% and 11.7 to 80.6\% and 13.3 when shortcuts are added. This is because the shortcut includes training on the central images after pretraining, which may damage the pretraining synthesizer. For methods without pretraining, the fidelity of synthetic images (measured by FLD) is consistently improved, while the improvement of utility (measured by Acc) fluctuates. These results show that ``$1 + 1 < 2$,'' indicating that a straightforward algorithmic combination does not always improve synthetic performance, and highlighting the need for further research into effectively integrating the strengths of diverse algorithms.}

\begin{table}[!t]
\setlength{\tabcolsep}{4.5pt}
\small
    \centering
    \caption{The FLD of synthetic {\tt CIFAR-10} images under $\epsilon = 10$, using the different subsets of {\tt ImageNet}. `5\%' means matching categories from {\tt CIFAR10} dataset in {\tt ImageNet} (50 out of the 1000 available categories) using PrivImage. `5\% rand.' refers to 5\% categories randomly selected. In each table, the first four methods above the horizontal line are GAN-based, while the remaining methods are diffusion-based. }
    \resizebox{0.48\textwidth}{!}{
    \begin{tabular}{l|cc|cc|c}
    \toprule
    \multirow{1}{*}{\textbf{FLD}} & 5\% & 5\% (rand.) & 5\%+5\% (rand.) & 10\% (rand.) & Whole \\
    \midrule
    DP-MERF & 30.2 & 31.6 & 29.8 & 29.7 & 28.4 \\
    DP-NTK & 50.0 & 50.5 & 55.9 & 53.6 & 49.8 \\
    DP-Kernel & 31.5 & 30.1 &  30.3 & 31.1 & 30.0 \\
    DP-GAN & 18.9 & 18.9 & 21.8 & 21.6 & 29.5 \\
    \hline
    PE & 4.5 & 6.5 & 12.2 & 10.6 & 8.5 \\
    DP-FETA & 10.7 & 12.4 & 9.3 & 12.8 & 10.6 \\
    PDP-Diffusion & 4.7 & 10.7 & 6.1 & 9.3 & 7.2 \\
    DP-LDM (SD) & 5.1 & 11.7 & 7.2 & 10.6 & 9.0  \\
    DP-LDM & 12.2 & 15.2 & 13.0 &  14.4 &  14.1 \\
    DP-LoRA & 8.1 & 11.5 & 8.0 & 11.4 &  9.3 \\
    \bottomrule
\end{tabular}}
\label{tab:select_public_dataset_whole}
\end{table}

\section{Error in DP-Promise}
\label{app:dppromise}

DP-Promise~\cite{dppromise} proposes reducing the noise introduced by DP during training for highly efficient training. We argue that \textit{the DP-Promise does not strictly satisfy DP}. In DMs, as introduced in Section~\ref{subsec:generative}, the image is gradually corrupted with noise until it becomes random noise, while a neural network is trained to reverse this process and recover the original image. DP-Promise proposes achieving DP by utilizing the noisy process in DM. Specifically, in the forward process of DMs, we add noise to the image $x_0$ through ${x_t} = \sqrt {{{\bar \alpha }_t}} {x_0} + e\sqrt {1 - {{\bar \alpha }_t}}$, where ${\bar \alpha }_t$\footnote{${\bar \alpha_t}$ is required for both training DMs and generating images.} is a fixed hyper-parameter, $e$ indicates the added noise, and $x_t$ is the noisy image. They theoretically prove that this process naturally satisfies DP with $\epsilon$ as the DP noise, and its privacy cost decreases as $t$ increases. Thus, when $t$ is sufficiently large, we do not need to introduce additional noise to the gradient. However, the analysis of DP-Promise overlooks the fact that, in addition to $x_t$, the training also has access to the noise $e$ (DP noise should be unknown) and the time step $t$ to calculate the objective function~\ref{eq:L_DM}. If $e$ and $t$ are not added with noise, we can easily reconstruct the original image $x_0$ via $x_0 = (x_t - e\sqrt{1 - {\bar \alpha_t}})/\sqrt{{\bar \alpha_t}}$ (by inverting Eq~(\ref{eq:dm:forward})). 
Therefore, we believe that DP-Promise violates DP, and we do not implement DP-Promise in \toolname{}.

\end{document}